\documentclass[prd, aps, nofootinbib, preprintnumbers, showpacs, superscriptaddress,twocolumn]{revtex4}

\usepackage{latexsym}
\usepackage{amssymb}
\usepackage{amsfonts}
\usepackage{amsmath}
\usepackage{bm}
\usepackage[dvips]{graphicx}

\def\de{\partial}
\def\lm{{\ell m}}
\def\e{({\rm e})}

\def\l{{\ell }}

\def\a{{\bar{a}_{\rm RR}}}
\def\vp{v_{\rm pole}}

\def\i{{\rm i}}

\def\F{{\cal F}}

\begin{document}

\title{Accurate Effective-One-Body waveforms of\\
       inspiralling and coalescing black-hole binaries}

\author{Thibault \surname{Damour}}
\affiliation{Institut des Hautes Etudes 
Scientifiques, 91440 Bures-sur-Yvette, France}
\affiliation{ICRANet, 65122 Pescara, Italy} 
\author{Alessandro \surname{Nagar}}
\affiliation{Institut des Hautes Etudes 
Scientifiques, 91440 Bures-sur-Yvette, France}
\affiliation{ICRANet, 65122 Pescara, Italy}
\affiliation{INFN, sez. di Torino, Via P.~Giuria 1, Torino, Italy}
\author{Mark \surname{Hannam}}
\affiliation{Theoretical Physics Institute, University of Jena, 07743, Jena,Germany}
\affiliation{Physics Department, University College Cork, Cork, Ireland}
\author{Sascha \surname{Husa}}
\affiliation{Theoretical Physics Institute, University of Jena, 07743, Jena,Germany}
\affiliation{Max-Planck-Institut f\"ur Gravitationsphysik,
  Albert-Einstein-Institut, Potsdam-Golm, Germany}
\author{Bernd \surname{Br\"ugmann}}
\affiliation{Theoretical Physics Institute, University of Jena, 07743, Jena,Germany}

\begin{abstract}
The Effective-One-Body (EOB) formalism contains several flexibility
parameters, notably $a_5$, $\vp$ and $\a$. We show here how to jointly
constrain the  values of these parameters by simultaneously best-fitting 
the EOB waveform to two, independent, numerical relativity (NR) simulations of
inspiralling and/or coalescing binary black hole systems: published 
Caltech-Cornell {\it inspiral} data (considered for gravitational wave frequencies
$M\omega\leq 0.1$) on one side, and newly computed 
{\it coalescence} data  on the other side. The resulting, approximately
unique, ``best-fit'' EOB waveform is then shown to exhibit 
excellent agreement with NR coalescence data for several mass ratios.
The dephasing between this best-fit EOB waveform and published
Caltech-Cornell inspiral data is found to vary between
$-0.0014$ and $+0.0008$ radians over 
a time span of $\sim 2464M$ up to gravitational wave 
frequency $M\omega= 0.1$, and between $+0.0013$ and 
$-0.0185$ over a time span of $96M$ after $M\omega=0.1$
up to $M\omega=0.1565$.
The dephasings between EOB and 
the new coalescence data are found to be smaller 
than: (i) $\pm 0.025$ radians over a time span of $730M$ (11 cycles) up to
merger, in the equal mass case, 
and (ii) $\pm 0.05$ radians over a time span of about $950M$ (17 cycles) 
up to merger in the 2:1 mass-ratio case.
These new results corroborate the aptitude of the EOB formalism to provide
accurate representations of general relativistic waveforms, which are
needed by currently operating gravitational wave detectors.
\end{abstract}

\date{\today}


\maketitle

\section{Introduction}
\label{intro}
The Effective-One-Body (EOB) 
formalism~\cite{Buonanno:1998gg,Buonanno:2000ef,Damour:2000we,Damour:2001tu}
is an analytical approach which aims at accurately describing both the motion 
of, and the radiation from, coalescing binary black holes. The EOB method
uses, as basic input, high-order post-Newtonian (PN) expanded results
(see~\cite{Blanchet:2002av} for a review of the PN-theory of gravitationally
radiating systems). However, one of the key ideas in the EOB method is to
avoid using PN results in their original `Taylor-expanded' form 
(symbolically $f^{\rm Taylor}(v/c) = c_0 + c_1 v/c + c_2 (v/c)^2+\cdots + c_n
(v/c)^n$), but, instead, to `re-package' them in some {\it resummed} form,
i.e., symbolically, to replace $f^{\rm Taylor}(v/c)$ by some non-polynomial
function $f^{\rm EOB}(v/c)$, defined so as to incorporate some of the expected
non-perturbative features of the (unknown) result. This re-packaging is crucial
for being able to bypass the strong limitations of PN results. Indeed, by
itself PN theory is unable to go beyond the (adiabatic) 
{\it early inspiralling} stage of black hole coalescence,\footnote{See Appendix~\ref{sec:T4} for 
a new confirmation of this fact}
while the EOB method is able to describe, in a continued manner, the full coalescence process:
adiabatic early inspiral, nonadiabatic late inspiral, plunge, merger and
ring-down. The EOB method comprises three, rather separate, parts:\\
1. a description of the conservative (Hamiltonian) piece of the dynamics of
two black holes;\\
2. an expression for the radiation-reaction force $\F_\varphi$ that supplements the
 Hamiltonian dynamics;\\
3. a description of the gravitational wave (GW) signal emitted by a
  coalescing binary system.

For each one of these parts, the EOB method uses special resummation techniques,
inspired by specific results going beyond perturbation theory. For instance,
the resummation of the EOB Hamiltonian (part 1.) was inspired by a specific
resummation of ladder diagrams used to describe positronium energy states in
Quantum Electrodynamics~\cite{Brezin:1970zr}.
The resummation of the radiation reaction force $\F_\varphi$ was inspired by the 
Pad\'e resummation of the flux function introduced in
Ref.~\cite{Damour:1997ub}. As for part 3., i.e. the EOB description of the
gravitational radiation emitted by a coalescing black hole binary, it was mainly
inspired by the classic work of Davis, Ruffini and Tiomno~\cite{Davis:1972ud},
which discovered the transition between the plunge signal and a ringing tail
when a particle falls into a Schwarzschild black hole.

Before the availability of reliable numerical simulations, the EOB method made
several quantitative and qualitative predictions concerning the dynamics of
the coalescence, and the corresponding GW radiation, notably: (i) a blurred
transition from inspiral to a `plunge' that is just a smooth continuation of
the inspiral, (ii) a sharp transition, around the merger of the black holes,
between a continued inspiral and the ring-down signal, and (iii) estimates of the
radiated energy, and of the spin of the final black hole (the latter estimates
were made both for nonspinning binaries~\cite{Buonanno:2000ef} and for
spinning ones~\cite{Buonanno:2005xu}). Those predictions have been broadly
confirmed by the results of recent numerical simulations performed by several
independent groups  (for a review of numerical relativity results 
see~\cite{Pretorius:2007nq}). The recent breakthroughs in numerical relativity
(NR)~\cite{Pretorius:2005gq,Campanelli:2005dd,Baker05a,
Gonzalez:2006md,Koppitz:2007ev} open the possibility of acquiring some knowledge about binary black hole
coalescence that goes beyond what either PN  theory, or its resummed avatars
(such as the EOB), can tell us. Actually, it was emphasized early 
on~\cite{Damour:2001tu,Damour:2002qh,Damour:2002vi} that the EOB method should
be considered as a {\it flexible} structure, containing several parameters
representing (yet) uncalculated results, that would need NR results (or real
observational data!) to be determined. For instance, Refs.~\cite{Damour:2001tu,Damour:2002qh} 
introduced a parameter (here denoted as $a_5$) representing uncalculated 4~PN,
and higher, contributions to the crucial EOB ``radial potential'' $A(R)$. 
Ref.~\cite{Damour:2002vi} introduced several more EOB 
{\it flexibility  parameters}, notably $\vp$ (entering the Pad\'e resummation
of the radiation reaction force) and a parameter (here replaced by $\a$)
describing uncalculated non quasi-circular (NQC) contributions to the
radiation reaction. Recently, Ref.~\cite{Damour:2007xr} augmented the list
of EOB flexibility parameters by introducing two parameters (here denoted as
$a$ and $b$) representing NQC contributions to the waveform, as well as two
parameters,  $t_m$ and $\delta$ (together with the choice of an integer $p$),
describing the ``comb'' used in matching the inspiralling and plunging
waveform to the ring-down one. Each one of these EOB flexibility parameters
($a_5$, $\vp$, $\a$, $a$, $b$, $t_m$, $\delta$, $p$) parametrizes a 
{\it  deformation}~\footnote{We use here the word {\it deformation} in the
mathematical sense. Ideally we would like the list of EOB flexibility
parameters to describe a kind of {\it versal deformation} of the original
EOB, i.e. a multi-parameter family which is general enough to encompass all
the physics contained in real GW coalescence signals, starting from the
originally defined EOB waveform, which was based on a rather coarse
representation of the coalescence waveform.} of the originally defined
EOB. Each direction of deformation, e.g., $\de/\de a_5$, hopefully adds some
``missing physics'' that either has not yet been calculated because of
technical difficulties~\footnote{For instance, the exact, general relativistic
value of $a_5$ (or, rather, of the $\nu$-dependent coefficient
$a_5(\nu)=\nu a_5 + \nu^2 a_5'+\cdots$ of $(GM/c^2 R)^5$ in $A(R)$) has not
yet been calculated simply because it would represent a huge technical
challenge, involving a 4~PN (and 4-loop) generalization of the rather involved
3~PN (and 3-loop) work that led to the unique determination of the lower-order
coefficient $a_4(\nu)$~\cite{Damour:2001bu,Blanchet:2002av}.}, 
or represent only an {\it effective} description of
a complicated, nonperturbative process which is not directly formalizable in a
calculable way. In both cases, the EOB programme aims at using NR results to
determine the ``best fit'' values of the flexibility parameters; i.e., the
values that, hopefully, allow an analytical EOB waveform to 
{\it accurately} represent the exact general
relativistic inspiralling and coalescing waveform.
Note that, in this paper, we will not use the terminology 
of  {\it faithful} (versus {\it effectual}) waveforms~\cite{Damour:1997ub}.
Indeed, this terminology refers to particular measures of the closeness of two
waveforms (called ``faithfulness'', ${\cal F}$, and ``effectualness'', 
${\cal E}$ in~\cite{Damour:2002vi}) which are based on specific ways of
maximizing normalized overlaps. These measures are not the best suited 
for our present purpose because they are detector dependent 
(through the use of the detector's spectral noise curve
$S_h(f)$ in the Wiener scalar product $<X,Y>$, see e.g., Eqs.~(6.1) and
(6.2) of~\cite{Damour:2002vi}). By contrast, we are interested here in
hopefully showing that EOB waveforms can be ``close'' to general relativistic
ones in a much stronger mathematical sense, say in the time-domain $L_\infty$
norm: ${\rm sup}_{t\in[t_1,t_2]}\left|h^{\rm EOB}(t)-h^{\rm Exact}(t)\right|<\varepsilon$.
Actually, the most important ``closeness'', for data analysis purposes, is the
closeness of the phases. Therefore we shall primarily consider the time-domain
{\it phase} $L_{\infty}$ norm: 
$||\Delta\phi||_{\infty}\equiv{\rm inf}_{\tau,\alpha}{\rm
  sup}_{t\in[t_1,t_2]}\left|\phi_{22}^{\rm EOB}(t+\tau)+\alpha-\phi^{\rm NR}_{22}(t)\right|$, 
where we minimize over the two arbitrary
parameters $\tau$ (time-shift) and $\alpha$ (phase-shift). 
When $||\Delta\phi||_{\infty}$ is smaller than $\varepsilon$
for most physically relevant intervals $[t_1, t_2]$, we shall say
that the (time-domain) EOB waveform $h^{\rm EOB}(t)$ is 
an $\varepsilon-accurate$ representation of $h^{\rm Exact}(t)$.

The programme of determining the ``best fit'' flexibility parameters by
comparing EOB predictions to NR results has been initiated in several 
works~\cite{Damour:2002qh,Damour:2007xr,Buonanno:2007pf,Damour:2007yf,Damour:2007vq}
(see also~\cite{Buonanno:2006ui,Damour:2007cb, Pan:2007nw} for other 
works involving the comparison of EOB waveforms to NR ones). For some
parameters,~\footnote{Note that several of the EOB flexibility parameters
listed above refer to the recently introduced resummed $3^{+2}$-PN accurate
EOB waveform~\cite{Damour:2007xr,Damour:2007yf} and to the ``comb'' matching
procedure of Ref.~\cite{Damour:2007xr}. The EOB dynamics and waveform used in
the works of Buonanno and collaborators differ in several ways from the
dynamics and waveform used by us, notably: (i) a radiation reaction force
of lesser PN accuracy, and  without NQC corrections, (ii) a waveform of
``Newtonian'' accuracy without NQC corrections, and (iii) a simpler matching
procedure between the plunge and the ring-down involving only three
(positive-frequency) quasi-normal modes (QNM) and an instantaneous matching
(as used in some of the original EOB papers~\cite{Buonanno:2000ef,Damour:2006tr}).}
it has already been possible to determine them, or, at least, to find a
rationale that allows one to fix them in a near-optimal manner. For instance, it
was found in Ref.~\cite{Damour:2007xr} that $p=2$, i.e. the use of $2p+1=5$
matching points and 5 corresponding positive-frequency QNMs was optimal from a
practical point of view, in the sense that smaller values led to visibly worse fits, 
while higher values led to only a rather marginal improvement. We 
shall therefore fix $p$ to the value $p=2$. Concerning the ``central matching
time'' $t_m$, previous work~\cite{Damour:2007xr,Buonanno:2007pf,Damour:2007vq}
has found that it was near optimal to choose (as advocated
in~\cite{Buonanno:2000ef}) $t_m$ to be the so-called ``EOB light-ring
crossing'' time, i.e. the EOB dynamical time when the EOB orbital frequency
$\Omega$ reaches its maximum. Concerning the matching-comb width parameter
$\delta=\Delta t/(2p)$ (where $\Delta t$ is the total width of the matching
interval), Refs.~\cite{Damour:2007xr,Damour:2007vq} found that 
$\delta=1.7 M_{\rm final}$ 
(corresponding to $\Delta t=4\delta=6.8M_{\rm  final}$) yielded a good
result. Here $M_{\rm final}$ denotes the mass of the final black hole. 
Here also, we fix $t_m=t_{\rm EOB}^{\rm light-ring}$, and $\delta=1.7 M_{\rm final}$.
Moreover, we shall discuss below a rationale allowing one to fix the 
parameters $a$ and $b$.

Summarizing: the only EOB flexibility parameters which have not yet been
uniquely determined are $a_5$, $\vp$ and $\a$. Some recent 
works~\cite{Buonanno:2007pf,Damour:2007yf,Damour:2007vq} have explored the
influence of these parameters on the EOB waveform and have made initial steps
towards determining `best fit' values for these parameters; i.e., values
leading to an optimal agreement with NR data.
In particular, Ref.~\cite{Buonanno:2007pf} found that
the {\it faithfulness} $\F$ (in the sense of Sec.~VIA of
Ref.~\cite{Damour:2002vi}) of restricted EOB waveforms against NASA-Goddard NR
coalescence waveforms was largest when~\footnote{Note that 
Ref.~\cite{Buonanno:2007pf} uses the notation $\lambda$ for $a_5$.} 
$a_5$ belongs to some rather wide interval, say $[20,100]$, centered around 
$a_5\sim 60$. On the other hand, Ref.~\cite{Damour:2007yf} found that 
the {\it accuracy} (in the sense of the $L_\infty$ norm of the phase difference)
of the resummed $3^{+2}$-PN EOB waveform\footnote{We refer to the PN accuracy
  of this waveform as $3^{+2}$PN because it includes not only the known
comparable mass 3~PN waveform corrections, but also the test-mass limit of the
4~PN and 5~PN waveform amplitude corrections~\cite{Damour:2007yf}.} 
with respect to the high-accuracy Caltech-Cornell (CC) NR long-inspiral waveform 
was at its best when $a_5$ belonged
to an interval $\sim[10,80]$ centered around $a_5\sim 40$. The influence of the
flexibility parameter $\vp$ was studied in Refs.~\cite{Damour:2007yf,Damour:2007vq}.
It was found that, for any given values of 
$a_5$ and $\a$, {\it and for any given NR waveform}, there existed a well
determined value of $\vp$ that minimized the phase difference between EOB and
NR (see below for a more precise formulation). However, those previous EOB-NR
comparisons limited themselves to considering one NR data set at a time (the
published Caltech-Cornell inspiral data for Ref.~\cite{Damour:2007yf}, and
some Albert Einstein Institute (AEI) coalescence data for Ref.~\cite{Damour:2007vq}).

The aim of the present paper is to go beyond this piece-meal consideration of
NR data and to best fit  ({\it in phase}) the flexed EOB waveform, $h^{\rm
  EOB}(a_5,\vp,\a;\,t)$, {\it simultaneously} to several independent NR
waveform data (namely inspiral and coalescence data produced by the Jena
group and reported here, and published inspiral Caltech-Cornell data). 
Our main result will be that the best fit values of the three remaining EOB flexibility parameters
$(a_5,\vp$, $\a)$ are approximately determined, in the sense that they must all
take values in relatively small, correlated, intervals. 
It is then found that the resulting,
approximately unique, best fitted EOB waveform exhibits a remarkable agreement
(modulo differences compatible with estimated numerical errors), both in phase
and in modulus, not only with the data that we use in the fit (i.e., {\it
  equal-mass} Jena data and {\it equal-mass} Caltech-Cornell data considered
for $M\omega\leq 0.1$), but also
with other NR data (namely, {\it unequal-mass} Jena data and Caltech-Cornell
data {\it after} $M\omega=0.1$).

Our work focusses on the comparison between the EOB predictions and NR data
because the EOB method is the only existing analytical approach which: (i)
incorporates, in an exact manner, all the theoretical knowledge acquired
through many years of post-Newtonian studies, (ii) provides waveforms covering
the full coalescence process from early inspiral to ring-down, and (iii) can
describe spinning binaries (see, in this respect Refs.~\cite{Damour:2001tu,Damour:2008qf}).
However, as some studies have emphasized the nice properties of one specific
PN approximant, called TaylorT4 in~\cite{Boyle:2007ft} (for consistency with
the T1, T2 and T3 Taylor approximants considered in~\cite{Damour:2000zb}), we
shall discuss it briefly in Appendix~\ref{sec:T4}, 
though it does not satisfy our requirements (ii) above, namely that of 
providing waveforms covering the full coalescence process.

This paper is organized as follows. In Sec.~\ref{sec:NR} we briefly describe
the numerical simulations, whose results we use in the following.  Section~\ref{sec:eob} summarizes the
definition of the $3^{+2}$-PN accurate EOB waveform that we use. Section~\ref{sec:calibrate}
is the central section of this work: it shows how the simultaneous comparison 
of EOB to two different NR data sets allows one to determine a small range of
`best fit' (correlated) EOB flexibility parameters $a_5$, $\vp$ and $\a$. 
Section~\ref{sec:best} selects central values for the best fit parameters and
discusses in detail the agreement between the EOB waveform and the Jena NR
waveform, for  various mass ratios. 
The paper ends with a concluding Section, followed by two  Appendices. 
Appendix~\ref{sec:Psi4toPsie}
is devoted to the issue of determining the metric waveform $h(t)$ from 
the curvature waveform $\psi_4(t)$, while Appendix~\ref{sec:T4} discusses
the TaylorT4 approximant.
Except when otherwise specified, we use in this paper units such that $G=c=1$.

\section{Numerical Relativity simulations}
\label{sec:NR}

Numerical simulations were performed with the BAM 
code~\cite{Bruegmann:2006at,Husa:2007hp}, which
evolves black-hole binaries using the ``moving-puncture''
approach~\cite{Campanelli:2005dd,Baker05a}. The relevant physical and 
numerical parameters for our simulations are presented in Table~\ref{tab:BHparameters};
note that the results from the equal-mass simulations were presented
in~\cite{Hannam:2007ik}, which also contains extensive error analysis
and comparison with standard post-Newtonian inspiral approximants. These
results are also in good agreement with those of~\cite{Boyle:2007ft} over
the shared frequency range. We shall present below an explicit comparison
of the phase of the waveform of Ref.~\cite{Boyle:2007ft} with the one
of our equal-mass simulation.

\subsection{Initial data}

Following the moving puncture approach we set up initial data containing 
two black holes via a
Brill-Lindquist-like wormhole construction \cite{Brill:1963yv}, where
the additional asymptotically flat end of each wormhole is
compactified to a point, or ``puncture''.
The entire $2$-wormhole topology can thus conveniently be represented
on $R^3$. It has long been understood how to set up such puncture initial data,
and in particular how to avoid working with divergent quantities
\cite{Beig94,Beig:1994rp,Brandt97b,Dain01a}. More recently it has
turned out that
the gauge conditions used in the moving puncture approach actually allow
a simpler representation of the black hole interior during the evolution:
the black-hole throat is pushed an infinite proper distance away from 
the horizon, and the initial puncture geometry is replaced by 
a new compactified asymptotics with a milder 
singularity~\cite{Hannam:2006vv,Hannam:2006xw,Hannam:2008sg}.

One key element of the simplicity of the moving puncture approach
is that black holes can be modeled on a Cartesian
numerical grid without the need to deal with black hole excision techniques.
Another is that the assumption of an initially conformally flat spatial
geometry yields a very simple way to generate any number of moving, 
spinning black holes \cite{Bowen80,Brandt97b}. Note however that
the puncture initial data are not restricted to conformal
flatness a priori~\cite{Beig94,Beig:1994rp,Dain01a}, and generalizations 
that better model
spinning black holes have been suggested~\cite{Dain00,Hannam:2006zt}.

Assuming conformal flatness for the initial data,
and assuming the extrinsic curvature of the initial slice to be within the class
of nonspinning Bowen-York solutions, the freedom in specifying initial data comprises the masses, 
locations and momenta of each black hole. 

The mass of each black hole, $M_i$ ($i=1,2$), is specified
in terms of the Arnowitt-Deser-Misner (ADM) mass at each puncture, which
is, to a very good
approximation~\cite{Schnetter:2006yt,Dennison:2006nq,Tichy:2003qi} 
equal to the irreducible mass~\cite{Christodoulou:1970wf,Christodoulou:1972kt} 
of the apparent horizon
\begin{equation} M_i = \sqrt{
    \frac{A_i}{16 \pi} }\,,
\end{equation}
where $A_i$ is the area of the apparent horizon. We identify this
mass with the mass (denoted below $m_i$) used in post-Newtonian theory. 
This assumption will only hold exactly in the limit where the
black holes are infinitely far apart and stationary, but we
consider any error in this assumption as part of the error due to
starting the simulation at a finite separation.

The constraint equations for black-hole binary puncture initial data 
are solved using a pseudo-spectral code
\cite{Ansorg:2004ds}, and resampled for our finite difference grid
using high-order polynomial interpolation \cite{Husa:2007hp}.

We want to specify initial data for non-spinning black holes in the 
center-of-mass frame, such that the trajectories correspond to quasicircular
inspiral, i.e. the motion is circular at infinite separation, and
the eccentricity vanishes. Following \cite{Husa:2007ec}, we obtain
the initial momenta of the black holes from a post-Newtonian inspiral
calculation, using a 3PN-accurate conservative 
Hamiltonian~\cite{Damour:2001bu}, and 3.5PN accurate
beyond leading order orbit-averaged radiation flux~\cite{Blanchet:2001ax,Blanchet:2004ek}.
We have measured the eccentricity from oscillations in the separation and frequency
as described in \cite{Husa:2007ec}, and have obtained the values $0.002$, $0.003$, $0.005$,
for mass ratios $q=1,2$ and $4$ respectively.

\subsection{Numerical evolution}

We use the BSSN formulation of the Einstein equations 
\cite{Shibata95,Baumgarte99} for time evolution, which
are formulated in terms of a conformal 3-metric
$\tilde\gamma_{ij}$, related to the physical metric
as 
\begin{equation}
\tilde\gamma_{ij} = \chi \gamma_{ij}.
\end{equation}
Representing the conformal factor by the quantity $\chi$ has
the advantage that, when dealing with puncture data,
the conformal factor $\chi$
conveniently vanishes at each puncture \cite{Campanelli:2005dd}. Details
of our implementation of the BSSN/moving-puncture system
are described in~\cite{Bruegmann:2006at}.
We also need to choose a lapse and shift during the evolution to determine
our coordinate gauge.
As is common in the moving puncture approach, 
we use the ``1+log'' slicing condition \cite{Bona95b}
\begin{equation}
\partial_0 \alpha = - 2 \alpha K \label{eqn:1plogfull}, 
\end{equation}
and  the $\tilde{\Gamma}$-driver
condition \cite{Alcubierre01a,Alcubierre02a}, 
\begin{eqnarray}
\partial_0 \beta^i & = & \frac{3}{4} B^i,  \label{eqn:driver1} \\
\partial_0 B^i & = & \partial_0 \tilde{\Gamma}^i - \eta B^i, 
\label{eqn:driver2}
\end{eqnarray} 
where $\partial_0 = \partial_t - \beta^i \partial_i$.
The parameter $\eta$ in the shift-condition effectively 
regulates the coordinate size of the apparent horizons, and is set
to $\eta = 2/M$ in our simulations.

The Einstein evolution equations are solved numerically with standard
finite-difference techniques as described 
in~\cite{Bruegmann:2006at,Husa:2007hp}.  Spatial derivatives are approximated
with sixth-order accurate stencils. First order derivatives 
corresponding to Lie derivatives with respect to the shift vector
are approximated with off-centered operators as described 
in ~\cite{Husa:2007hp}, all other derivatives are approximated with
centered finite difference operators. Kreiss-Oliger artificial dissipation 
operators which converge to zero at fifth order are applied
as described in ~\cite{Bruegmann:2006at,Husa:2007hp}.
Time evolution is performed
with a fourth-order Runge-Kutta integration. Our box-based 
mesh refinement is described in ~\cite{Bruegmann:2006at}.
Time interpolation errors in the mesh-refinement algorithm
converge only at second order, but do not seem to contribute
significantly to the error budget, as does the  Runge-Kutta time integration.

The grid configurations we have used for our equal mass runs are described
in \cite{Hannam:2007ik}. For the unequal mass runs, we have used the 
56,64,72--gridpoints
configurations of \cite{Hannam:2007ik}, adding two further refinement levels
to push the outer boundary further out by roughly a factor of four. In order 
to be able to re-use
our equal mass grid configurations, we always choose the mass of the smaller black
hole, which determines our resolution requirements, at $M_1 = 0.5$. 

\subsection{Wave extraction}
\label{sbsc:we}

The gravitational wave signal is extracted at different surfaces of constant radial
coordinate by means of the Newman-Penrose Weyl
tensor component $\psi_4$~\cite{Newman62a,Stewart:1990uf} which is a
measure of the outgoing transverse gravitational radiation in an
asymptotically flat spacetime. At finite distance to the source the
result depends on the coordinate gauge and the choice of a null tetrad.
Our choice of tetrad and details of the wave extraction algorithm are described in 
detail in \cite{Bruegmann:2006at}. We choose our extraction surfaces at 
40, 50, 60, 80, and  90 $M$. In~\cite{Hannam:2007ik} we extrapolated the waveform
amplitude (though not its phase) to the value that would be observed at
infinity; in this work we deal with the raw numerical data at the farmost
extraction radius, but use some  extraction-radius-extrapolated results
to provide uncertainty estimates. See next subsection.

The analysis carried out in this paper will use, as approximate asymptotic
waveform, the curvature perturbation extracted at radius  $90M$, without
any extrapolation (neither with respect to extraction radius, nor with 
respect to resolution). The comparisons between numerical data and analytical
predictions done below will make use of {\it metric} (by contrast to
curvature) waveforms. We discuss in Appendix~\ref{sec:Psi4toPsie} the
integration procedure we used to compute the numerical metric waveform 
from the raw curvature waveform output of the simulations.
In this paper, we focus on the $\l=m=2$ ``quadrupolar'' waveform.

\begin{table*}[t]
\caption{\label{tab:table1} Details of the simulations discussed in the
 test. From left to right, the columns report: mass ratio $q=m_2/m_1$;
 symmetric mass ratio $\nu=m_1 m_2/(m_1+m_2)^2$;  initial coordinate 
 separation $D$ of the punctures; inital ADM mass; initial tangential 
 ($p_t$) and radial ($p_r$) momentum of the black holes;
 mass and dimensionless spin parameter $j_{\rm f}=J_{\rm f}/M_{\rm f}^2$ 
 of the final black hole; mass and dimensionless spin parameter 
 $j_{\rm f}^{\rm ring}$ and $M_{\rm f}^{\rm ring}/M$ of the final black 
 hole obtained only from the ringdown; dominant (quasi-normal-mode) frequency
 of the ringdown. Quantities are scaled by the total initial black hole mass
 $M=m_1+m_2$ as indicated.}
\begin{center}
  \begin{ruledtabular}
  \begin{tabular}{cccccccccccc}
    $q$ & $\nu$  &$D/M$ & $M_{\mathrm{ADM}}/M$ & $\vert p_t/M\vert$ & $10^3 \times
    \vert p_r/M\vert$ & $M_{\rm f}/M$ & $j_{\rm f}$ & 
    $M_{\rm f}^{\rm ring}$ &
    $j_{\rm f}^{\rm ring}$ & $M\sigma^+_{2220}$ \\ 
    \hline \hline
  1 & $1/4$  &12 & 0.991225 & 0.085035 & 0.053729 & $0.9514 \pm 0.0016$ & $0.687 \pm 0.002$ & $0.962$  & 0.690 & $0.0850+ \i\, 0.5521$ \\    
  2 & $2/9$  &10 & 0.990901 & 0.085599 & 0.794821 & $0.96   \pm 0.003$  & $0.625 \pm 0.004$ & $0.977$  & 0.635 & $0.0856+ \i\, 0.5214$ \\  
  4 & $4/25$ &10 & 0.993522 & 0.061914 & 0.043332 & $0.978  \pm 0.003$  & $0.472 \pm 0.004$ & $0.990$  & 0.487 & $0.0874+ \i\, 0.4683$ \\
  \end{tabular}
\end{ruledtabular}
\end{center}
\label{tab:BHparameters}
\end{table*}%

\subsection{Accuracy}
\label{sbsc:acc}

\subsubsection{The equal-mass case}
A detailed error analysis was performed for the equal-mass 
waveforms in~\cite{Hannam:2007ik}. In this section, we will first 
summarize the results of that error analysis, and then complete it by 
more carefully quantifying the uncertainty in the phase. 
As we shall see, our refined estimate of the uncertainty in the phase will
end up being {\it significantly lower} than the upper bound, 0.25 radians, 
quoted in~\cite{Hannam:2007ik}.

The amplitude and phase of the equal-mass nonspinning waveforms show 
sixth-order convergence with respect to 
numerical grid resolution prior to merger, with a small drop in convergence order 
around merger time. Higher-accuracy results were constructed by Richardson
extrapolation with respect to numerical resolution, and this procedure also 
allowed an estimation of the contribution to the uncertainty in the amplitude
and phase from discretization error.  
The discretization error in the amplitude was found to be 
below 0.5\%, while the discretization error in the phase was estimated to be below 0.01 radians. 
These are conservative error estimates obtained by observing the numerical errors
over the course of the entire simulation. See in particular Fig.~4 in~\cite{Hannam:2007ik}.
If we look at that figure we may conclude that the error estimate of the phase is extremely 
conservative, but one should also be aware that the quoted numerical phase error takes into 
account only instantaneous differences in the value of the waveform phase, but not secular
drifts. When the analysis for~\cite{Hannam:2007ik} was performed, the authors
hoped that the conservative value quoted would account for any phase
drifts. We shall see below that, however, there might remain sources of
secular drifts that are not yet well understood. 

In addition to the discretization error, there is also an error due to measuring the 
waveform at a finite distance from the source. For both the waveform amplitude and
phase, it was found in \cite{Hannam:2007ik} that finite extraction radii errors were
much larger than discretization errors. Prior to merger, the error in the amplitude was
found to fall off as $1/R_{ex}^2$, where $R_{ex}$ was the radiation extraction radius, 
and this observation allowed a clean extrapolation to $R_{ex} \rightarrow \infty$, and, once
again, an estimate of the uncertainty in the amplitude. The uncertainty in the 
{\it extrapolated} amplitude was at most 2\% before merger. Around merger time,
the amplitude error fall-off is dominated by a $1/R_{ex}$ term, and the uncertainty
in the extrapolated amplitude grows to around 5\%. However, in this paper we use
the raw data calculated at the extraction radius $R_{ex} = 90M$, and as such the
uncertainties are larger, as much as 5\% over the entire simulation. The largest 
uncertainties in the  finite-extraction-radius amplitude are at early times, when the 
amplitude is small, and around merger, when the dynamics are strongest.

In~\cite{Hannam:2007ik} the total phase uncertainty accumulated on a time interval
of duration $1400M$ extending up to gravitational wave frequency 
$M\omega =0.1$ was quoted as being 0.25 radians. 
This large value was an upper bound which was quoted in view of the difficulty in finding 
a robust method to extrapolate the phase to infinite extraction radius. 
These difficulties were related to the specific phase alignment method which
was used in~\cite{Hannam:2007ik}. There, one was first choosing some frequency
at which to line up the phases and frequencies of waves
from different extraction radii, and then attempting to perform an extrapolation. Although
it is entirely valid to time- and phase-shift any number of waveforms to perform a comparison
between them, it turned out that  this is not an efficient way to perform a
consistent extraction-radius extrapolation.

By contrast, for the purpose of the present paper we have performed a new
study of the extraction-radius extrapolation which follows the strategy
proposed in~\cite{Boyle:2007ft}.
More precisely we used two similar, but different, phase alignment methods.
The first one consists of simply introducing the ``Newtonian retarded time'', 
at the coordinate extraction radii $R_{ex}$, $u_{\rm N}=t-R_{ex}$ 
and study the waves as function of $u_{\rm N}$.
Then, when attempting extrapolation with respect to $R_{ex}$, 
we find a clear $c_0(u_{\rm N} )+c_2(u_{\rm N}  )/R_{ex}^2$ fall-off in the error, 
and are able to make a clean extrapolation to infinity. 
We have also repeated the analysis with an extra $+c_3(u_{\rm N})/R_{ex}^3$ 
term in the fit.

\begin{figure}[t]
  \begin{center}
    \includegraphics[width=95 mm]{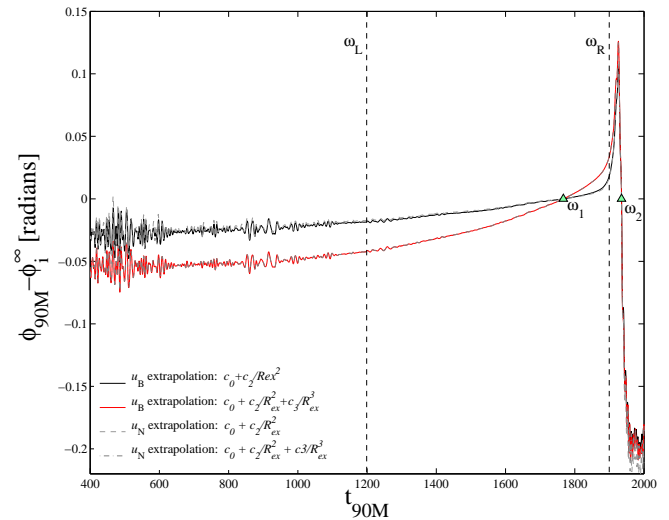}
  \end{center}
\vspace{-4mm}
  \caption{\label{fig:ExtrapolationComparison} 
  Differences between the phase extracted
  at $R_{ex} = 90M$, and the phase extrapolated to infinity based 
  on two choices of the retarded time and on two choices of the 
  extrapolating polynomial, as described in the text. 
  The choice of retarded time makes little difference to the result.}
\end{figure}

The second method consists of using, inspired by the result
in~\cite{Boyle:2007ft}, the a priori more 
accurate definition of retarded time, $u_{\rm B} = t - r_*$, where the
(approximate) Regge-Wheeler tortoise coordinate $r_*$
is (following~\cite{Boyle:2007ft}) defined as $r_* = R_{ex}  +
M_{\rm ADM} + 2M_{\rm ADM} \log[ (R_{ex} + M_{\rm ADM})/(2M_{\rm ADM}) - 1]$. 
This improved choice of retarded time allows us again to perform a clean extrapolation to
infinity. 
As when using $u_{\rm N}$, we use two different fits:
$c_0(u_{\rm B} )+c_2(u_{\rm B}  )/R_{ex}^2$ and $c_0(u_{\rm B} )
+c_2(u_{\rm B}  )/R_{ex}^2+c_3(u_{\rm B}  )/R_{ex}^3$.
We then estimate the uncertainty in the phase of the farmost unextrapolated data,
extracted at $R_{ex}=90M$, by comparing the following five phases:
(i) the raw phase $\phi_{90M}$ measured at $R_{ex} = 90M$, (ii) the phase
$\phi_{\rm N2}^{\infty}$ extrapolated 
using $u_{\rm  N}$ and assuming a $1/R_{ex}^2$ fall off, 
(iii) the phase $\phi_{\rm N3}^{\infty}$ extrapolated 
using $u_{\rm  N}$ and assuming a 
$1/R_{ex}^2 + 1/R_{ex}^3$ fall  off,
(iv) the phase  $\phi_{\rm B2}^{\infty}$  extrapolated using $u_{\rm B}$ 
and assuming a $1/R_{\rm ex}^2$ fall off
and (v) the phase  $\phi_{\rm B3}^{\infty}$ extrapolated using $u_{\rm B}$ and 
assuming a $1/R_{ex}^2 + 1/R_{ex}^3$ fall off.
The differences between the phases are computed after they have been aligned
by using the two-times pinching technique of Ref.~\cite{Damour:2007vq}
(which is reviewed in Sec.~\ref{sec:calibrate} below).
For consistency with our EOB-NR matching discussed in Sec.~\ref{sec:best}
below  we use as ``pinching'' gravitational wave frequencies 
$\omega_1\approx 0.1$ and $\omega_2=0.4717$. Note that these 
frequencies bracket the merger time.
The four phase differences $\phi_{\rm 90M}-\phi_i^{\infty}$ 
where $i\in\{{\rm N2,N3,B2, B3}\}$ are exhibited 
as functions of the
numerical relativity coordinate time at $90M$, in 
Fig.~\ref{fig:ExtrapolationComparison}. 
The triangles in the figure indicate the two times corresponding
to the two pinching frequencies $(\omega_1, \omega_2)$, while
the vertical dashed lines indicate the  time
interval $[1200,1900]\approx [t_{\rm L},t_{\rm R}]$ which will turn out
to be crucial for our analysis in Sec.~\ref{sec:best} below.
Several conclusions can be drawn from Fig.~\ref{fig:ExtrapolationComparison}:
First, the choice of retarded time, $u_{\rm N}$ or $u_{\rm B}$, does not
matter much for the extrapolation procedure. Second, though the phase 
differences over the entire span of the simulation can reach 
values $\sim +0.13$ radians around merger time (peak at $t\approx 1930M$) 
and/or $\sim -0.2$ radians (during ringdown), they stay quite small during the
time interval $[t_{\rm L}, t_{\rm R}]$ that we shall focus on
in our analysis below.~\footnote{Note also that with the above choice of pinching
times $(\omega_1, \omega_2)$ the phase differences stay quite small, 
namely $-0.06$ radians, during the entire inspiral. However, this result 
sensitively depends on the way the phases have been matched. For instance,
when using pinching frequencies $\omega_1=\omega_2=0.1$ one observes
maximum phase differences of $\sim +0.07$ radians at merger and $\sim -0.45$
radians during ringdown, while they stay between (-0.03,0) radians 
during the inspiral. On the other hand, when using pinching frequencies
around merger, i.e., $\omega_1=0.36$ and $\omega_2=0.38$, one gets
quite small phase differences during merger and ringdown, but one
observes large dephasings at early times, that build up to $-0.6$ radians.} 
Most importantly for the following the maximum phase differences over
the interval $[t_{\rm L},t_{\rm R}]$ stay within the 
rather small interval $(-0.042,+0.032)$ radians.

\begin{figure}[t]
  \begin{center}
    \includegraphics[width=95 mm]{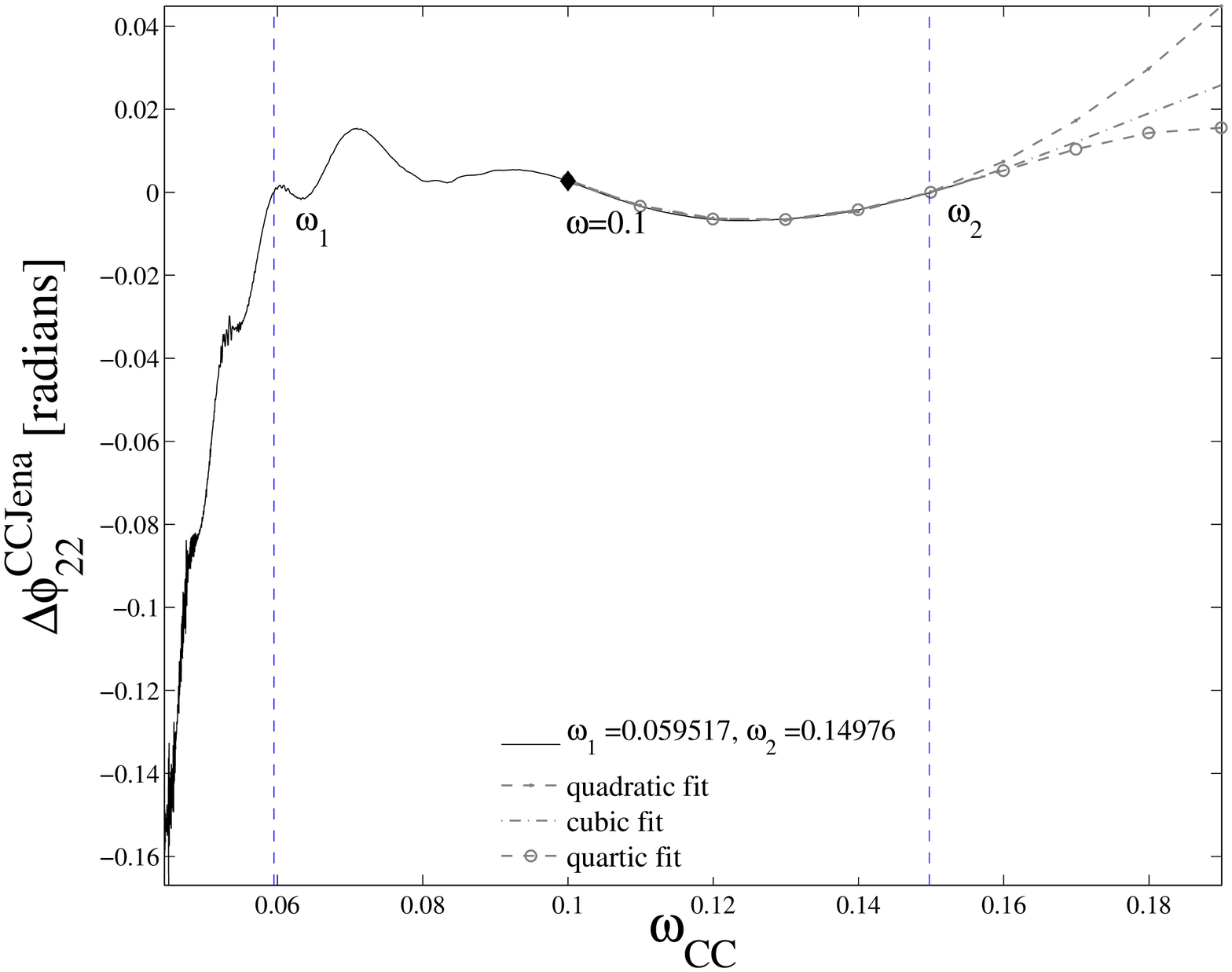}
  \end{center}
\vspace{-4mm}
  \caption{\label{fig:CCJena}Comparison between Caltech-Cornell and Jena
  actual numerical data: 
  the phase difference $\Delta\phi_{22}^{\rm CCJena}={\phi^{\rm CC}-\phi_{90M}^{\rm Jena} }$ 
  is shown versus Caltech-Cornell GW frequency $\omega_{\rm CC}$.}
\end{figure}

After this ``internal'' way of trying to estimate the numerical errors in the
phase of our equal-mass simulation, let 
us consider an ``external'' way which consists in directly comparing
the unextrapolated, $90M$ raw phase $\phi_{90M}(t)$ to the phase 
computed by Boyle et al.~\cite{Boyle:2007ft} and kindly communicated
to us.
In Fig.~\ref{fig:CCJena} we are directly comparing two phases:
our unextrapolated $\phi_{90M}(t)$ and the 
resolution- and radius-extrapolated Caltech-Cornell
curvature phase $\phi_{\rm CC}(t)$. The phase difference
$\Delta\phi^{\rm CCJena}_{22}=\phi_{\rm CC}-\phi_{90M}$ was plotted versus
the Caltech-Cornell (curvature) frequency $\omega_{\rm CC}$.
This phase difference was computed in the following way. First, we used the
two-pinching frequencies $\omega_1=0.059517$ and $\omega_2=0.14976$ 
(indicated by two dashed vertical lines in the figure)
to determine the time and phase shifts $(\tau,\alpha)$, see below, then
the result 
$\Delta\phi^{\rm CCJena}_{22}(t_{\rm CC})=\phi_{\rm CC}(t_{\rm CC})-(\phi_{90M}(t_{\rm CC}+\tau) + \alpha)$
is plotted  versus $\omega_{\rm CC}$ instead of $t_{\rm CC}$.
In addition, since the Caltech-Cornell simulation extends only up
to $\omega^{\rm max}_{\rm CC}\sim\omega_2\approx 0.15$, we have
estimated three different possible extrapolations of the phase difference
$\Delta\phi^{\rm CCJena}_{22}$ beyond $\omega_2$ and up to $\omega_{\rm R}=0.1898$.
\footnote{Note that $\omega_1=\omega_{\rm L}$ corresponds to the lower 
limit $t_{\rm  L}$ of the crucial EOB-NR comparison interval
used in Sec.~\ref{sec:calibrate}, while
$\omega_{\rm R}$ corresponds to its upper limit $t_{\rm R}$.}
These three different extrapolations were obtained by fitting
$\Delta\phi(\omega_{\rm CC})$ over the interval $[0.1,0.15]$ 
by three different polynomial functions of $\omega_{\rm CC}$:
quadratic, cubic and quartic.
As we see on Fig.~\ref{fig:CCJena}, the quadratic fit is the 
one which gives the worst possible phase difference over the interval 
$[\omega_{\rm L},\omega_{\rm R}]$. We use this worst case as estimate
of the maximum phase difference between Caltech-Cornell and Jena phasings
over $[t_{\rm L},t_{\rm R}]$.
More precisely, while the minimum value of $\Delta\phi_{22}^{\rm CCJena}$ 
over the interval is $-0.0068$ radians, its maximum is $+0.04484$ 
radians (at $\omega_{\rm R}$,
i.e., at the extreme right of Fig.~\ref{fig:CCJena}). This corresponds
to a two-sided CC-Jena phase difference 
(in the sense of footnote~12 of Ref.~\cite{Damour:2007vq} )
$\pm 1/2(0.04484-(-0.0068))= \pm 0.026$ radians over the interval 
$[\omega_{\rm L},\omega_{\rm R}]$.
As this difference a priori comprises many 
possible ``noise sources'' coming from comparing 
two very different simulations, with different wave extraction procedures, 
we consider that this is our best present estimate of the unknown ``real'' 
error on the difference $\phi_{90M}-\phi_{\rm CC}$.
In addition, as a recently published 
{\it upper limit} on the {\it total accumulated phase error} 
in the Caltech-Cornell data of Ref.~\cite{Boyle:2007ft} is 0.01 
radians~\cite{KidderJena} (which is rather small), 
we shall consider in the following that $\pm 0.026$ radians provides
our best current estimate of the real error on the equal mass $\phi_{90M}$
over the interval $[\omega_{\rm  L},\omega_{\rm R}]$.
Note, in passing, that the internal error analysis procedure discussed
above was giving a roughly comparable error estimate, namely a two-sided
phase difference $\pm 1/2(0.032-(-0.042))\sim \pm 0.037$ radians.
However, we cannot rely on this internal analysis because 
it fails to explain the origin of a striking feature of Fig.~\ref{fig:CCJena},
which is that, before the plateau of very small phase differences reached
between frequencies 0.06 and 0.15, there is a steeper phase gradient which reaches
$-0.16$ radians at $\omega= 0.04445$, roughly corresponding to the 
beginning of the Jena simulation.
Part of this error may be due to residual eccentricity --- a quick 
comparison with post-Newtonian results using the techniques described 
in~\cite{Husa:2007ec} suggests that the phase error
from a residual eccentricity of $e \sim 0.002$ could be as much as 0.05 radians.
We feel, however, that most of the error comes from some secular drift
at early times which is not yet well understood.

\subsubsection{The unequal-mass cases}
For the unequal-mass case 2:1, we find similar results, namely, that the finite 
extraction radii dominate the error, and the amplitude error is below 5\% prior 
to merger.
As for the accumulated phase error in the $700M$ time span up to
$M\omega=0.1$, we did not carry out the radius extrapolation analysis
discussed above in the unequal mass case. As a rough upper limit
we quote an accumulated phase error of 0.15 radians.
In contrast to the equal-mass case, the fall-off 
in the amplitude error with respect to radiation extraction radius is not so clean
around merger time, preventing us from performing an accurate extrapolation to 
infinity. As such, we would conservatively give an uncertainty estimate of 10\%
of the amplitude at merger and later.

In the unequal-mass case 4:1, the case is different again: here the discretization
error dominates the phase error, suggesting that higher-resolution simulations are 
needed. Our estimate for the accumulated phase uncertainty up to $M\omega = 0.1$ is 0.25
radians, based entirely on discretization error. For the amplitude we estimate that the 
uncertainty is similar to that 
in the 1:2 case, i.e., around 5\% prior to merger, and 10\% after that time. 

\subsection{Final parameters of the black hole}
\label{sec:params}

The final mass of the black hole is obtained by subtracting the radiated energy
from the initial mass. While the initial mass (the ADM mass) is known very
accurately from the solution of the constraints with spectral 
methods~\cite{Ansorg:2004ds}, the radiated energy is less accurate 
and dominates the errors of the final mass and Kerr spin parameter.
The radiated energy is not very accurate, due to the loss of accuracy
in the wave signal at merger time for the equal mass
case (leading to a conservative error estimate of $4\%$), 
and the problems of extrapolation in radius and gridspacing for the unequal 
mass cases, which lead us to a conservative error estimate of $10\%$ in those 
cases.

The error in the radiated energy also dominates computing the quantity 
$j_f=J/M^2$, where we either compute $J$
from a surface integral as in~\cite{Bruegmann:2006at} 
and $M$ as described above, or we calculate $j_{\rm f}$ itself from the ringdown.
The error in computing the angular momentum $J$ from a surface integral falls
off very quickly
with separation. The dominant error in this quantity comes from 
high-frequency numerical noise in the integrals, which is however much 
smaller than the error in $j_{\rm f}$ resulting from errors in the final mass.

To determine the mass and spin parameter of the final black hole from 
the ringdown, we have performed two types of fits to the dominant mode. 
First, the quality factor has been obtained from a fit to the dominant
quasi-normal mode\footnote{In the notation introduced
in Sec.~\ref{sec:eob} below, the dominant mode corresponds to the labels
$(\pm,\l,\l',m,n)=(+,2,2,2,0)$.}
of the {\it complex} ringdown waveform. This fit was performed by 
a non-linear least-squares Gauss-Newton method, 
using $\exp(-\sigma t + \rho)$  as a parameter--dependent 
template (with two {\it complex} parameters ($\sigma,\rho$)), 
and an appropriate time interval during the ringdown 
(chosen by minimizing the post-fit residual).
Then, from the best-fit  value of $\sigma$ (i.e., the QNM dominant
complex frequency  $\sigma^+_{2220}$ ), 
we computed the values of $(M^{\rm ring}_{\rm f},j^{\rm ring}_{\rm f})$
of the final black hole by using the interpolating fits 
given in Appendix~E of~\cite{Berti:2005ys}. The triplets 
$(M^{\rm ring}_{\rm f}/M,j^{\rm ring}_{\rm f},M\sigma^{+}_{2220})$
are listed in Table~\ref{tab:table1}.

This method does not require knowledge of the final mass, but
is actually not well conditioned due to the shape of the curve 
$j(\omega)$.
Better accuracy is obtained by only using the real part of 
the frequency, then again, the error in $j$ is dominated by
the error in the radiated energy. The values are consistent with the values
obtained from the surface integrals for the angular momentum $J$. 
The numbers $M_{\rm f}$ and $j_{\rm f}$  quoted in Table~\ref{tab:BHparameters}
are consistent with both methods, and with the analytical
fit for $j_{\rm f}$ published for shorter and less accurate 
waveforms in~\cite{Berti:2007fi}.
By contrast $(M_{\rm f}^{\rm ring},j_{\rm f}^{\rm ring })$,   
are ``best-fit'' values that will be used below to compute
the EOB ringdown waveform.

\section{The EOB waveform}
\label{sec:eob}
We shall not review here the EOB formalism, which has been
described in several recent 
publications~\cite{Buonanno:2007pf,Damour:2007yf,Damour:2007cb,Damour:2007vq,Damour:2008yg}.
We refer to these papers, and notably to Refs.~\cite{Damour:2007cb,Damour:2007yf}, 
for detailed definitions of the dynamics and of the waveform. Let us only
indicate here a few of the crucial elements of the EOB implementation that 
we use here. We recall below the main ingredients of the EOB formalism,
focusing on the appearance of the various EOB flexibility parameters.
\begin{itemize}
\item The EOB Hamiltonian $H_{\rm real}$ describes the conservative part
of the relative two-body dynamics. We use for the crucial ``radial potential''
$A(r)$ entering this Hamiltonian the $P^1_4$ Pad\'e resummation of
\begin{equation}
A^{\rm Taylor}(a_5, \nu;\,u) = 1-2u +2\nu u^3 + a_4 \nu u^4 + a_5\nu u^5,
\end{equation}
where~\cite{Damour:2000we,Damour:2001bu}
\begin{equation}
a_4 = \dfrac{94}{3} - \dfrac{41}{32}\pi^2,
\end{equation}
where $a_5$ is the 4~PN flexing parameter introduced
in~\cite{Damour:2001tu}, and where~\footnote{Except when said otherwise, we henceforth 
systematically scale dimensionful quantities by means of 
the total rest mass $M\equiv m_1 + m_2$ of the binary system. 
For instance, we use the dimensionless EOB radial coordinate 
$r\equiv R_{\rm EOB}/M$, with $G=1$. Note also that 
$\nu \equiv \mu/M$ with $\mu \equiv m_1 m_2/M$.}  $u=1/r$. 
\item The EOB {\it radiation reaction force} ${\cal F}_{\varphi}(\vp,\a,\nu)$,
that we shall use here, has the form
\begin{equation}
{\cal F}_{\varphi}(\vp,\a,\nu)=f^{\rm NQC}_{\rm RR}(\a) {\cal F}_{\varphi}^0(\vp,\nu),
\end{equation}
where ${\cal F}_{\varphi}^0(\vp,\nu)$ is defined as a 
Pad\'e resummation~\cite{Damour:1997ub} of its Taylor expansion.
See Eq.~(17) of~\cite{Damour:2007xr} where
$f_{\rm DIS}$ is the $P^{4}_{4}$ 
Pad\'e resummation of $(1-v/v_{\rm  pole})\hat{F}^{\rm Taylor}(v;\nu)$.
In addition, the factor $f^{\rm NQC}_{\rm RR}$ is a non quasi-circular (NQC)
correction factor of the form 
\begin{equation}
\label{fnqc}
f^{\rm NQC}_{\rm RR}(\a)=\left( 1+ \bar{a}^{\rm RR}
\dfrac{p^2_{r_*}}{(r\Omega)^2+\epsilon_{\rm RR}}\right)^{-1}.
\end{equation}
This factor was introduced in Refs.~\cite{Damour:2007xr,Damour:2007vq} 
(see also Ref.~\cite{Damour:2002vi}). We fix the value of $\epsilon_{\rm RR}$ 
to $\epsilon_{\rm RR}=0.2$ as in~\cite{Damour:2007vq}.
\item We use {\it improved ``post-post-circular''} EOB dynamical initial data
  (positions and momenta) as in~\cite{Damour:2007yf,Damour:2007vq}.
\item We use the {\it resummed} $3^{+2}$PN  accurate ``inspiral-plus-plunge''
Zerilli-Moncrief normalized metric waveform introduced 
in Ref.~\cite{Damour:2007xr,Damour:2007yf}. It has the form
\begin{equation}
\label{hinsplunge}
\Psi_{22}^{\rm insplunge}(a,b;\,\nu,t)=-4\sqrt{\dfrac{\pi}{30}}\nu
     (r_{\omega}\Omega)^2 f_{22}^{\rm NQC}(a,b) F_{22}(\nu) e^{-2{\i}\Phi}.
\end{equation}
Here $\Phi(t)$ is the EOB orbital phase, $\Omega=\dot{\Phi}$ is the
EOB orbital frequency, $r_{\omega}\equiv r[\psi(r,p_\varphi)]^{1/3}$ is a 
modified EOB radius, with $\psi$ being defined as 
\begin{align}
\psi(r,p_\varphi)&=\dfrac{2}{r^2}\left(\dfrac{dA(r)}{dr}\right)^{-1}\nonumber\\
                 &\times\left[1+2\nu\left(\sqrt{A(r)\left(1+\dfrac{p_\varphi^2}{r^2}\right)  }-1\right)\right],
\end{align}
which generalizes the 2PN-accurate Eq.~(22) of 
Ref.~\cite{Damour:2006tr}. 
The factor $F_{22}$ is a resummed, 
$3^{+2}$-PN-accurate complex amplitude correction valid during 
the (adiabatic) inspiral (see~\cite{Damour:2007yf}), 
and $f_{22}^{\rm NQC}(a,b)$ is the following extra complex  correcting factor, aimed at
taking care (in an effective way) of  various non quasi-circular 
effects during the plunge
\begin{equation}
\label{f22NQC}
f_{22}^{\rm NQC}(a,b) = \left[ 1 + a \frac{ p_{r_*}^2}{(r\Omega)^2 + \epsilon_a}\right]
e^{\i b \frac{p_{r_*}}{r\Omega}},   
\end{equation} 
where $p_{r_*}$ is the momentum conjugate to the EOB-tortoise radial
coordinate $r_*$. Here we shall fix $\epsilon_a=0.12$. 
In these equations, we have only indicated the {\it explicit} appearance
of the waveform flexibility parameters $(a,b)$. In addition, the waveform
is, evidently, implicitly depending on $a_5$, which enters the Hamiltonian,
as well as on $\vp$ and $\a$, that enter the radiation reaction force.
\item We use a {\it ringdown waveform}, 
\begin{equation}
\label{hringdown}
\Psi^{\rm ringdown}_{22}(t) = \sum_N C_N^{+} e^{-\sigma_N^{+} t}
\end{equation}
where the label $N$ actually refers to a set of indices $(\l, \l', m, n)$, 
with $(\l,m) = (2,2)$ being the Schwarzschild-background multipolarity 
degrees of the considered $\Psi_{\l m}$  waveform 
with $n=0,1,2,...$ being the ``overtone number'' of the considered 
Kerr-background Quasi-Normal Mode (QNM; $n=0$ denoting the fundamental mode),
and $\l'$ the degree of its associated spheroidal harmonics $S_{\l' m}(a \sigma, \theta)$.
In addition  $\sigma_N^{+}= \alpha_N^{+} + {\i}\omega_N^{+}$ refers to the
positive   complex QNM frequencies  ($\alpha_N^{+} >0$ and
$\omega_N^{+}>0$  indicate the inverse damping time and the oscillation frequency of 
each mode respectively). The sum over $\l'$ comes from the fact that an
ordinary spherical harmonics $Y_{\l m}(\theta, \phi)$ (used as expansion basis 
to define $\Psi_{\l m}$) can be expanded in the spheroidal harmonics 
$S_{\l' m}(a \sigma, \theta) e^{{\i} m \phi}$ characterizing the angular
dependence of the Kerr-background QNMs~\cite{PT1973}.
As explained in Sec.~III of Ref.~\cite{Damour:2007vq}, we use five 
positive frequency QNMs computed starting from the values of
$M_{\rm f}^{\rm ring}/M$ and $j_{\rm f}^{\rm ring}$ listed in Table~\ref{tab:table1}.
\item We {\it match} the inspiral-plus-plunge waveform to the ring-down one,
on a ($2p+1)$-tooth ``comb'' $(t_m - p \delta, t_m - (p-1) \delta,\ldots,
 t_m -  \delta, t_m , t_m + \delta, \ldots, t_m + p \delta)$, of total 
length $\Delta t=2p\delta$, which is centered around some ``matching'' 
time $t_m$. We fix the integer $p$ to the value $p=2$, corresponding 
to five matching points. As mentioned above, we follow 
previous work~\cite{Buonanno:2000ef,Buonanno:2006ui,Damour:2007vq} in
fixing the ``matching time'' $t_m$ to coincide with the so-called
``EOB light-ring'', i.e.\ the instant when the orbital frequency $\Omega(t)$
reaches its maximum (this defines, within the EOB approach, the merger time). 
As in~\cite{Damour:2007vq}, we fix $\delta=1.7M_{\rm f}^{\rm ring}$, which
corresponds to a total width for the matching interval $\Delta
t=4\delta=6.8M_{\rm f}^{\rm ring}$. 

\item Finally, we define the complete EOB matched waveform (from $t=- \infty$ to $t=+ \infty$) as
\begin{align}
\label{hmatched}
&\Psi^{\rm EOB}_{22}(a_5,\vp,\a,a,b,t_m,\delta;\,\nu,t)\nonumber\\
&\equiv\theta(t_m -t) \Psi^{\rm insplunge}_{22}(t)+ \theta(t -t_m ) \Psi^{\rm ringdown}_{22}(t),
\end{align}
where $\theta(t)$ denotes Heaviside's step function. 

This {\it metric} EOB waveform then defines a corresponding {\it curvature}
waveform, simply (modulo a factor $r$ and normalization conventions) 
by taking two time derivatives of~\eqref{hmatched}, namely
\begin{equation}
r\psi_4^{\l m} = \dfrac{d^2 }{dt^2}(r h_{\l m})= N_{\l}\dfrac{d^2}{dt^2}(\Psi_{\l m}), 
\end{equation}
where $N_\l\equiv \sqrt{(\l+2)(\l+1)\l(\l-1)}$ (see Appendix~\ref{sec:Psi4toPsie}).
Note, however, that in view of the imperfect 
smoothness~\footnote{A partial cure to this problem would consist in replacing the discontinuous
step function $\theta(t -t_m)$ in Eq.~\eqref{hmatched} by one
of Laurent Schwartz's well-known smoothed step functions 
(or ``partitions of unity'')  ${\theta_{\varepsilon}}((t-t_m)/(2p\delta))$.} 
of the EOB matched metric waveform~\eqref{hmatched} around $t=t_m$, we find it
more convenient, when comparing EOB to numerical data that include the merger,
to work with 
the metric waveform without taking any further time derivatives. We discuss
in Appendix~\ref{sec:Psi4toPsie} the procedure that we use to compute 
from the numerical relativity curvature waveform a corresponding metric 
waveform by two time integrations.
\end{itemize}

\subsection{Fixing the $a$ and $b$ flexibility parameters}
In this brief subsection we discuss a rationale for choosing
two of the EOB flexibility parameters mentioned above, namely $a$ and $b$,
that enter the NQC waveform correction factor~\eqref{f22NQC}.

Ref.~\cite{Damour:2007xr} found that it was near optimal to fix the NQC
parameter $a$ entering the modulus of the 
waveform~\footnote{Here $a$ and $b$ denote the parameters called $a'$ and $b'$
in footnote 9 of~\cite{Damour:2007xr}} so as to ensure that the maximum of the
modulus of the EOB quadrupolar metric waveform sits on top of that of the EOB
orbital frequency, i.e., at the ``EOB 
light-ring''~\footnote{Note that this coincidence in the locations of the
maximum of $\left|h_{22}(t)\right|$ and of $\Omega$ is automatically
ensured when one uses (as advocated in~\cite{Buonanno:2000ef}) a ``restricted''
EOB waveform $\Psi_{22}(t)\propto \Omega^{2/3}\exp[-2\i\Phi(t)]$. It is, however, a
non trivial fact that NR results show (both in the test-mass limit~\cite{Damour:2007xr}
and in the equal-mass case~\cite{Damour:2007vq}) that the maximum of
$\left|\Psi_{22}(t)\right|$ does occur very near the maximum of the
(corresponding, best matched during inspiral) EOB orbital frequency $\Omega(t)$.
This can be considered as another successful prediction of the EOB formalism.
Note that this property does not apply to the maximum of the modulus of other 
GW quantities, such as the instantaneous energy flux or the modulus of
quadrupole {\it curvature} waveform $r\psi_4^{22}(t)$, which occur
significantly after the EOB light-ring~\cite{Buonanno:2006ui}.}. We shall
therefore ``analytically'' determine the value of the waveform NQC parameter
$a$, as a function of the symmetric mass ratio $\nu=\mu/M=m_1m_2/(m_1+m_2)^2$ 
by imposing the following {\it requirement}: that the  maximum 
of $\left|\Psi_{22}^{\rm  EOB}(t)\right|$ be on top of the
$\Omega(t)$. 

In principle, the determination of $a$ by this requirement depends on
the choice of the other EOB flexibility parameters. In other words, 
the satisfaction of this condition will determine $a$ as a function
of all the parameters entering the EOB dynamics and inspiral
waveform: $a=a(a_5,\a,\vp,\nu)$. In practice, however, and as a first
step towards a fully consistent choice of all the EOB flexibility 
parameters, we fixed $a$ in the following way.
In previous work it was found both analytically 
(when $\nu\ll 1$, see  Ref.~\cite{Damour:2007xr}) and numerically 
(when $\nu=1/4$, see Ref.~\cite{Damour:2007vq}) that the value 
$a=0.5$, together with $\epsilon_a=0.12$, led to a sufficiently
accurate solution of the above requirement. For the present
work, we partially took into account the parameter dependence
of $a$ by fixing $(a_5,\vp,\a)$ to the central best-fit values
that we will select below and by then numerically finding the
optimal value of $a$ as a function of $\nu$ only. In particular,
we identified the following pairs $(\nu,a)$ of near-optimal values:
$(0.25,0.44)$, $(0.2222,0.49)$, $(0.16,0.64)$, $(0.05,0.905)$ 
and $(0.01,0.985)$. These are the values that we shall use in 
this work. Note also that the  $\nu$-dependence can be approximately
represented by a simple linear fit, namely $a(\nu)=1.019-2.345\nu$.

As for the NQC parameter $b$ entering the
phase of the (quadrupolar) waveform, previous work~\cite{Damour:2007vq} has 
found that it had a very small effect (when using the new, $3^{+2}$-PN
accurate EOB waveform which already includes the leading NQC phase correction)
and that it could simply be set to $b=0$. We shall also do so here.

\section{Selecting best-fit EOB flexibility parameters}
\label{sec:calibrate}

As recalled in the Introduction, and in the previous section, 
the only EOB flexibility parameters
whose best-fit values are still quite indeterminate are 
$a_5$, $\vp$ and $\a$. In this section we shall show how to
remedy this situation by combining information coming from
various NR data, namely, on the one hand, from published 
Caltech-Cornell data, and, on the other hand, from 
recently computed Jena data (reported here).

\subsection{Using Caltech-Cornell published data to determine
$\vp$ and $\a$ as functions of $a_5$}

To start with, let us recall that Ref.~\cite{Damour:2007yf} 
had fixed $\a=0$ and had then showed that imposing one 
constraint relating the EOB waveform and Caltech-Cornell {\it inspiral} 
data, namely $\rho^{\rm bwd}_{\omega_4}(a_5,\vp)=1$,
(see Eq.~(35) in~\cite{Damour:2007yf}), implied a rather
precise functional relationship between $\vp$ and $a_5$ 
(see Fig.~3 there). More recently, Ref.~\cite{Damour:2007vq}
compared the same type of EOB waveform with NR waveforms, 
computed with the CCATIE code of the Albert Einstein Institute,
and suggested that it might be useful to flex the EOB waveform
by introducing a nonzero value of $\a$, i.e.\ a non quasi-circular 
correcting factor $f^{\rm NQC}$, Eq.~\eqref{fnqc}, in the radiation
reaction.
Here we shall combine these two strategies by starting from 
an EOB waveform depending on the three a priori independent 
parameters $(a_5,\,\vp,\,\a)$ and by imposing 
{\it two independent constraints} relating the EOB waveform 
to published Caltech-Cornell data.
These constraints have the form
\begin{align}
\label{constraints}
\rho^{\delta t_{\omega_4}}_{\omega_4}(a_5,\a,\vp)=1, \\ 
\label{constraints2}
\rho^{\delta t_{\omega_4}'}_{\omega_4}(a_5,\a,\vp)=1, 
\end{align}
where 
\begin{equation}
\label{eq:ratio_rho}
\rho_{\omega_m}^{\delta t_{\omega_m}}(a_5, \a, \vp)\equiv\dfrac{\Delta^{\omega_m}
               \phi_{\rm T4EOB}\left( t^{\omega_m}_{\rm NR}+\delta
               t_{\omega_m}\right)}{\delta_m} \ .
\end{equation}
Here $\Delta^{\omega_m} \phi_{\rm T4EOB}\left( t^{\omega_m}_{\rm NR}+\delta
t_{\omega_m}\right)$ is the value at the time  $t^{\omega_m}_{\rm NR}+\delta t_{\omega_m}$
of the {\it curvature} waveform ($\psi_4^{22}$)  phase difference 
between T4 and EOB when the two waveforms are matched at the gravitational 
wave frequency $\omega_m$ (in the sense of~\cite{Boyle:2007ft}).
The $\delta_m$'s appearing in Eq.~\eqref{eq:ratio_rho} are estimates
of the value of the phase difference between TaylorT4 and 
Caltech-Cornell numerical relativity data at various times differing from the
matching time $t^{\omega_m}_{\rm NR}$ by $\delta t_{\omega_m}$, 
as measured by us on the left panel of  Fig.~19 of~\cite{Boyle:2007ft}.
Following the procedure outlined in Sec.~IV of~\cite{Damour:2007yf},
we use the matching frequency $\omega_4\equiv\omega_m=0.1$.
Then, we consider two of the measured values which have been used
to produce the empty circles appearing in Fig.~5 of~\cite{Damour:2007yf}.
These two values are 
\begin{align}
\label{delta4a}
\delta_4 &= 0.055\quad   {\rm corresponding \quad to\quad} \delta t_{\omega_4}=-1809M ,\\
\label{delta4b}
\delta_4'&= 0.04\;\;\quad  {\rm corresponding \quad to\quad} \delta t'_{\omega_4}=-529M .
\end{align}
The data point $( \delta t_{\omega_4},\delta_4)$ corresponds to the leftmost
empty-circle on the top panel of Fig.~5 of~\cite{Damour:2007yf}, while
the point  $( \delta t'_{\omega_4},\delta_4')$ corresponds to the next
to next empty circle on the right of $( \delta t_{\omega_4},\delta_4)$.
The former data point was used in Ref.~\cite{Damour:2007yf} as the
``main backward'' $\omega_4$ data. Note that the new data point that
we use here is also ``backward'' (with respect to $\omega_m=\omega_4=0.1$),  
though it is less ``backward'' by about a factor three. 
We use these two points here because we think they
represent the best ``lever arms'' to exploit the
approximate~\footnote{As a measure of the accuracy of the approximate data
points quoted in Eqs.~\eqref{delta4a}-\eqref{delta4b} above, let us mention
that we have, since, directly determined from the Caltech-Cornell numerical
data provided to us the values $\delta_4 = 0.05497$ corresponding to 
$\delta t_{\omega_4}=-1809M$, and $\delta_4'=0.03957$ corresponding 
to $\delta t'_{\omega_4}=-529M$ (with a numerical relativity time 
$t_{\omega_4}=3782.1489M$ corresponding to $\omega_4=0.1$). } 
numerical data represented in Fig.~5 of~\cite{Damour:2007yf}. 
In particular, we do not use any ``forward'' data point because 
the accuracy with which we could measure them is more uncertain.
Let us emphasize that, as a consequence of this choice, 
our determination of the functional relationships $v_{\rm pole}(a_5)$ and $\bar{a}_{\rm RR}(a_5)$ 
exhibited below only relies on Caltech-Cornell data up to gravitational
wave frequency $M\omega\leq 0.1$.
\begin{figure}[t]
  \begin{center}
    \includegraphics[width=90 mm]{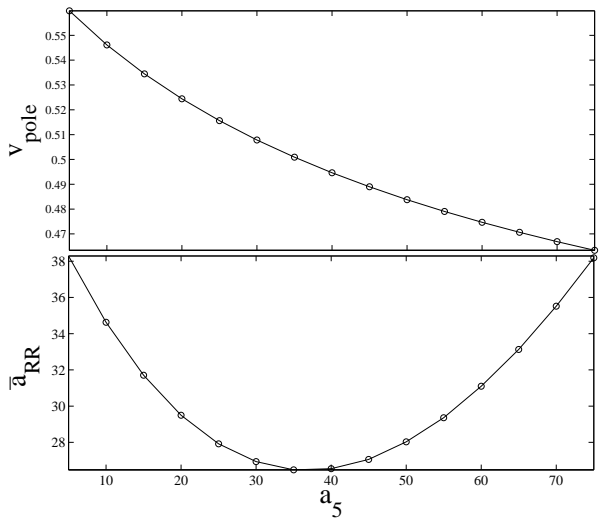}
  \end{center}
\vspace{-4mm}
  \caption{\label{label:fig4} Functional relationships linking 
    $\vp$ and  $\a$ to $a_5$ obtained by imposing the two 
    constraints~\eqref{constraints}-\eqref{constraints2} based on
    published Caltech-Cornell {\it inspiral} waveform data.}
\end{figure}

  \begin{table}[t]
    \caption{\label{table:versus_a5} Explicit values of the EOB 
      effective parameters $\a$ and $v_{\rm pole}$ for a certain
      sample of $a_5$. These values  correspond to imposing 
      the two constraints
      $\rho_{\omega_4 }^{\delta t'_{\omega_4}}
      \simeq 1\pm 10^{-4}\simeq \rho_{\omega_4}^{\delta t_{\omega_4}}$.}    
    \begin{ruledtabular}
      \begin{tabular}{ccc}
	$a_5$ & $\a$ & $v_{\rm pole}$ \\
	\hline
	  5.0000 &    38.286713287  &    0.559878668 \\
	 10.0000 &    34.630281690  &    0.546122851 \\
	 15.0000 &    31.708633094  &    0.534478193 \\
	 20.0000 &    29.496402878  &    0.524422704 \\
	 25.0000 &    27.919708029  &    0.515629404 \\
	 30.0000 &    26.940298507  &    0.507845655 \\
	 35.0000 &    26.484962406  &    0.500903097 \\
	 40.0000 &    26.545801527  &    0.494646066 \\
	 45.0000 &    27.057692308  &    0.488978922 \\
	 50.0000 &    28.031496063  &    0.483798488 \\
	 55.0000 &    29.360000000  &    0.479064301 \\
	 60.0000 &    31.097560976  &    0.474690707 \\
	 65.0000 &    33.130252101  &    0.470660186 \\
	 70.0000 &    35.517241379  &    0.466908044 \\
	 75.0000 &    38.189655172  &    0.463416027 \\
      \end{tabular}
    \end{ruledtabular}
  \end{table}

The two constraints~\eqref{constraints}-\eqref{constraints2} 
were solved by numerical
Newton-Raphson iteration in $\vp$ starting from a grid of values
of $(\a,a_5)$. The iteration was stopped when the constraints were
satisfied to better than the $10^{-4}$ level.
The result of this procedure consists of two separate functional 
relations linking, on the one hand, $\vp$ to $a_5$ and, on the other
hand, $\a$ to $a_5$. These two functional relations are plotted 
in Fig.~\ref{label:fig4}. The upper panel of the figure is a modified
version of the $\vp(a_5)$ functional relationship represented 
in the upper panel of Fig.~3 of~\cite{Damour:2007yf}.
The latter curve was drawn by fixing $\a$ to zero
and by imposing only the first constraint, 
$\rho^{\delta  t_{\omega_4}}_{\omega_4}(a_5,0,\vp)=1$.
By constrast, the curve $\vp(a_5)$ in the upper panel of 
Fig.~\ref{label:fig4} was obtained by {\it simultaneously} tuning 
$\vp$ and $\a$ so as to satisfy the {\it two} 
constraints~\eqref{constraints}-\eqref{constraints2}. 
The numerical data behind the plots of Fig.~\ref{label:fig4} are also
given in explicit numerical form in Table~\ref{table:versus_a5}.
[The many digits quoted there are only given for comparison purposes.]

\begin{figure}[t]
  \begin{center}
    \includegraphics[width=90 mm]{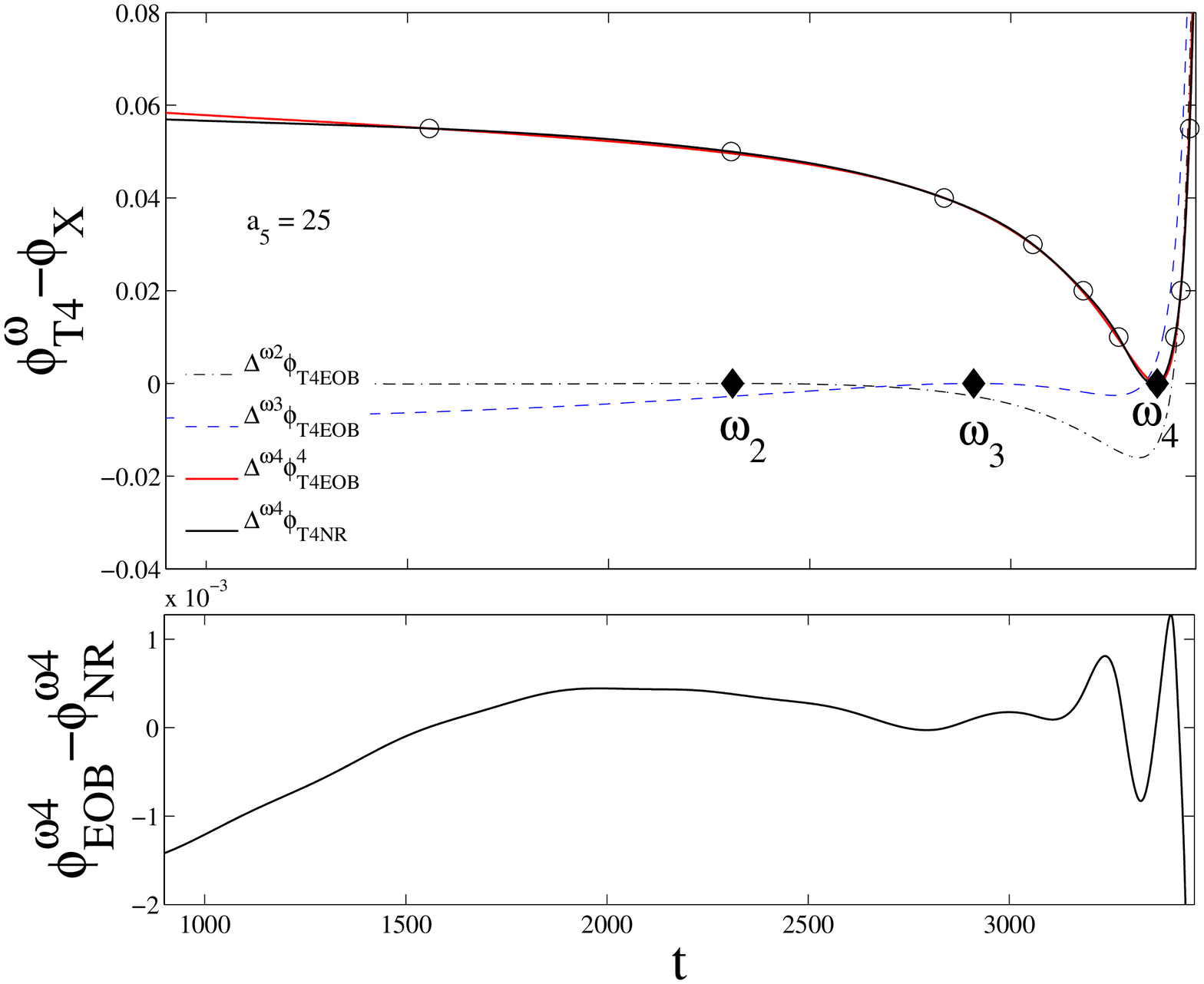} 
  \end{center}
  \caption{\label{label:figCC}Top panel: near-perfect agreement between T4-EOB and T4-NR 
   phase differences when $a_5=25$, $\a=27.9197$ and $\vp=0.51563$. Here NR
   refers to the published results of the Caltech-Cornell inspiral simulation.
   The corresponding EOB-NR phase difference (bottom panel) is of the order of
   $10^{-3}$ radians over the 30 GW cycles of the Caltech-Cornell inspiral simulation.}
\end{figure}

In the upper panel of Fig.~\ref{label:figCC} we exhibit, 
for the particular value
$a_5=25$ (and, correspondingly, $\a=27.9197$ and $\vp=0.51563$) 
the near-perfect agreement between the two $\omega_4-matched$ 
phase differences $\phi_{\rm T4}-\phi_{\rm EOB}$ and 
$\phi_{\rm T4}-\phi_{\rm NR}$. 
[Our choice of the particular value $a_5=25$ will be motivated in the next subsection].
For completeness, we have also included in the upper panel 
(see dash and dash-dot curves) the analogous phase differences 
matched at the matching frequencies $\omega_2=0.05$ and $\omega_3=0.063$
instead of $\omega_4=0.1$. The visual agreement between these
three phase-difference curves and the corresponding ones displayed
in the left panel of Fig.~19 in Ref.~\cite{Boyle:2007ft} is
striking. [As in Fig.~19 of~\cite{Boyle:2007ft},   
we use here TaylorT4 3.5/2.5; see Appendix~\ref{sec:T4} for its precise definition].
The lower panel of Fig.~\ref{label:figCC} plots the $\omega_4$-matched 
phase difference $\phi_{\rm EOB}-\phi_{\rm NR}=[\phi_{\rm T4}-\phi_{\rm NR}] -
[\phi_{\rm T4}-\phi_{\rm EOB}]$, i.e., the difference between 
the two solid curves (red online and black)  in the upper panel.

Note that this phase difference varies 
between
$-0.0014$ and $+0.0008$ radians over 
the time span (of $\sim 2464M$ up to $M\omega= 0.1$)
which was used in our EOB-CC fitting procedure.
A study of the continuation of the curve
exhibited in the bottom panel of Fig.~\ref{label:figCC} then
shows that, after $M\omega=0.1$ and up to a final frequency 
$M\omega=0.1565$, this phase difference
varies  between $+0.0013$ and $-0.0185$ radians over a time span 
of $96M$. Note that a recent report~\cite{KidderJena} has indicated
that a refined estimate of the total phasing error in the Caltech-Cornell
simulation was of the order of 0.01 radians over the entire span of the
simulation. Therefore the accuracy of our EOB-fit is consistent with such
an error estimate.  

Summarizing so far: by best fitting the three-parameter flexed EOB 
waveform $\Psi_{22}^{\rm EOB}(a_5,\vp,\a;\,t)$ to 
published~\footnote{Since 
we had had recently access to the actual Caltech-Cornell data we 
could and did check the reliability of the results obtained from 
the published data. In particular, when computing the phase difference
$\phi_{\rm EOB}-\phi_{\rm CC}^{\rm actual}$ we essentially recovered
the results quoted in the text. For instance, we find that the 
actual phase difference varies between: $-0.002$ radians at Caltech-Cornell
time $600M$ and $-0.01766$ radians at the end of the simulation 
($M\omega=0.1565$), passing through zero at $M\omega=0.1$.
The number of GW cycles between $t_{\rm CC}=600M$ and $t_{\rm CC}=3782$
($M\omega=0.1$) is $22.10$, while the number of GW cycles in the final part
of the simulation (after frequency 0.1) is 2.16.}
Caltech-Cornell {\it inspiral} data before $M\omega=0.1$ (in the sense of imposing 
the two constraints Eqs.~\eqref{constraints}-\eqref{constraints2})
we have reduced the number of independent unknown EOB flexibility 
parameters to only one, namely the ``4~PN'' EOB parameter $a_5$.
The basic physical reason behind the difficulty of determining 
$a_5$ by means of inspiral data only (especially when relying, as we did
above) on data below GW frequency 0.1, is the fact that $a_5$ starts
significantly affecting the EOB dynamics (and waveform) only 
during the late inspiral, when the dynamics becomes strongly 
nonadiabatic. Our next step will be to constrain $a_5$
by best fitting the EOB waveform to numerical data covering 
more of the late-inspiral dynamics.

\subsection{Using numerical data covering late-inspiral and plunge
 to constrain the ``4~PN'' EOB flexibility parameter $a_5$}

In this subsection we shall fulfill, at least in first approximation, the aim
of the EOB-NR comparisons initiated in 
Refs.~\cite{Damour:2002qh,Buonanno:2007pf,Damour:2007yf,Damour:2007vq};
i.e., to determine an essentially unique set of ``best-fit'' EOB 
flexibility parameters $(\vp,\a,a_5)$.
In view of the results of the previous subsection, we now need to best-fit
the {\it one-parameter flexed} EOB waveform 
\begin{equation}
\label{eob_psi_a5}
\Psi^{\rm EOB}_{22}(a_5;\,t)\equiv \Psi^{\rm EOB}_{22}[\vp(a_5),\a(a_5),a_5;\,t],
\end{equation}
where $\vp(a_5)$ and $\a(a_5)$ are the functional relationships illustrated
in Fig.~\ref{label:fig4} above, to a numerical waveform smoothly connecting, 
without interruption, the nonadiabatic late-inspiral to the early-inspiral and
to the subsequent plunge.
Here we shall make use of recently computed numerical data 
(see Sec.~\ref{sec:NR}) that cover (for the equal mass case) about 
20 GW cycles of inspiral and plunge up to merger. As we shall see,
for the purpose of determining $a_5$, we will mainly use the signal
only up to the plunge.

As quantitative measure of the EOB-NR agreement we shall consider 
here the following $L_{\infty}$ norm of the 
$a_5$-dependent EOB-NR phase difference (using the EOB
metric waveform, Eq.~\eqref{eob_psi_a5} above)
\begin{align}
\label{Linf_norm}
&||\Delta\phi||_{\infty}^{\rm EOBNR}(a_5;\,t_1, t_2;\,t_{\rm L}, 
t_{\rm R})\equiv\nonumber\\
& {\rm sup}_{t\in[t_{\rm L},t_{\rm R}]}
\left|\phi_{22}^{\rm EOB}(a_5;\,t+\tau_{12})+\alpha_{12} -\phi^{\rm
  NR}_{22}(t)\right|.
\end{align}
Here $[t_{\rm L},t_{\rm R}]$ denotes the time interval on which one
computes the $L_{\infty}$ norm of the phase difference.
In addition, $(t_1,t_2)$ denote two ``pinching'' times which are used 
to determine some time and phase shifts, $\tau_{12}=\tau(t_1,t_2)$ and
$\alpha_{12}=\alpha(t_1,t_2)$, needed to compare the EOB and
NR phase functions (which use different time scales and phase references).

Let us recall the ``two-pinching-times'' procedure, introduced
in~\cite{Damour:2007vq}, for determining the time and 
phase shifts $\tau$ and $\alpha$.
First, the two waveforms being complex numbers, we decompose them
in amplitude and phase: $\Psi_{22}^{X}=A_{X}\exp(-\i\phi^X)$ 
where the label X can be either ``EOB'' or ``NR''. 
The corresponding instantaneous (metric) GW frequencies are then defined
as $\omega^X(t)\equiv d\phi^X/dt$.
We start by fixing two ``pinching'' times  $(t_1, t_2)$ on the NR time scale
$t$. We then define the time-shift $\tau$ by solving the equation 
$\phi^{\rm NR}(t_2)-\phi^{\rm NR}(t_1)=\phi^{\rm EOB}(\tau+t_2)-\phi^{\rm EOB}(\tau+t_1)$.
Then, we define the phase shift $\alpha$ such that 
$\phi^{\rm NR}(t_1)=\phi^{\rm EOB}(t_1+\tau)+\alpha$.
In the limiting case where the corresponding GW 
frequencies $\omega_1=\omega^{\rm  NR}(t_1)$ and 
$\omega_2=\omega^{\rm  NR}(t_2)$ are nearly coincident,
$\omega_1\approx\omega_m\approx\omega_2$, this procedure coincides
with the one introduced in Ref.~\cite{Boyle:2007ft} and based on
the choice of a single matching frequency $\omega_m$.

\begin{figure}[t]
  \begin{center}
    \includegraphics[width=95 mm]{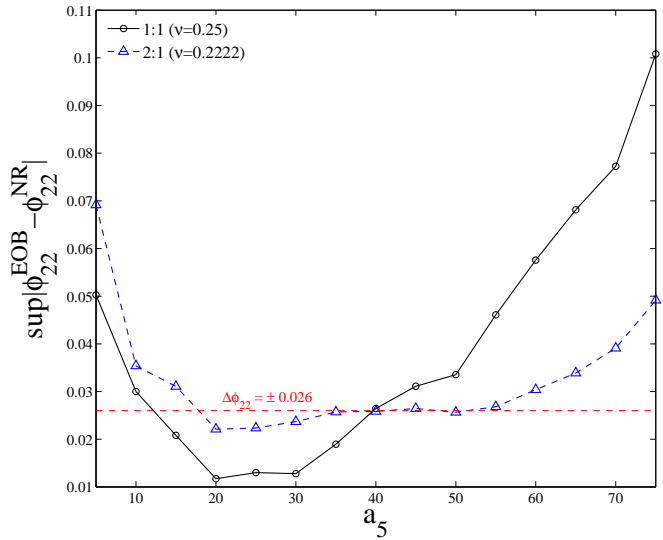} 
  \end{center}
  \caption{\label{label:fig6} $L_{\infty}$ norm of the EOB-NR
    late-inspiral ($[t_{\rm L},t_{\rm R}]$)
    phase difference, as a function of $a_5$ for $\nu=0.25$ 
    (1:1 mass ratio) and $\nu\simeq 0.2222$ (2:1 mass ratio).
    NR refers to results of Jena coalescence simulations reported here.}
\end{figure}
\begin{table}[t]
 \caption{\label{table:pinch} Pinching NR times and corresponding NR
    gravitational wave frequencies used to perform the EOB-NR 
    comparison of Fig.~\ref{label:fig7}.}
    \begin{ruledtabular}
      \begin{tabular}{ccccc}
	$\nu$ & $t_1$ & $t_2$ & $\omega_1^{22}$ & $\omega_2^{22}$ \\
	\hline
	  0.25  &    1764.9  &    1940.1 & 0.0998 & 0.4716\\
	 0.2222 &     893.9  &    1071.9 & 0.1005 & 0.4542\\
	 0.16   &    1297.6  &    1476.3 & 0.1051 & 0.4189\\
     \end{tabular}
   \end{ruledtabular}
\end{table}
We shall first consider the equal-mass case, $\nu=1/4$. For this case
we choose the following NR pinching times: $t_1=1764.9$ and $t_2=1940.1$ 
(corresponding to NR gravitational wave frequencies $\omega_1^{22}=0.0998$ 
and $\omega_2^{22}=0.4717$).
These times {\it bracket the merger time}. This is done to 
optimize the EOB-NR agreement 
over the physically most crucial (and possibly numerically most accurate)
part of the waveform, i.e. the late-inspiral, plunge, merger and ringdown.
Concerning the choice of the interval $[t_{\rm L},t_{\rm R}]$ 
used to compute the $L_{\infty}$ norm, we selected it with the following
criteria in mind: as $a_5$ is most important during late-inspiral and
plunge, but is somewhat uncorrelated to the way EOB approximates the
plunge-ringdown matching, we chose $[t_{\rm L},t_{\rm R}]$ to cover the
crucial stage of the late inspiral. More precisely, we have fixed 
$t_{\rm R}$ such that the NR
gravitational wave phase is approximately 7.6 radians (i.e. 1.21 GW cycles) 
smaller than the phase when the EOB waveform modulus reaches its maximum 
(which is close to merger time in view of the discussion of Sec.~\ref{sec:eob}).
Then, $t_{\rm L}$ was chosen such that 
$\phi^{\rm NR}(t_{\rm L})_{22}=\phi^{\rm  NR}_{22}(t_{\rm R})-58.5$.
Their numerical values are $t_{\rm L}=1198.8M$ and $t_{\rm R}=1899.6M$,
while the corresponding NR gravitational wave frequencies
are $\omega_{\rm L}^{22}= 0.05952$ and $\omega_{\rm R}^{22}= 0.1898$ respectively. 
Using these specified values we have computed the $L_{\infty}$ norm 
of the EOB-Jena phase difference, Eq.~\eqref{Linf_norm}, as a function of
$a_5$. The result is plotted, as a solid line (1:1 mass ratio), in Fig.~\ref{label:fig6}.
This figure shows that the limited range of values $20\lesssim a_5\lesssim 30$
is preferred in that it yields a {\it minimum} of the largest EOB-NR phase 
difference  $||\Delta\phi||_{\infty}^{\rm EOBNR}$ over the 
$[t_{\rm L},t_{\rm R}]$ interval specified above.
This minimum phase difference is on the order of 0.01 radians.
We note, in passing, that this late-inspiral interval partially overlaps (frequency-wise) 
with the range of the published Caltech-Cornell data as we used it
above (i.e., focusing on frequencies $M\omega\leq 0.1$), but crucially 
extends to frequencies reaching roughly as high as the EOB adiabatic LSO 
frequency ($\omega_{\rm LSO}^{\rm EOB}= 0.2114$).
Though Fig.~\ref{label:fig6} is qualitatively similar to 
the $L_{\infty}$ norm of the EOB/Caltech-Cornell phase difference 
displayed in Fig.~4 of Ref.~\cite{Damour:2007yf}, it is important 
to remark that in the latter figure the $L_{\infty}$ norm varied
by only about a factor 2 over the entire $a_5$ range, 
$0\leq a_5\leq 100$. By contrast, in the current Fig.~\ref{label:fig6}
the $L_{\infty}$ norm varies by about a factor 2 in the much smaller 
interval $15\leq a_5\leq 35$ and then increases by almost
a factor 10 over the entire $a_5$ range, $5\leq a_5\leq 75$.
We can now use the ``uncertainty level'' 
$\pm0.026$ radians in $\phi^{\rm NR}_{22}\equiv\phi_{22}^{\rm Jena}$ 
(determined in Sec.~\ref{sbsc:we} above by comparing it with Caltech-Cornell
data), as indicated by the horizontal line in the figure, to determine
a corresponding interval of ``best-fit'' values of $a_5$.
Though this uncertainty level is admittedly rather uncertain at this stage,
it suggests that the ``real''~\footnote{Note that though $a_5$ is, to start with,
a theoretically well defined quantity within the EOB framework, its
``experimental measurement'' obtained by comparing specifically resummed
versions of the EOB waveforms with numerical data partially transforms it 
into an ``effective parameter'' describing a complicated nonperturbative process.}
value of $a_5$ probably lies in the interval $12 \lesssim a_5 \lesssim 40$.
To firm up our conclusion, we have also considered numerical data concerning 
the 2:1 mass ratio case. In that case we considered again the $L_{\infty}$
norm, Eq.~\eqref{Linf_norm}, and we made similar choices both for the pinching
times and for the extremities of the $L_{\infty}$ interval. In particular,
$t_{\rm R}$ was chosen to sit 7.6 radians before the maximum modulus while
we kept the left-right phase difference to the same value as above,
namely $\phi_{22}^{\rm NR}(t_{\rm L})=\phi_{22}^{\rm  NR}(t_{\rm R})-58.5$.
The resulting $L_{\infty}(a_5)$ function is plotted as a dashed line
in Fig.~\ref{label:fig6}. Though the minimum of this curve is much more
shallow than before, the important fact is that the 1:1 preferred $a_5$
range is consistent with the 2:1 $L_{\infty}$ result.
Let us observe (without wishing to attribute any deep significance to this
fact) that the preferred range for $a_5$ happens to be close to
the ``special'' $a_5$ value for which the ``EOB-horizon'' decreases, when $\nu$ 
increases up to $1/4$, 
down to a vanishing EOB radial coordinate. 
Indeed the $P^{1}_{4}$ Pad\'e approximant that
we use here to define the $a_5$-flexed EOB radial potential
$A(u)=P^1_4[A^{\rm Taylor}(u)]$ has the structure 
$A(a_5;\,u)=(1-r_H u)/D_4(u)$ where $D_4(u)$ is a 4th-order 
polynomial in $\nu$ (see Eq.~(3.10d) of Ref.~\cite{Damour:2002vi}), and where
\begin{equation}
  r_H(a_5,\nu) = 4\;\dfrac{768-(3584 -123\pi^2)\nu-24a_5\nu}{1536-(3776-123\pi^2)\nu} .
\end{equation} 
Here $r_H$ is the radial location of the ``EOB horizon'', in the
sense that $A(u)$ vanishes for $r\equiv 1/u=r_H$ (at least when $r_H$ is positive).
For any given positive $a_5$, $r_H$ is a decreasing function of $\nu$. If we
require that $r_H$ stays positive for all values of $\nu\in[0,1/4]$, we find
that $a_5$ must be smaller than the ``special'' value
\begin{equation}
a_5^*=\dfrac{123\pi^2-512}{24}=29.2484.
\end{equation}
Note, however, that there is nothing a priori wrong with higher values of 
$a_5$. In that case the radial function $A(r)$, considered versus $r$, 
has anyway a third-order zero at $r=0$.

Summarizing: by combining the comparison of the EOB waveform with, 
on the one hand, published Caltech-Cornell inspiral data and, on 
the other hand, our coalescence data, we have been able to select
a preferred small region of the EOB flexibility parameters. 
This region is made of (approximately) correlated triplets
$(a_5, v_{\rm pole}(a_5),\a(a_5))$, and is located between 
the second and the 8th lines of Table~\ref{table:versus_a5}.

\section{Detailed EOB-NR waveform comparisons for $\mathbf{a_5=25}$}
\label{sec:best}
To confirm the validity of the conclusions reached in the previous 
section, we shall now study in detail the performance of the center
of the above selected interval, namely $a_5=25$ together with the 
corresponding values of $\a$ and $\vp$ listed in Table~\ref{table:versus_a5}.
\begin{figure*}[t]
\begin{center}
    \includegraphics[width=85 mm]{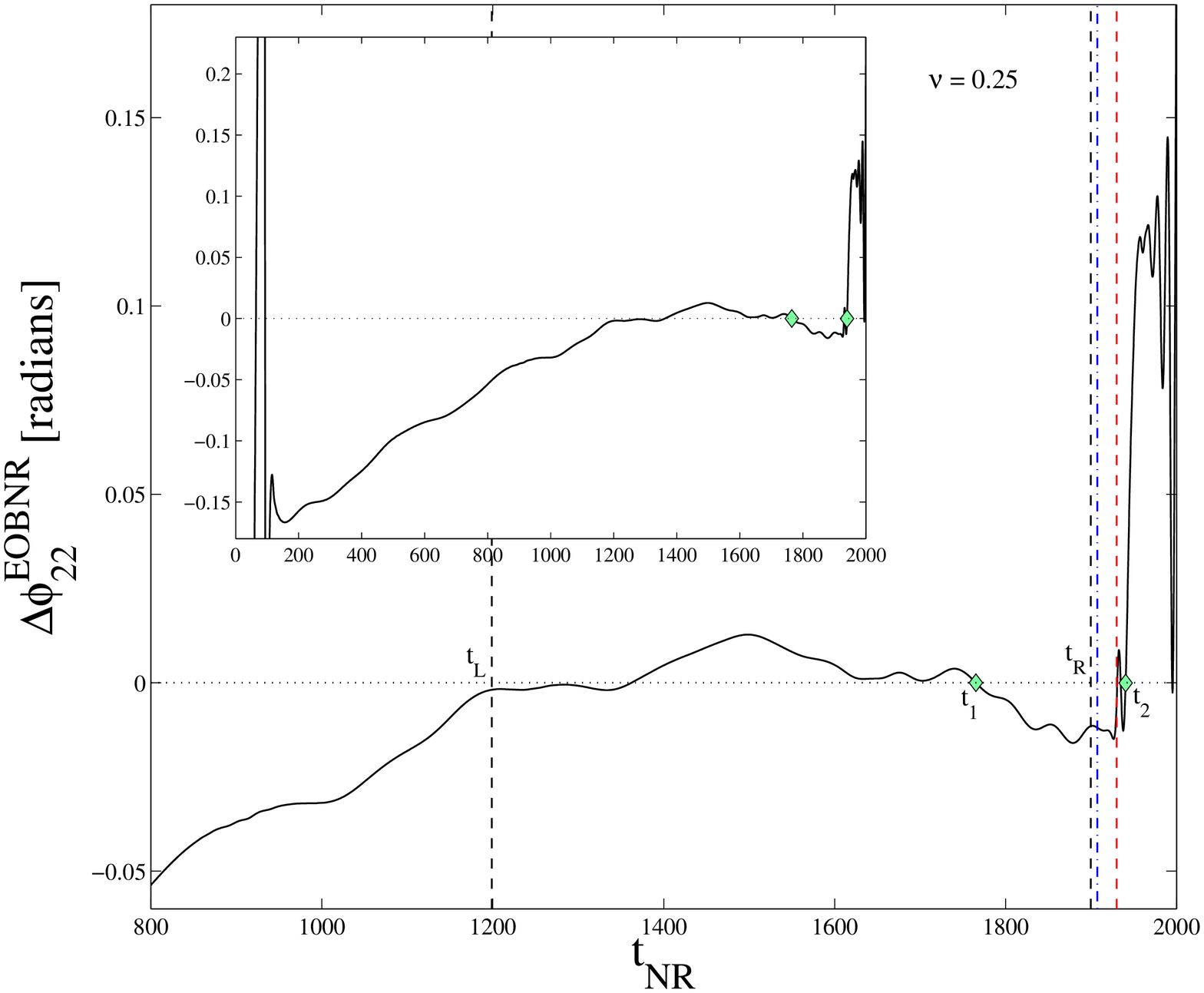} 
    \includegraphics[width=85 mm]{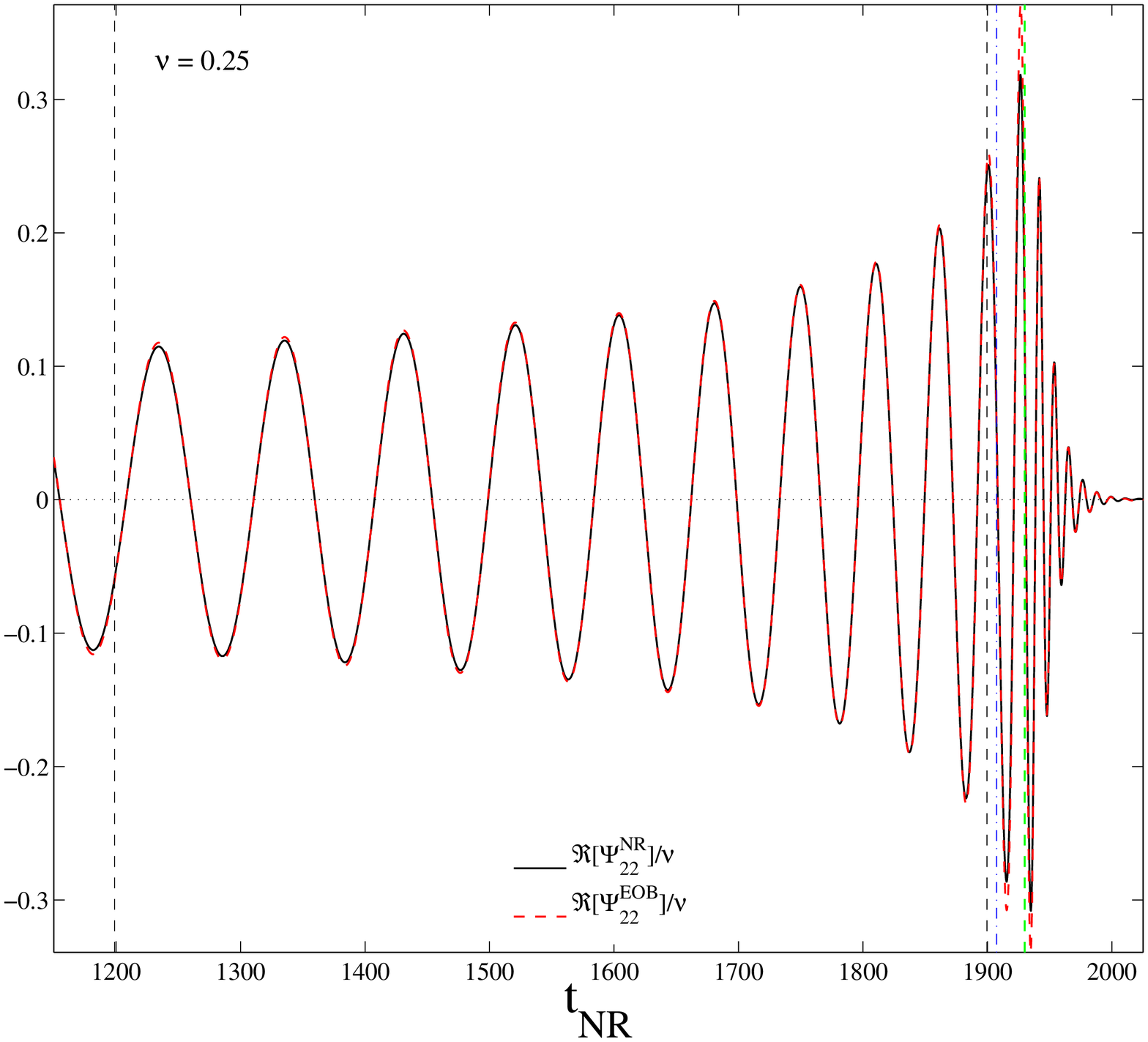}\\ 
    \includegraphics[width=85 mm]{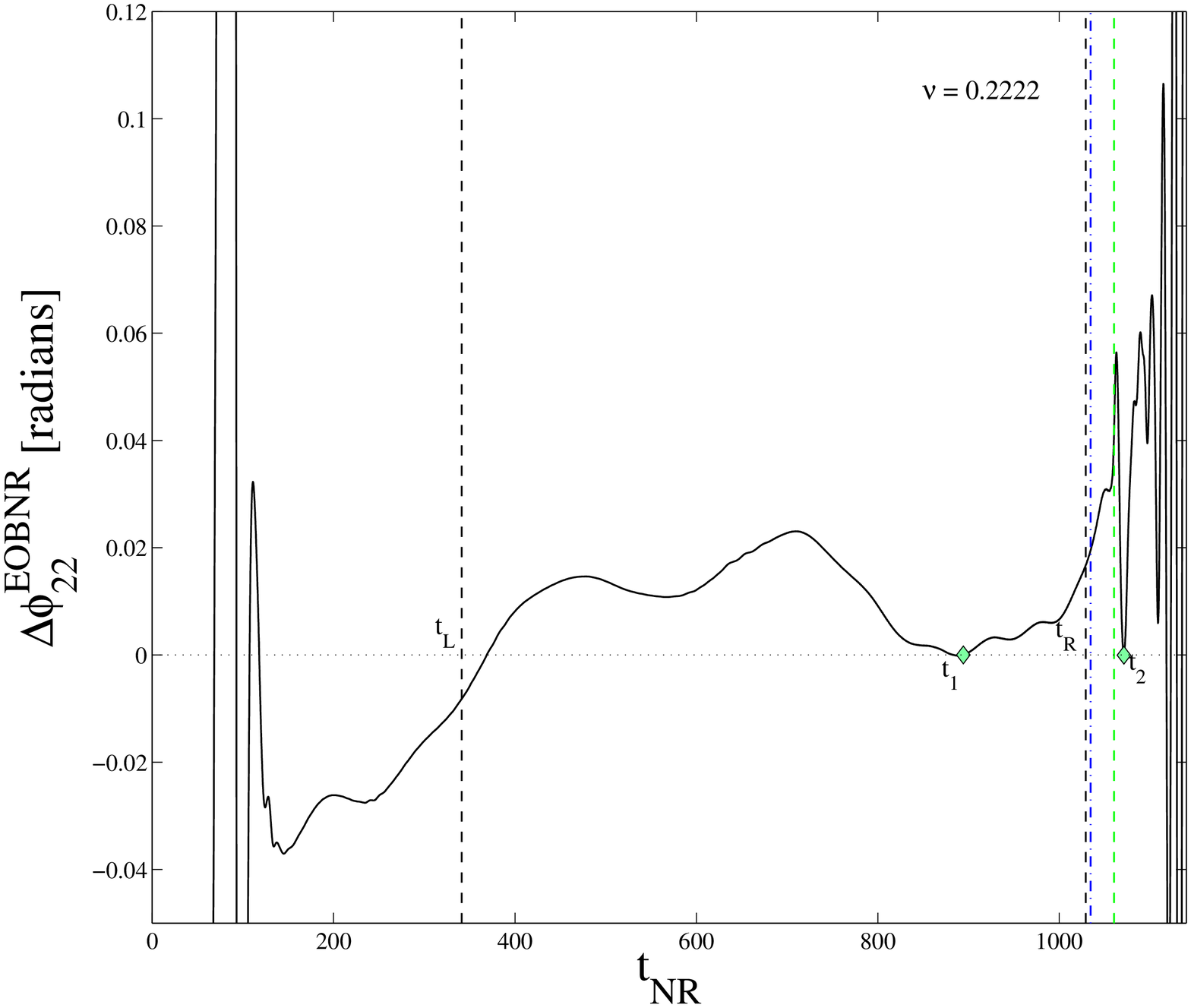} 
    \includegraphics[width=85 mm]{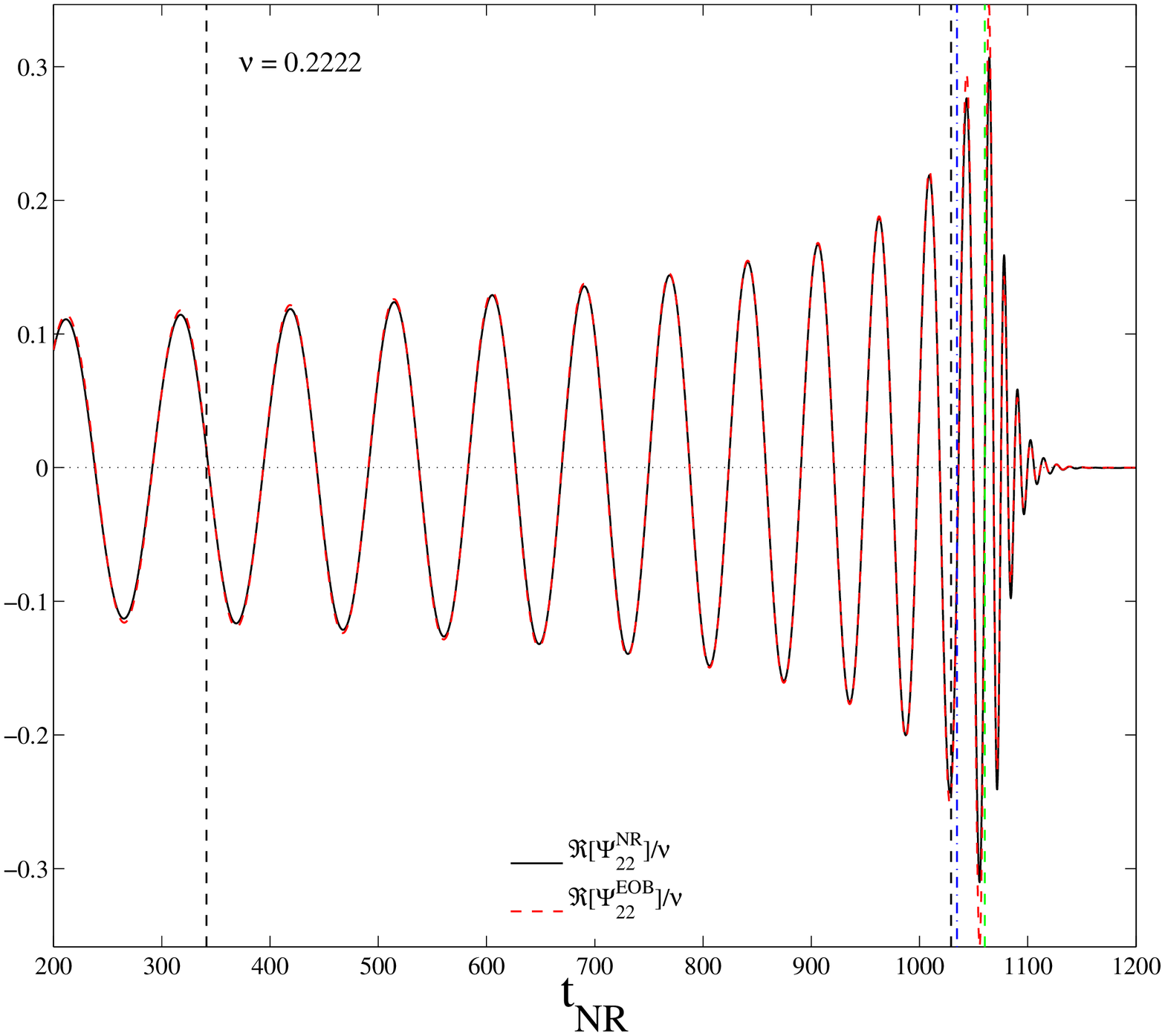}\\
 \caption{\label{label:fig7}Comparison between NR and EOB waveforms for
  $\nu=0.25$ (top), $\nu=0.2222$ (bottom). The left
  panels depict the EOB-NR phase difference; the right panels show the
  real part of the metric waveforms. Here, NR refers to the full results of
  Jena coalescence simulations, from early-inspiral to ringdown, by contrast
  to the $L_{\infty}$ norm of Fig.~\ref{label:fig6} which concerned a late
  inspiral stage $[t_{\rm L},t_{\rm R}]$. This interval is indicated on
  the figures. The pinching times $(t_1,t_2)$ of Table~\ref{table:pinch}
  are also shown. The dash-dot and the dash vertical lines at the extreme
  right of the figures mark the location of the EOB adiabatic LSO and the 
  ``EOB light-ring'' respectively.}
\end{center}
\end{figure*}
\begin{figure*}[t]
\begin{center}
    \includegraphics[width=85 mm]{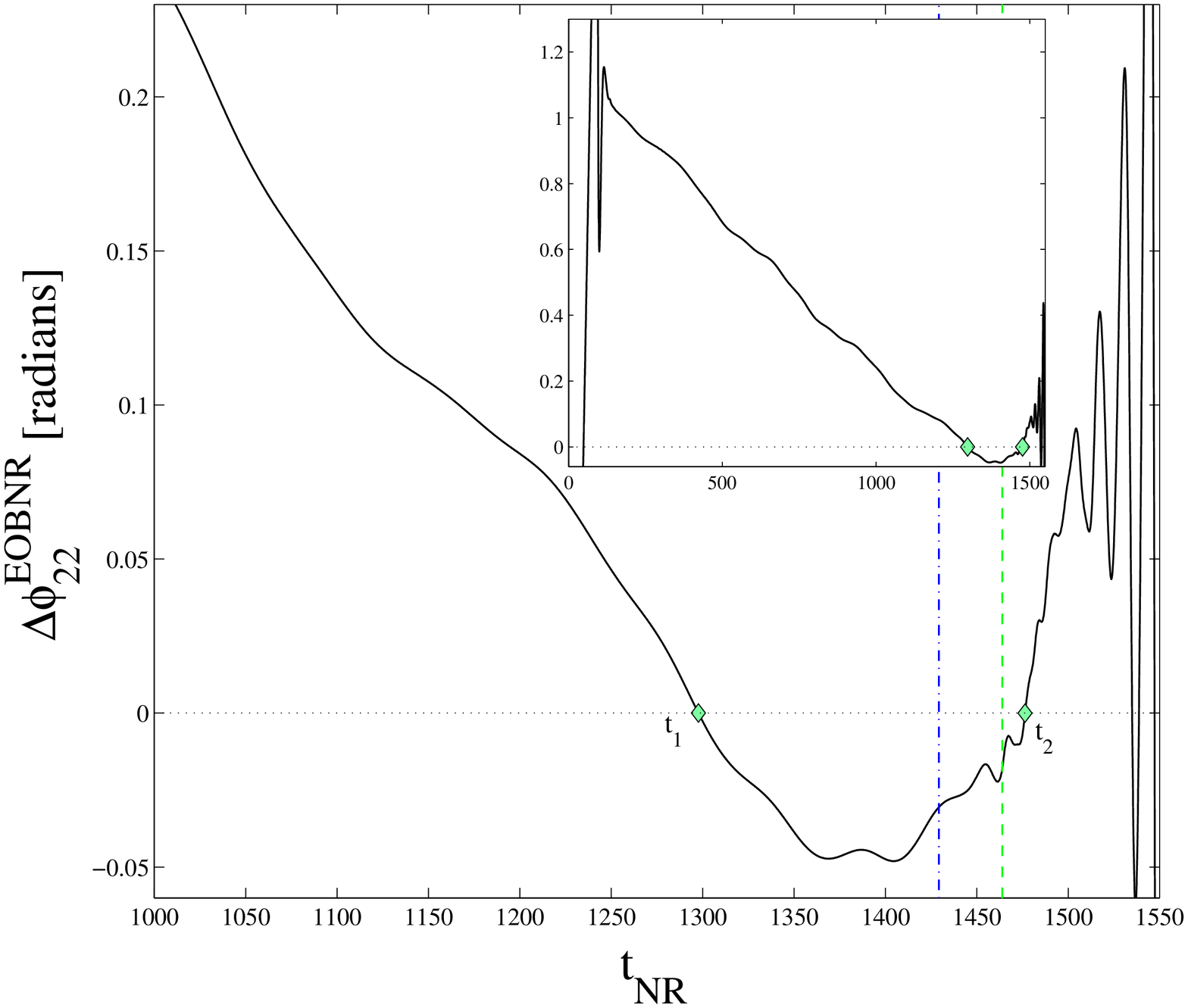} 
    \includegraphics[width=85 mm]{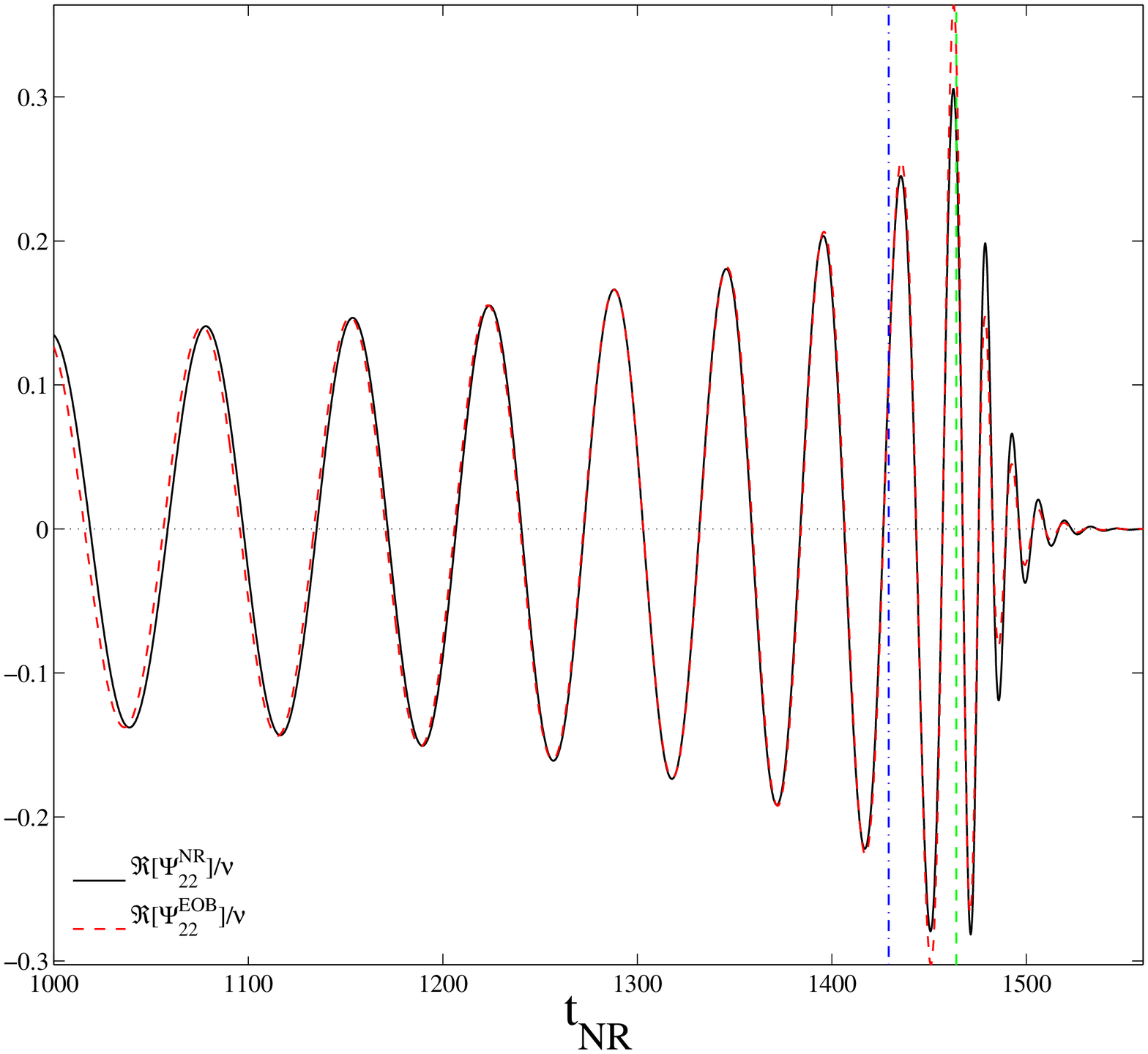} 
 \caption{\label{label:fig016}Comparison between NR and EOB waveforms for
  $\nu=0.16$. The left
  panel depicts the EOB-NR phase difference; the right panels show the
  real part of the metric waveforms. The pinching times $(t_1,t_2)$ of 
  Table~\ref{table:pinch} are also shown.
  The dash-dot and the dash vertical lines at the extreme
  right of the figures mark the location of the EOB adiabatic LSO and the 
  ``EOB light-ring'' respectively.}
\end{center}
\end{figure*}

In this section we shall consider numerical waveforms 
for {\it three} different values of $\nu$, namely $\nu=0.25$,
$\nu=2/9=0.2222$ and $\nu=0.16$ (corresponding respectively
to the mass ratios 1:1, 2:1 and 4:1) extracted from the 
simulations of Table~\ref{tab:table1}. Note that our best-fit procedure
outlined above essentially relied only on the 1:1 mass-ratio case
so that the other cases that we consider here will test the
ability of the EOB formalism to capture the NR waveforms. 
The EOB flexibility parameters used for the various mass ratios are the ones 
listed in the fifth row of Table~\ref{table:versus_a5}. In view of the proximity of the
``best-fit'' $\vp$ value  $\vp^{\rm best}(\nu=0.25)=0.5156$ 
to the ``best-fit'' $\vp$ found  (following the 
strategy of~\cite{Damour:2007yf}), in the test mass 
limit, $\vp^{\rm best}(\nu=0)=0.52655$ 
(for the $P^4_4$ 4~PN-accurate flux), we made no attempt 
at interpolating $\vp(\nu)$ between the two values of $\nu$.

To compare EOB and NR waveforms we follow the procedure indicated above.
This procedure involves choosing two ``pinching'' times $t_1$ and $t_2$ 
(which should not be confused with the $L_{\infty}$ times $t_{\rm L}$ and
$t_{\rm R}$ which will play no role in this section). We summarize in 
Table~\ref{table:pinch} the ``pinching'' times we use, together with
the corresponding frequencies. Note that in all cases the lowest 
pinching frequency is around 0.1 while the highest one (reached after
the merger) is roughly $10\%$ lower than the main ringdown frequency.

The results of the detailed EOB-NR comparison are presented in 
Fig.~\ref{label:fig7} and~\ref{label:fig016}. For completeness, 
we have used the full numerical waveforms including the burst of 
junk radiation it contains at the beginning. 

The two upper panels of Fig.~\ref{label:fig7} refer to the equal-mass 
case ($\nu=0.25$). On the left, we plot the ``pinched'' EOB-NR phase
difference (in radians) over the full simulation time (see inset).
Note that the total simulation covers about $\sim 146$ radians of 
GW phase; i.e., 23.24 GW cycles (starting from the beginning of 
the inspiral, when $t_{\rm NR}\sim 110M$, to the middle of the 
ringdown, up to $t_{\rm NR}=1980M$).
We see that the EOB-NR phase 
disagreement stays quite small during most of the inspiral.
More precisely $\Delta\phi^{\rm EOBNR}$ stays in the range
$[-0.04,\,0.01]$ all over the time interval 
$1200\lesssim t_{\rm NR}\lesssim 1930$. This corresponds to
a ``two-sided'' (in the sense of footnote 12 of Ref.~\cite{Damour:2007vq})
EOB-NR phase difference smaller than $\pm 0.025$ radians, or
$\pm 0.004$ GW cycles over $730M$. 
As in previous analysis, the jump in the phase difference 
around $t_{\rm NR}\approx 1930$ is connected to the rather
coarse way in which the EOB formalism represents the merger.
Still, the accumulated phase difference over the transition
between plunge and ringdown is only of the order of $0.15$ radians;
i.e., 0.02 GW cycles. Note that over the full simulation time 
(see inset in top-left panel) there is an accumulated phase
difference of about -0.2 radians. In view of the discussion 
on the accuracy of the numerical simulations in Sec.~\ref{sec:NR},
it is quite possible that this difference is mainly due to 
effects related to the use of finite extraction radii. 
Similarly, part of the phase disagreement around the merger 
might come from numerical inaccuracies.
The upper right panel of the figure compares
the real part of the two metric waveforms. The visual agreement between
the two is striking, apart from the amplitude disagreement ($\sim 20\%$, see
below)  localized around the merger. In view of the discussion in
Sec.~\ref{sec:NR}, part of this difference might also have a numerical origin.

The bottom panels of Fig.~\ref{label:fig7} refer to the 2:1 
mass ratio case ($\nu=2/9=0.2222$). Here the phase agreement (left panel)
is even better than before. Over the nearly full time interval 
$143\lesssim t_{\rm NR}\lesssim 1100$ the EOB-NR (two-sided) 
phase difference is smaller than $\pm 0.05$ 
radians; i.e., $\pm 0.008$ GW cycles. The corresponding middle-right
panel compares the real part of the two metric waveforms. Again,
the agreement is striking apart from a~$\sim 20\%$ amplitude 
disagreement localized around the merger (see below). 

Finally, Fig.~\ref{label:fig016} deals with the
4:1 mass ratio case ($\nu=0.16$). Here the agreement is still quite
good, though it is noticeably less good than in previous cases. 
Consistently with the discussion of numerical accuracy in Sec.~\ref{sec:NR},
this less compelling accordance is likely to have its origin 
in numerical discretization errors. A clarification of this issue 
would need higher-accuracy simulations.

\begin{figure}[t]
  \begin{center}
    \includegraphics[width=90 mm]{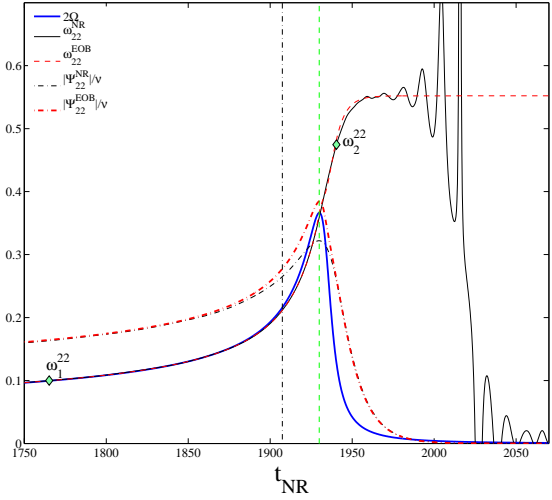} 
  \end{center}
  \caption{\label{label:fig8a} Comparison between EOB and NR instantaneous 
   gravitational wave frequencies (and moduli) for the equal mass 
   case, $\nu=0.25$. Here, as in Fig~\ref{label:fig8}, NR refers to the Jena
   coalescence simulation. The dash-dot vertical line indicates the EOB
   adiabatic LSO, while the dash one the ``EOB light-ring''.
   The pinching frequencies $(\omega^{22}_1,\omega^{22}_2)$ of
   Table~\ref{table:pinch} are also indicated.}
\end{figure}

Figure~\ref{label:fig8a} completes the comparison between EOB and NR waveforms, 
for the equal-mass ratio case, by simultaneously displaying, versus time: 
(i) the two GW frequencies~\footnote{For clarity we add in several figures a 
subscript $22$ to the gravitational wave frequency or phase as a reminder of
the fact that we compare quadrupolar $\l=m=2$ waveforms.} , (ii) twice the EOB 
orbital frequency $\Omega$, and (iii) the two moduli.
The leftmost (dashed) vertical line indicates the location of the EOB adiabatic
LSO, while the rightmost one refers to the ``EOB-light-ring''.
Though this figure exhibits the approximate nature of the EOB matching
procedure (notably visible in the small differences in the GW frequencies), 
it also illustrates how the apparently coarse EOB-matching procedure is 
able to effectively reproduce, with high accuracy, the overall 
time variation of the GW frequency through the merger onto the ringdown.
We have obtained similarly good agreements for the other mass ratios. 

We conclude this section by showing in Fig.~\ref{label:fig8} the fractional amplitude
differences, for the three mass ratios considered here, between 
EOB and NR waveforms. The solid line in the figure plots the quantity
$\Delta A/A\equiv (A_{\rm EOB}-A_{\rm NR})/A_{\rm NR}$ versus NR time for 
$\nu=0.25$.
It is quite possible
that  the approximately linear trend visible on this (solid) line is
due to effects related to the finite extraction radius; the decrease in
amplitude disagreement as we go to later inspiral times is consistent with the 
decrease in amplitude uncertainty (as discussed in Sec.~\ref{sec:NR})
as the amplitude rises. If this is 
the case, the minimum value, before the merger, might be indicative 
of the actual EOB-NR amplitude agreement. For $\nu=0.25$ this minimum is 
${\rm min}[\Delta A/A]\approx +5\times10^{-3}$.
The jump in $\Delta A/A$ during merger is of the order of $20\%$. Though
part of this jump might have a numerical origin, we think that most of 
it comes from the EOB approximate matching procedure around merger.
Let us recall, in this respect, that in Ref.~\cite{Damour:2007vq}
$\Delta A/A$, for $\nu=0.25$, was of order $\pm 1\%$ during inspiral 
and rose to a maximum of $+18\%$ at merger. 
The leftmost curve (dashed line) on Fig.~\ref{label:fig8} refers to
the $\nu=0.2222$ case, while the middle curve (dash-dot line) refers
to the $\nu=0.16$ case. For the same reasons as above it is likely
that the approximate linear trends (which are smaller by a factor $\sim2$
than before) are of numerical origin. The minimum values before merger
of $\Delta A/A$ are ${\rm min}[\Delta A/A]\approx +7\times10^{-3}$ (for
$\nu=0.2222$) and ${\rm min}[\Delta A/A]\approx+5\times10^{-3}$ 
(for $\nu=0.16$). Note that the jumps in $[\Delta A/A]$ around merger
are quite similar to the $\nu=0.25$ case, namely about $\sim +20\%$.

\section{Conclusions}
\label{sec:conclusions}

We have compared the ``flexed''~\cite{Damour:2002vi,Damour:2007xr}
resummed $3^{+2}$PN-accurate~\cite{Damour:2007yf} Effective-One-Body (EOB)
waveform to two, independent, numerical relativity (NR) data on
inspiralling and/or coalescing binary black hole systems: on the one hand, 
published Caltech-Cornell {\it inspiral} data~\cite{Boyle:2007ft} (mainly
used by us only up to $M\omega\lesssim 0.1$) 
and, on the other hand, newly computed {\it coalescence} 
data using the BAM code~\cite{Bruegmann:2006at,Husa:2007hp}.

We effected this EOB-NR comparison with a strategy allowing us to
locate a ``best-fit spot'' in the space of the three main 
EOB flexibility parameters $(a_5,\vp,\a)$.
This strategy is multi-pronged: 
\begin{itemize}
\item We selected two measurements of  published Caltech-Cornell
      equal-mass inspiral data concerning the TaylorT4-NR phase 
      differences at two different times, approximately spanning 
      the GW frequency interval $0.04\lesssim\omega\lesssim 0.1$.
\item We imposed two constraints requiring that these NR phase differences 
      be equal to two corresponding analytical TaylorT4-EOB phase differences,
      see Eqs.~\eqref{constraints}-\eqref{constraints2}. This gave us two
      equations for the three main flexibility parameters $(a_5,\vp,\a)$.
      By numerically solving these two equations we determined two functional
      relationships linking, separately, $\vp$ to $a_5$ and $\a$ to $a_5$.
      See Fig.~\ref{label:fig4} and Table~\ref{table:versus_a5}.
\item Having in hands these ``Caltech-Cornell-preferred'' functional 
      relationships $\vp(a_5)$ and $\a(a_5)$, we selected from our newly
      computed coalescence simulation (again for 
      the equal-mass case~\footnote{When best-fitting the EOB flexibility
      parameters, we use $\nu=0.25$ data because these are more sensitive to
      $a_5$.}) a time interval $[t_{\rm L},t_{\rm R}]$ corresponding to
      the following GW frequency interval $0.060\lesssim\omega\lesssim0.19$. 
      On this 
      time interval we compared the Jena numerically simulated phase evolution
      to the $a_5$-dependent analytical EOB one, and we computed the
      $L_{\infty}$ norm of their difference, i.e. (see
      Eq.~\eqref{Linf_norm} for more details)
     \begin{align}
      ||\Delta\phi&||_{\infty}^{\rm EOBNR}(a_5;\,t_{\rm L},t_{\rm R})  
      \nonumber\\\equiv\
      &{\rm sup}_{t\in[t_{\rm L},t_{\rm R}]} \left|\phi_{22}^{\rm EOB}(a_5;\,t)-\phi^{\rm NR}_{22}(t)\right|.
     \end{align}
%
\begin{figure}[t]
  \begin{center}
    \includegraphics[width=90 mm]{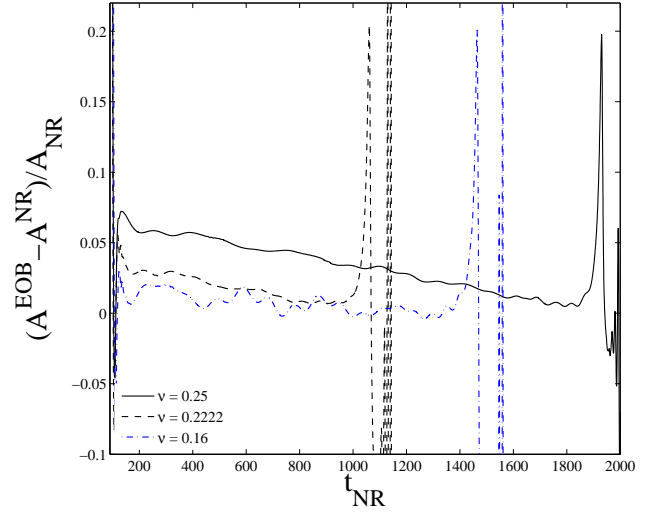} 
  \end{center}
  \caption{\label{label:fig8}Fractional EOB-NR differences in the
  gravtitational wave (metric) amplitudes, $A\equiv|\Psi_{22}|$,
  versus NR time for different mass ratios.}
\end{figure}
%
\item We plotted (as a solid line) in Fig.~\ref{label:fig6} $||\Delta\phi||_{\infty}^{\rm EOBNR}$ 
      as a function of $a_5$. We found that this $L_{\infty}$ norm has a
      rather well localized minimum around the 
      region $20\lesssim a_5\lesssim 30$.
      To transform this fact into an actual ``error-bar'' on the value of
      $a_5$ we would need to have in hands a precise measure of the level
      of the errors present in the (Jena) numerical data over the time
      interval $[t_{\rm L}, t_{\rm R}]$ on which the $L_\infty$ norm is
      computed.
      At this stage we do not have at our disposal a reliable measure of this error.
      However, in Sec.~\ref{sbsc:we} we have given what we think is our 
      current best estimate of this error level by directly comparing, on the crucial time
      interval $[t_{\rm L},t_{\rm R}]$, the Jena phase data to the actual 
      Caltech-Cornell data.
      This current best estimate is $\pm 0.026$ radians and, according to
      Fig.~\ref{label:fig6}, would correspond to the following confidence interval
      for $a_5$:  $12\lesssim a_5 \lesssim 40$. More work is needed to nail down
      in a more precise way the error level in the Jena phase (see in
      particular our discussion above on the internal error estimate
      based on comparing various radius extrapolations methods).
      In addition, for any value of $a_5$ in such an allowed confidence level,
      we would conclude that the corresponding triplets of correlated values 
      $a_5$, $\vp(a_5)$ and $\a(a5)$ obtained from
      Table~\ref{table:versus_a5}, determine preferred best-fit
      values of the EOB flexibility parameters
      \footnote{Note
      that the functional relationships $\vp(a_5)$ and $\a(a_5)$ discussed
      above have no invariant physical meaning and are just intermediate tools in converging
      on the looked-for best-fit point in the three dimensional EOB flexibility 
      parameter space.} $(a_5,\vp,\a)$.
      In other words, our current preferred values of the EOB parameters
      $(a_5,\vp,\a)$ lie between the second an the 8th lines of 
      Table~\ref{table:versus_a5}.
\item The present implementation of this strategy is, however, certainly
      somewhat affected by numerical noise. A possible indication of this
      fact is that the computation of a similarly selected $L_{\infty}$
      norm pertaining to the 2:1 mass ratio simulation gives results
      (plotted as a dashed line in Fig.~\ref{label:fig6}) which, though
      they are fully consistent with the 1:1 mass ratio case, exhibit
      a more shallow minimum versus $a_5$. For the 2:1 $L_{\infty}$
      diagnostic to select an interval of preferred values of 
      $a_5$ we would need a reliable estimate of the numerical error level in
      the 2:1 phase data. However, at this stage we do not have such
      an estimate. The rough error level quoted in Sec.~\ref{sbsc:acc} is just
      a very conservative upper limit which, moreover, does not concern
      the specific time interval $[t_{\rm L}, t_{\rm R}]$ we are interested
      in.
      Let us emphasize that, anyway,
      even if $a_5$ is allowed to vary in the full interval $5\leq a_5\leq 75$ 
      that we explored, the maximum EOB-NR phase disagreement (on the considered
      late-inspiral interval, which corresponds to about 58.5 radians before
      crossing the last stable orbit) is below 0.1 radians, i.e. 0.015 GW cycles.
\item We think that it would be necessary to devote a special effort toward
      having very high-accuracy numerical simulations covering the crucial
      late-inspiral, corresponding to the frequency range 
      $0.1\lesssim \omega\lesssim 0.2$, for several mass ratios. Pending the
      availability of such simulations, we provisionally conclude that our
      current ``best-bet'' choice of EOB flexibility parameters is at the
      center of the above-selected interval; i.e., it is given
      by $a_5\simeq 25$ together with the correlated values of $\vp$ and $\a$
      listed in Table~\ref{table:versus_a5}. In Sec.~\ref{sec:best} we presented 
      evidence that these values of $(a_5,\vp,\a)$ lead  to an excellent 
      agreement between EOB and NR for several mass ratios and for the
      entire time-interval covering inspiral, late-inspiral, plunge, merger
      and ringdown. In particular, we found that the dephasing between EOB and
      our new coalescence data are smaller than: (i) $\pm 4\times 10^{-3}$ GW
      cycles over $730M$ (11 cycles), in the equal mass case, and (ii) $\pm 8\times 10^{-3}$
      GW cycles over about $900M$ (17 cycles) in the 2:1 mass-ratio case.
      In addition, we recall that the phase difference between our current
      ``best-bet'' EOB and both {\it published} and {\it actual}
      Caltech-Cornell data stays within 0.018 radians over the entire span of
      the simulation. Such a phase inaccuracy is comparable with the current,
      updated estimate of the numerical errors of the waveforms of
      Ref.~\cite{Boyle:2007ft}, namely 0.01 radians~\cite{KidderJena}.
\item As a contrast to the EOB performance, we also study in Appendix~\ref{sec:T4}
      the performance of the TaylorT4 approximant. Our analysis shows that the
      apparently good performance of TaylorT4 during the inspiral is due
      to a lucky compensation between two effects going in opposite
      directions: (i) the bad convergence of the adiabatic PN expansion and
      (ii) the fact that the T4 approximant does not take into
      account nonadiabatic effects. This compensation causes an 
      ``enhancement'' in the domain of validity of T4. However, we show that
      this enhancement holds only for a limited range of values of the 
      mass ratio. This is consistent with the finding of~\cite{Hannam:2007wf}
      that the enhanced validity of T4 is fragile and is undone by spin
      effects.

\end{itemize}

In conclusion we think that the results presented here corroborate the aptness
of the EOB formalism to provide accurate representations of general
relativistic waveforms. We suggest that the specific $3^{+2}$PN-accurate
resummed EOB waveform (with the current ``best-bet'' values of the flexibility
parameters determined here) be used in constructing banks of waveform
templates for currently operating gravitational wave detectors.

\acknowledgments
We are grateful to M.~Boyle, D.A.~Brown, L.E.~Kidder, A.H.~Mrou\'e, H.P.~Pfeiffer,
M.A.~Scheel, G.B.~Cook, S.~Teukolky and L.~Lindblom for communicating to
TD and AN some of the data published in~\cite{Boyle:2007ft}.  
A. Nagar is supported by INFN.
S. Husa is a VESF fellow of the European Gravitational Observatory
(EGO).
For part of this work M. Hannam
was supported by FSI grant 07/RFP/PHYF148. 
This work was supported in part by DFG grant SFB/Transregio~7
``Gravitational Wave Astronomy'' and the DLR (Deutsches Zentrum f\"ur
Luft- und Raumfahrt) through ``LISA Germany''.
Computations were performed at LRZ Munich (supported by a grant from
LRZ Munich) and the Doppler and Kepler clusters at the
Theoretisch-Physikalisches Institut, Friedrich-Schiller-Universit\"at
Jena.

\appendix
\section{Computing metric waveforms from curvature waveforms}
\label{sec:Psi4toPsie}

This first Appendix is devoted to the discussion of an appropriate 
way of choosing the integration constants that enter the metric waveform
$h(t)$ when deriving it by double time-integration from a given
(numerical) curvature waveform $\psi_4(t)$.

Our conventions are as follows: for reasons of continuity with 
the recent papers~\cite{Nagar:2006xv,Damour:2007xr,Damour:2007vq} 
we use the normalization factor $N_{\l} = \sqrt{(\l+2)(\l+1)\l(\l-1)}$
in the metric waveform to get the so-called Zerilli-Moncrief normalized 
waveform that we shall denote $\Psi_{\lm}^{(\rm e/o)}$ (for even and odd-parity)
as used in Ref.~\cite{Nagar:2005ea}. The metric waveform is expanded in
spin-weighted spherical harmonics of spin-weight $s=-2$ as
\begin{align}
h_+ - {\i} h_\times=\sum_{\l=2}^{\infty}\sum_{m=-\l}^{\l}h^{\ell m}{}_{-2}Y^{\ell m}(\theta,\phi)
\end{align}
where the link between the multipolar metric waveform $h_{\lm}$ 
(as used for instance in~\cite{Kidder:2007rt}) is
\begin{equation} 
\label{eq:gi}
h^{\ell m} = \dfrac{N_\l}{r}
\left(\Psi^{(\rm e)}_{\ell m} + \i \Psi^{(\rm o)}_{\ell m}\right).
\end{equation}
The raw output of the numerical simulation used here 
is the Newman-Penrose scalar $\psi_4$. This is decomposed in harmonics as
\begin{equation}
\ddot{h}_+ - \i\ddot{h}_\times = \psi_4 = \sum_{\l=2}^{\infty}\sum_{m=-\l}^{\l}\psi_4^{\ell m}{}_{-2}Y^{\ell m}(\theta,\phi).
\end{equation}
The computation of the Zerilli metric multipoles from its curvature 
correspondant $\psi_4^{\l m}$ requires a double time integration.
Various ways of fixing the two integration constants entering this process
have been discussed in the
literature~\cite{Koppitz:2007ev,Pollney:2007ss,Berti:2007fi,Schnittman:2007ij}. 
We focus here on the $\l=m=2$ multipole 
of the  Zerilli-Moncrief normalized metric waveform $\Psi^{(\rm e)}_{22}$.

\begin{figure}[t]
  \begin{center}
    \includegraphics[width=95 mm]{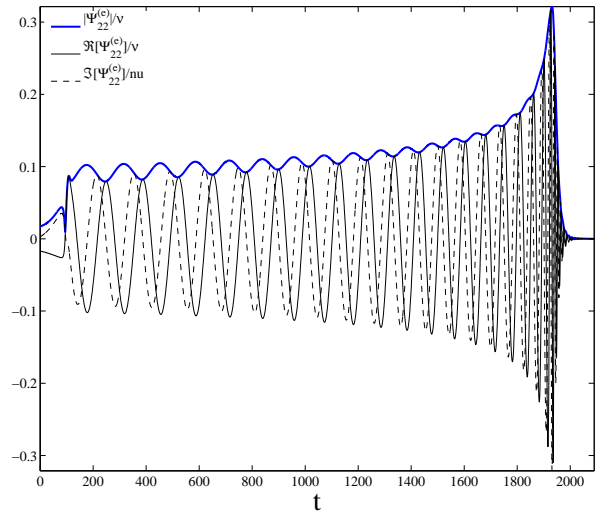} 
  \end{center}
  \vspace{-4mm}
  \caption{\label{label:fig9}Computation of the Zerilli normalized metric waveform
    $\Psi^{\e}_{22}$ from $r\psi_4^{22}$ via two  backward time integrations
    starting with zero integration constants at the final time. The data 
    refer to the 1:1 mass ratio ($\nu=0.25$) numerical simulation. Unphysical
    oscillations in the modulus are quite visible at early times.}
\end{figure}

We wish to emphasize that the choice of integration constants is
particularly delicate when dealing with the metric waveform $h(t)$, by
contrast to dealing with the quantity $\dot{h}(t)$ which is most prominent
in other applications, such as the computation of recoil.
For instance, Ref.~\cite{Schnittman:2007ij} suggested to integrate backward 
in time starting with zero integration constants at $t=+\infty$. This
procedure leads to a rather accurate $\dot{h}(t)$. However,
we found that the resulting $h(t)$ is not accurate enough for the 
purpose of high-accuracy waveform comparison discussed in this paper.
This is exemplified in Fig.~\ref{label:fig9}. This figure shows the 
metric waveform obtained by such a backward integration. The important
point is that the modulus of the complex waveform exhibits quite visible
unphysical oscillations at early times (during inspiral).  

\begin{figure}[t]
  \begin{center}
    \includegraphics[width=95 mm]{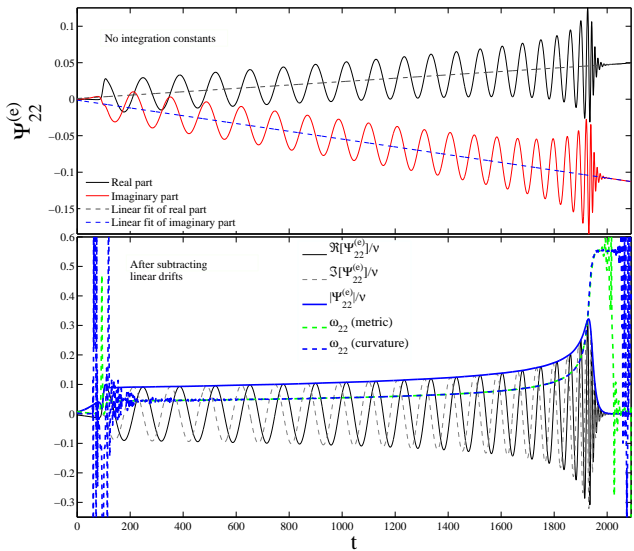} 
  \end{center}
  \vspace{-4mm}
  \caption{\label{label:fig10}Computation of the metric waveform
  $\Psi^{\e}_{22}$ from $r\psi_4^{22}$ via two time integrations
  starting at $t=0$. The upper panel has zero integration constants
  and exhibits clear linear drifts $\alpha t+\beta$.
  The bottom panel shows the result of subtracting the linear drift
  of the waveform obtained by fitting the upper panel over the entire
  time interval starting at $t=0$.}
\end{figure}

By contrast, we found that the following procedure (related to some of the 
suggestions of Ref.~\cite{Berti:2007fi}) gave reliably accurate results.
We start by computing (e.g., separately for the real and imaginary parts, 
or directly for the complex quantity) the first and  second forward time integrals 
(using e.g. Simpson's rule) of $r\psi_4^{\l m}$, starting at $t=0$ with 
zero integration constants, i.e., we define
\begin{align}
\dot{h}_0(t) & = \int_0^t dt'r\psi_4^{\l m}(t'), \\
h_0(t)       & = \int_0^t dt'\dot{h}_0(t'). 
\end{align}
Then, we fit over the full simulation time interval
(separately for the real and imaginary parts) 
the second integral $h_0(t)$ to a linear function of $t$, 
say $h_0^{\rm lin-fit}(t) = \alpha t + \beta$ where $\alpha$ 
and $\beta$ are complex quantities.
Finally, we define the metric waveform as
\begin{equation}
h(t) \equiv h_0(t)-h_0^{\rm lin-fit}(t)=h_0(t)-\left(\alpha t + \beta\right).
\end{equation}
Note that this also defines the time-derivative of the metric waveform as
\begin{equation}
\dot{h}(t) \equiv \dot{h}_0(t)- \alpha. 
\end{equation}
The result of this procedure is shown in Fig.~\ref{label:fig10}.
Here we applied the procedure explained above  
to the $r\psi_4^{22}$ waveform coming from the 1:1 
mass-ratio simulation extracted at $r=90$.
The top panel shows the real and imaginary parts of $h_0(t)$ (divided by the
normalization factor $N_2$) together with their best linear fits, i.e. 
the real and imaginary parts of $\alpha t+\beta$. 
The bottom panel shows the final waveform $h(t)$, i.e. the difference
between $h_0(t)$ and the best linear fit $\alpha t +\beta$. 
The important point is to notice that the modulus of $h(t)$ (the blue 
line in the bottom panel) is monotonically increasing with $t$ during inspiral
without exhibiting any of the unphysical oscillations that were present in the
previous figure.\footnote{This 
is a good indication that the integration constants have been computed
correctly and that the real and the imaginary parts of the waveform are
dephased by $\pi/2$ with very good approximation.}
We show on the same plot also the real and imaginary parts of the complex 
quantity $\Psi^{\e}_{22}$ (which correspond to the $h_+$ and $h_\times$ 
polarizations of the wave after division by $r$ and multiplication 
by the spin-harmonic ${}_{-2}Y^{22}$) as well as the gravitational wave 
frequency $\omega_{22}$ obtained from the metric
waveform $\Psi^{\e}_{22}$  and the gravitational wave 
frequency obtained from the curvature waveform $\psi_4^{22}$.

\begin{figure}[t]
  \begin{center}
    \includegraphics[width=90 mm]{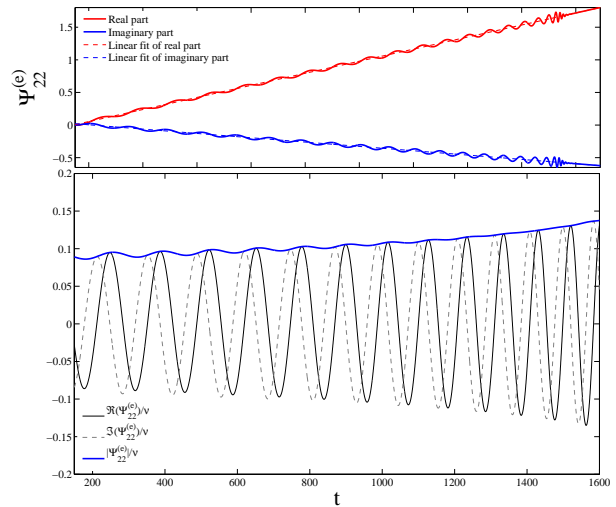} 
  \end{center}
  \vspace{-4mm}
  \caption{\label{label:fig11}Same as Fig.~\ref{label:fig10} except that the
  integration and the linear fit have been done starting at time $t\sim 150$,
  i.e. at the beginning of the inspiral signal. Note the oscillations in the
  modulus of the bottom panel, and the fact that the linear drifts (visible in
  the upper panel) are much larger than in Fig.~\ref{label:fig10}.}
\end{figure}

In addition, let us emphasize that for this procedure to work it is
important to start the integration from the absolute beginning of 
the numerical simulation, by which we really mean $t=0$, i.e. before 
any signal reaches the observer. One might have thought that 
it is better to start the integration after the junk radiation, at the 
beginning of the inspiral signal. This is not the case, as it is 
illustrated in Fig.~\ref{label:fig11}.
This figure shows the worsened result we obtain when we
use exactly the procedure explained above, but on the time interval $t\geq 150$,
i.e. starting at the beginning of the inspiral signal instead of starting
at $t=0$. Note the oscillations in the modulus of $\Psi^{\e}_{22}$.
By contrast, even if we blow up the corresponding graph in
Fig.~\ref{label:fig10} the oscillations are practically absent.
Note also that the linear drifts are now much larger than before. This is
part of the reason why the results are less good in this case.
By contrast to the first case where, starting at $t=0$ meant starting
with extremely small initial values of $r\psi_4^{22}$ , starting at the 
beginning of the inspiral means starting with much larger values of
$r\psi_4^{22}$: this effect enlarges the linear floors and therefore
the errors on the determination of the linear floors to be subtracted.

\section{Shortcomings of the TaylorT4 PN approximant}
\label{sec:T4}
To contrast with the EOB-NR comparison done in the text,
we consider in this appendix the comparison between
the so-called TaylorT4 post-Newtonian 
approximant~\cite{Baker:2006ha,Buonanno:2006ui,Pan:2007nw,Boyle:2007ft,Gopakumar:2007vh,Hannam:2007wf}. 
and various NR data.
This approximant  is  defined by two successive prescriptions: 
the first concerns the computation of a ``T4 orbital phase'' $\Phi_{\rm T4}(t)$ 
while the second concerns  the definition of a ``T4 metric waveform''.
Here we shall focus only on the $\l=m=2$ quadrupolar waveform.
The ``T4 orbital phase'' $\Phi_{\rm T4}(t)$ is defined by
integrating the ordinary differential equations 
\begin{align}
\dfrac{d\Phi_{\rm T4}}{dt} &= x^{3/2}, \\
\label{TaylorT4}
\dfrac{dx}{dt}             &= \dfrac{64\nu}{5}x^5 a_{3.5}^{\rm Taylor}, 
\end{align}
where $a_{3.5}^{\rm Taylor}$ is the 3.5~PN Taylor approximant, 
for any given value of $\nu$, to the Newton-normalized ratio 
(flux-function)/(derivative of energy function)=$\hat{F}(x)/\widehat{E(x)}$
where $E'(x)=dE/dx$.
As in the text, we scale dimensionful 
quantities by the total ``bare'' mass $M=m_1+m_2$. This is 
for instance the case for the time variable $t$ in the above equations.
The explicit expression of $a^{\rm Taylor}_{3.5}(x)$
reads~\cite{Buonanno:2006ui,Pan:2007nw} (for the nonspinning case)
\begin{align}
&a^{\rm Taylor}_{3.5}(x)=1-\left(\dfrac{743}{336}+\dfrac{11}{4}\nu\right)x + 4\pi x^{3/2}\\
                    & +\left(\dfrac{34103}{18144}+\dfrac{13661}{2016}\nu\right)x^2
                     -\left(\dfrac{4159}{672}+\dfrac{189}{8}\nu\right)\pi x^{5/2}\nonumber\\
                   & +\bigg[\dfrac{16447322263}{139708800}-\dfrac{1712}{105}\gamma
                    -\dfrac{56198689}{217728}\nu \nonumber \\
                   & +\dfrac{541}{896}\nu^2 - \dfrac{5605}{2592}\nu^3 +
                    \dfrac{\pi^2}{48}\left(256 +
                    452\nu\right)-\dfrac{856}{105}\log(16
                    x)\bigg]x^3\nonumber\\
		   &+\left(-\dfrac{4415}{4032} + \dfrac{358675}{6048}\nu +
                    \dfrac{91495}{1512}\nu^2\right)\pi x^{7/2} .
\end{align}
This phasing evolution is completed by a quadrupolar waveform which 
is known (for any given value of $\nu$) at the 3~PN accuracy 
level~\cite{Arun:2004ff,Berti:2007fi,Damour:2007yf,Kidder:2007rt}.  
Following~\cite{Kidder:2007rt,Boyle:2007ft} we define the  3~PN-accurate
T4 waveform by dropping all the $\ln(x/x_0)$ terms in Eq.~(79) 
of Ref.~\cite{Kidder:2007rt}. We display it explicitly here 
to clarify which waveforms we use in our T4 studies.
The explicit expression of the $\l=m=2$ Zerilli-normalized 
metric waveform reads
\begin{widetext}
\begin{align}
&\Psi_{22}^{\rm T4} = -4\nu\sqrt{\dfrac{\pi}{30}}e^{-2\i\Phi} x
  \bigg\{ 1-x\left(\dfrac{107}{42}-\dfrac{55}{42}\nu\right)
         +2\pi x^{3/2} -
         x^2\left(\dfrac{2173}{1512}+\dfrac{1069}{216}\nu-\dfrac{2047}{1512}\nu^2\right)
         -x^{5/2}\left[\left(\dfrac{107}{21}-\dfrac{34}{21}\nu\right)\pi+24\i\nu\right]\nonumber\\
%
& +x^3\left[\dfrac{27027409}{646800} +\dfrac{2}{3}\pi^2
  +\dfrac{428}{105}\left[\i\pi - 2\gamma_E - \ln(16x)\right] 
-\left(\dfrac{278185}{33264}-\dfrac{41}{96}\pi^2\right)\nu-\dfrac{20261}{2772}\nu^2
    + \dfrac{114635}{99792}\nu^3\right]
\bigg\}.
\end{align}
\end{widetext}
where $\gamma_E= 0.57721\dots$ is Euler's constant. The taylorT4 3.5/2.5
waveform (used in most of our comparisons) is obtained by dropping the 
terms $\propto x^3[a\ln(x)+b]$ on the r.h.s. of this equation.

Thorough comparisons between the TaylorT4 3.5/2.5 waveform (i.e., 3.5~PN
accuracy for phase and 2.5~PN accuracy {\it only} for amplitude) 
and NR waveforms were performed, for the equal mass case, $\nu=0.25$, 
in~\cite{Baker:2006ha,Boyle:2007ft,Hannam:2007wf}. 
Ref.~\cite{Boyle:2007ft} concluded that
this approximant yields
an ``astonishingly good'' agreement with numerical data during 
the inspiral, i.e.\ a dephasing smaller than 0.05 radians over 
$\sim 30$ GW cycles before reaching the GW frequency $M\omega_{22}=0.1$. 
On the other hand, Ref.~\cite{Hannam:2007wf} showed that the 
inclusion of spins on the black holes had the effect of considerably
worsening the agreement between T4 and NR data. Here we shall study
the effect of varying the mass ratio (for nonspinnig black holes).
We shall also go beyond the analyses of~\cite{Baker:2006ha,Boyle:2007ft}
in discussing the behaviour of T4 for GW frequencies above 0.1.
Let us first compare 
the~\footnote{Here, to facilitate the comparison with previous work,
we use a T4 approximant with 2.5~PN accurate amplitude. Our main 
conclusions would be similar had we used the 3~PN accurate amplitude.} 
$\nu=0.25$ TaylorT4 3.5/2.5 quadrupolar 
waveform $\Psi^{\rm T4}_{22}$ with equal-mass NR waveforms 
computed by the Jena group. 
As discussed in Sec.~\ref{sec:NR}, the BAM code outputs the 
Newman-Penrose curvature scalar $\psi_4(t,r,\theta,\varphi)$
at various extraction radii $r$. This angular-dependent curvature scalar is 
then: (i) decomposed on the basis of spin-weighted spherical 
harmonics and then (ii) integrated twice over time to yield 
the metric waveform $\Psi^{\rm NR}_{22}$. The choice of integration constants
in this integration procedure was done according to the procedure outlined in
Appendix~\ref{sec:Psi4toPsie}. To compare the two waveforms 
$\Psi_{22}^{\rm NR}$ and $\Psi_{22}^{\rm T4}$, as functions
of their respective time variables, we choose a relative 
time-shift $\tau$ and a relative phase-shift $\alpha$
by following the same two-pinching-time procedure used in the text.

\begin{figure}[t]
  \begin{center}
    \includegraphics[width=90 mm]{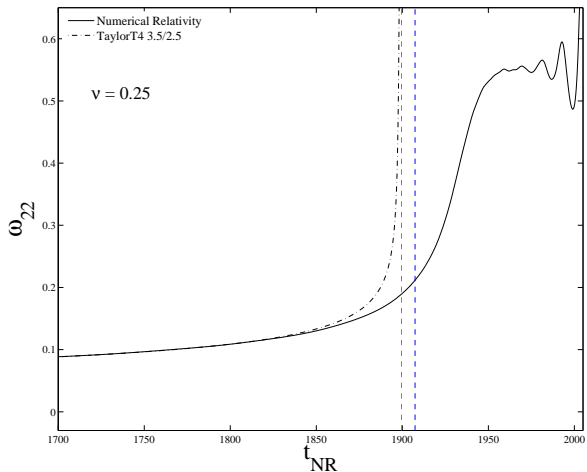} 
  \end{center}
  \vspace{-4mm}
  \caption{\label{label:fig1}Numerical relativity (Jena) and TaylorT4: comparison
    between the instantaneous gravitational wave frequencies for the
    equal mass case ($\nu=0.25$).}
\end{figure}

Figure~\ref{label:fig1} compares the gravitational 
wave frequency $\omega_{22}$ computed from 
the numerical data and plotted as a function of the NR time scale (solid line) 
with that of the TaylorT4 3.5/2.5 approximant  plotted as a 
function of the shifted T4 time-scale (dash-dot line). 
The two waveforms have been ``pinched'' at the NR times $t_1=1299.9$ 
and $t_2=1399.8$, corresponding to NR frequencies $\omega_1=0.062643$
and $\omega_2=0.066292$, respectively (which approximate the matching
frequency $\omega_m=\omega_3=0.063$ of~\cite{Boyle:2007ft}).

\begin{figure*}[t]
  \begin{center}
    \includegraphics[width=85 mm]{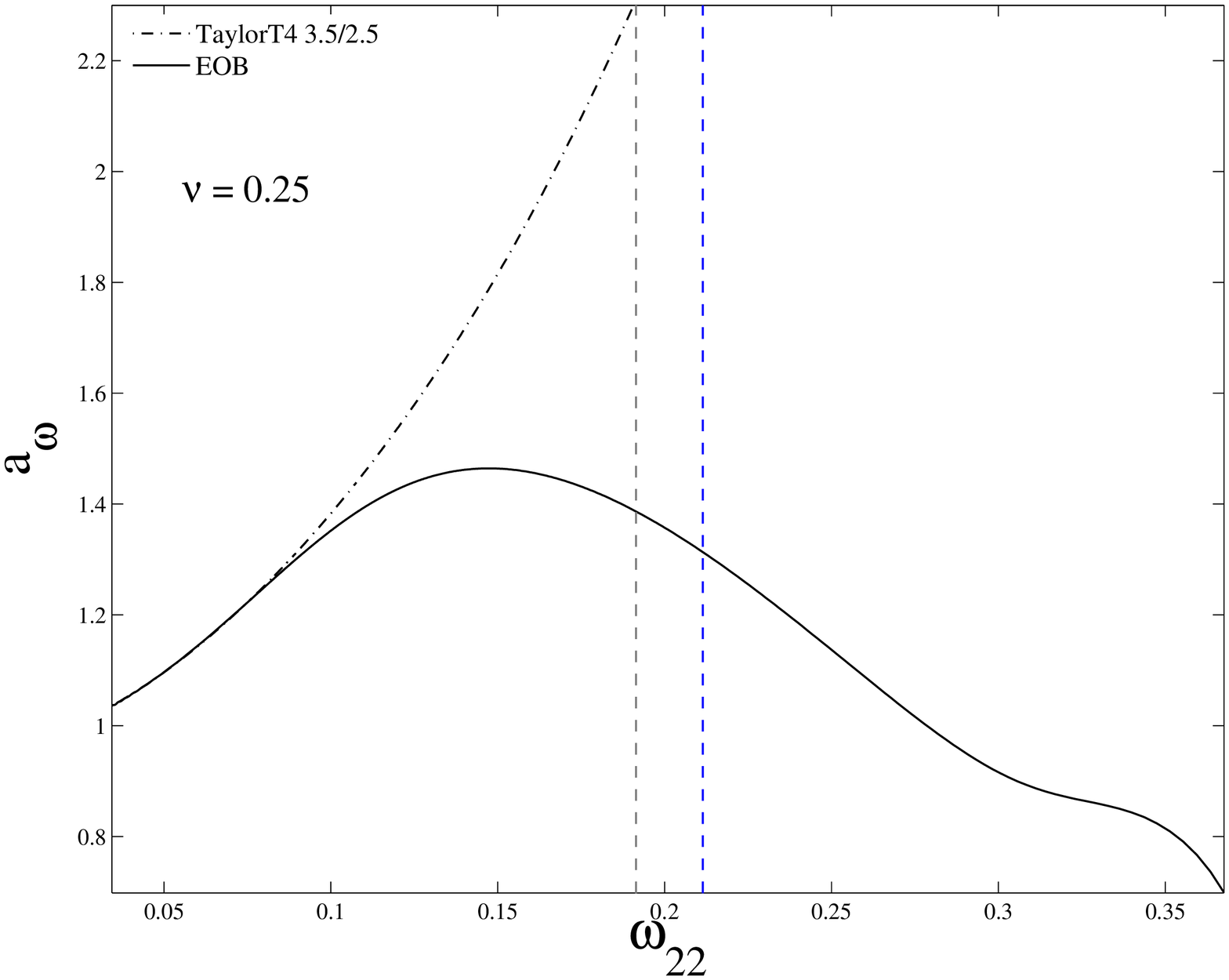} 
    \includegraphics[width=85 mm]{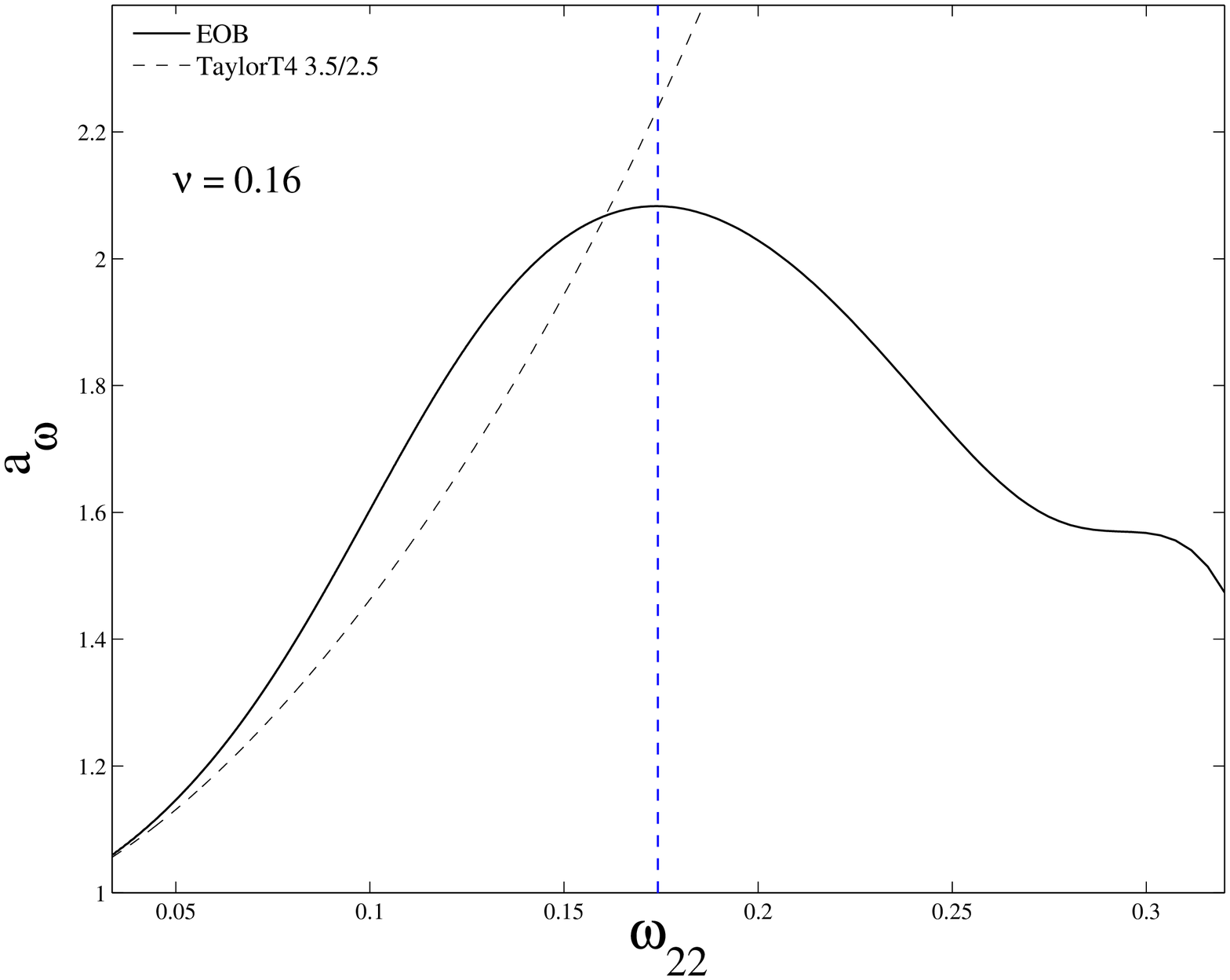} 
  \end{center}
  \vspace{-4mm}
  \caption{\label{label:fig2}Comparison between phase  acceleration curves
 $a_{\omega}$ of EOB (solid line) and TaylorT4 (dash-dot line) for $\nu=0.25$
 (left panel) and $\nu=0.16$ (right panel). The rightmost dashed vertical line 
 indicates the location of the EOB adiabatic LSO. The leftmost vertical line
 on the left panel indicates the EOB GW frequency corresponding the the
 instant when T4 blows up.}
\end{figure*}

We see on this figure that there is a very good agreement between 
the two frequencies during the inspiral, up to, say, the NR 
time $t_{\rm NR}=1850$, where $\omega^{\rm NR}_{22}=0.1301$ and
$\omega_{22}^{\rm T4}$ differ by about $2.2\%$. Then, soon after,
$\omega_{22}^{\rm T4}$ starts deviating very strongly from 
$\omega^{\rm NR}_{22}$ and {\it blows up to infinity} at the finite
time $t_{\rm blowup}=1899.5$ (indicated by the leftmost vertical dashed line in the
figure). This blow-up time, considered on the NR time-scale, corresponds to a NR frequency 
$\omega^{\rm NR}_{\rm blowup}\simeq 0.1889$. Note that this 
frequency is {\it smaller} than the
Effective-One-Body GW frequency at the adiabatic Last Stable Orbit (LSO), 
which is equal to $2\Omega_{\rm LSO}^{\rm EOB}=0.2114$ (corresponding 
to an EOB radial coordinate $r^{\rm EOB}_{\rm LSO}=4.4729$).
The rightmost vertical dashed line in the figure indicates 
the ``$\omega-{\rm LSO}$'', in the sense of Ref.~\cite{Buonanno:2000ef},
i.e.\ the time when the (NR) GW frequency $\omega^{\rm NR}_{22}$ equals 
the adiabatic LSO frequency. Here we consider the case $\nu=0.25$ 
and $a_5=25$ and we compute the LSO frequency within the EOB approach. 
Therefore, in the equal mass case, the TaylorT4 approximant breaks 
down already during late inspiral, before the EOB LSO and before the plunge.

The fact that the T4 approximant blows up at a finite time is
a simple mathematical consequence of the structure of the differential
equation~\eqref{TaylorT4}, given that, $a_{3.5}^{\rm Taylor}(x)$ is found
to remain positive for every $x\geq0$. Indeed, one can even easily 
analytically compute the blow-up time as being $t_{\rm blowup}=t_0+\Delta t$
where $t_0$ is any given ``reference'' time on the $T4$ time scale 
(corresponding to a frequency parameter $x(t_0)=x_0$), and where
$\Delta t$ is given by the following {\it convergent} integral
\begin{equation}
\label{eq:Deltat}
\Delta t = \int_{x_0}^\infty \dfrac{dx}{C_\nu x^5 a_{3.5}^{\rm Taylor}},
\end{equation}
where $C_\nu= 64\nu/5$. 

After having compared the T4 approximant to 
NR data (in the equal-mass case) let us compare the T4 approximant 
to the EOB one. As emphasized in Ref.~\cite{Damour:2007yf}, a convenient 
way of comparing two waveforms 
(which avoids the issue of finding suitable time shifts and phase shifts)
consists in considering the following shift-invariant ``phase-acceleration''
function
\begin{equation}
\label{aomega}
a_{\omega}(\omega)=\dfrac{\dot\omega}{c_\nu\omega^{11/3}} \qquad
c_{\nu}=\dfrac{12}{5}2^{1/3}\nu.
\end{equation}
Note that in the present paper we consider the frequencies of the
{\it metric} waveforms (by contrast to the frequencies of the {\it curvature}
waveforms considered in~\cite{Damour:2007yf}).

In the left-panel of Fig.~\ref{label:fig2} we compare the phase acceleration
curves of T4 3.5/2.5 (dash-dot line) and EOB (solid line) for the equal mass case.
The leftmost vertical line indicates the EOB frequency $\approx0.19$ 
corresponding to the T4 blow-up time (computed by Eq.~\eqref{eq:Deltat}). 
The rightmost vertical line indicates the adiabatic EOB 
LSO frequency, $2\Omega_{\rm LSO}^{\rm EOB}\approx 0.21$ as above.
We terminated the horizontal axis at $\omega_{22}=\omega^{\rm LR}_{22}=0.3676$ 
which corresponds to the EOB time when the EOB orbital frequency reaches 
its maximum; i.e., the so-called ``EOB light-ring'', which defines
the ``merger time'' within the EOB approach. Note that this
figure shows the metric waveform analogue of the EOB (curvature) 
phase acceleration curve of Fig.~2 of Ref.~\cite{Damour:2007yf}
and extends it up to the merger time.
As was already emphasized in~\cite{Damour:2007yf}, the figure shows
that the T4 acceleration curve strongly diverges away from the EOB
one for frequencies $\omega_{22}\gtrsim 0.1$, i.e.\ during the late inspiral,
before reaching the LSO.

The right-panel of Fig.~\ref{label:fig2} illustrates the case where the
mass ratio is $4:1$, i.e.\ $\nu=0.16$. The vertical dashed line indicates
the adiabatic EOB LSO frequency $2\Omega_{\rm LSO}^{\rm EOB}\approx 0.17$.
For this value of $\nu$, the blow-up frequency, computed as above, turns
out to be larger than the EOB light-ring frequency $\omega^{\rm LR}_{22}=0.3201$.
We see on this plot that, contrary to the equal mass case, the T4 acceleration
curve starts to deviate significantly from the EOB one for frequencies
$\omega_{22}\gtrsim 0.05$. Note, however, that because the two curves 
cross again just before the LSO, we expect that the phase difference 
between T4 and EOB will remain, on average, rather small 
up to the LSO. However later on the T4 phasing will drastically 
deviate from the EOB one. 

\begin{figure}[t]
  \begin{center}
    \includegraphics[width=90 mm]{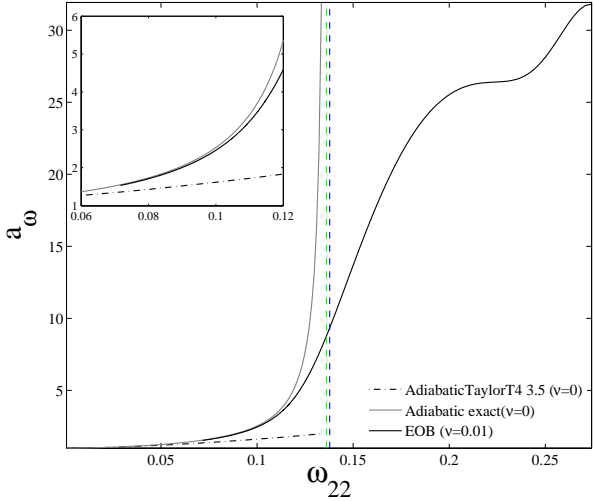} 
  \end{center}
  \vspace{-4mm}
  \caption{\label{label:fig3}This figure (done for $\nu\ll 1$) illustrates, in
  particular, the fact that the domain of validity of TaylorT4 when
  $\nu\lesssim 0.16$ reduces to the ``normal'' 3.5~PN one, namely
  $\omega\lesssim 0.06$.}
\end{figure}
Finally, Fig.~\ref{label:fig3} considers the test-mass limit 
($\nu\to 0$). Here we compare three acceleration curves: 
(i) the {\it adiabatic} limit of the T4 acceleration curve,
given simply by $a_{\omega}(\omega)=a^{\rm Taylor}_{3.5}(x)$ 
computed in the limit $\nu\to 0$ and with $x=(\omega/2)^{2/3}$;
(ii) the EOB $a_{\omega}$ curve computed for $\nu=0.01$ and
(iii) the exact adiabatic limit of the test-mass acceleration 
curve, i.e.\ the Newton-normalized ratio $\hat{F}(x)/\widehat{E'(x)}$  
(see e.g. Ref.~\cite{Damour:1997ub}). Here the flux function $\hat{F}(x)$
is the one computed numerically in Ref.~\cite{Cutler:1993vq,Poisson:1995vs}.
The two vertical lines in the figure refer to the $\nu=0$ limit 
of the adiabatic LSO frequency (leftmost line, $2\Omega=0.1361$ ) 
and to the $\nu=0.01$ EOB adiabatic LSO frequency 
(rightmost line, $2\Omega_{\rm LSO}=0.1378$).
This figure illustrates two facts: first, the T4 approximant 
starts strongly deviating from the exact result early 
on (say for $\omega_{22}\lesssim 0.06$, see inset); 
second, one needs to consider $\nu<0.01$ to ensure that the 
usual adiabatic approximation is satisfactory up to frequencies
close to the LSO one. This is consistent with the analytical estimate
obtained in~\cite{Buonanno:2000ef} according to which the deviations
from adiabaticity become important when the 
frequency fractionally deviates from the LSO frequency 
by $\delta \omega/\omega_{\rm LSO}\sim \nu^{2/5}$. The presence 
of the $2/5$ power means that we need $\nu\leq 3\times 10^{-3}$ 
to be approximately adiabatic up to $90\%$ of the LSO frequency.

Summarizing, the main results of the present Appendix 
(and of complementary investigations of the different ``speeds'' 
with which the T4 and EOB waveforms ``move'' as $\nu$ varies)
are: (i) we predict that the T4 approximant will define 
an {\it effective} phasing template for the 
{\it inspiral} waveform only up to some 
$\nu$-dependent upper GW frequency, say  $\omega^{\rm T4}_{\rm max}(\nu)$,
having the following properties;
(ii) for $\nu=0.25$, $\omega^{\rm T4}_{\rm max}(0.25)\approx 0.14$
(consistently with~\cite{Boyle:2007ft}) which is significantly
above the expected range of validity of a normal PN approximant,
but still significantly below the 
(EOB-estimated~\footnote{For the reasons discussed above, our
EOB  estimates here use $a_5=25$.}) gravitational wave LSO frequency
$\omega_{\rm LSO}^{\rm EOB}=0.2114$. For $\nu=0.16$, the upper bound
{\it increases} to $\omega^{\rm T4}_{\rm max}(0.16)\approx 0.17$, 
i.e.\ around the corresponding  LSO frequency. 
For intermediate values $0.16\lesssim\nu\lesssim 0.25$,
the situation smoothly interpolates between these two cases;
(iii) by contrast, as $\nu$ gets smaller than about 0.16, 
$\omega^{\rm T4}_{\rm max}(\nu)$ will decrease
down to values of order of 0.05, which are typical of the expected 
upper frequency of validity for a normal PN 
approximant~\footnote{Indeed,  Ref.~\cite{Brady:1998du} 
estimated the (3~PN accurate) ``PN failure point'' to be 
around $\Omega^{1/3}\approx 0.3$ which corresponds to
$\omega_{22}=2\Omega=0.054$.}; (iv) in all cases the range of 
validity of T4 is limited to the inspiral and, contrary to the EOB,
does not include the plunge; (v) in all cases, T4 exhibits a blow-up
of the frequency at a finite time. However, this blow up is not 
always the main reason limiting the validity of the approximant.
For instance, this is the case when $\nu\approx 0.25$, but not when 
$\nu\lesssim 0.16$.
Let us finally emphasize that the ``enhancement'' in the domain of
validity of T4 when $0.16\lesssim\nu\lesssim 0.25$ with respect to
the normal expected PN validity is due to a {\it lucky compensation} (which
does not take place when $\nu\lesssim 0.16$) between
two effects going in opposite directions: on the one hand, the bad 
convergence of the adiabatic PN expansion; on the other hand, 
the fact that the T4 approximant does not take into account 
nonadiabatic effects (which are quite significant as 
emphasized in~\cite{Buonanno:2000ef} and displayed in Fig.~2 of~\cite{Damour:2007yf}).
Our present result clarifies the theoretical underpinnings of 
the result found in~\cite{Hannam:2007wf}, namely
that ``deformation'' of the T4 approximant by spin effects removes
the accidental nice agreement between T4 and NR. Indeed, one should not expect
such a chance compensation to be stable under any deformation of the
underlying physics (such as additional spins or a varying mass ratio).


\bibliography{paper_resub}

\begin{thebibliography}{76}
\expandafter\ifx\csname natexlab\endcsname\relax\def\natexlab#1{#1}\fi
\expandafter\ifx\csname bibnamefont\endcsname\relax
  \def\bibnamefont#1{#1}\fi
\expandafter\ifx\csname bibfnamefont\endcsname\relax
  \def\bibfnamefont#1{#1}\fi
\expandafter\ifx\csname citenamefont\endcsname\relax
  \def\citenamefont#1{#1}\fi
\expandafter\ifx\csname url\endcsname\relax
  \def\url#1{\texttt{#1}}\fi
\expandafter\ifx\csname urlprefix\endcsname\relax\def\urlprefix{URL }\fi
\providecommand{\bibinfo}[2]{#2}
\providecommand{\eprint}[2][]{\url{#2}}

\bibitem[{\citenamefont{Buonanno and Damour}(1999)}]{Buonanno:1998gg}
\bibinfo{author}{\bibfnamefont{A.}~\bibnamefont{Buonanno}} \bibnamefont{and}
  \bibinfo{author}{\bibfnamefont{T.}~\bibnamefont{Damour}},
  \bibinfo{journal}{Phys. Rev.} \textbf{\bibinfo{volume}{D59}},
  \bibinfo{pages}{084006} (\bibinfo{year}{1999}), \eprint{gr-qc/9811091}.

\bibitem[{\citenamefont{Buonanno and Damour}(2000)}]{Buonanno:2000ef}
\bibinfo{author}{\bibfnamefont{A.}~\bibnamefont{Buonanno}} \bibnamefont{and}
  \bibinfo{author}{\bibfnamefont{T.}~\bibnamefont{Damour}},
  \bibinfo{journal}{Phys. Rev.} \textbf{\bibinfo{volume}{D62}},
  \bibinfo{pages}{064015} (\bibinfo{year}{2000}), \eprint{gr-qc/0001013}.

\bibitem[{\citenamefont{Damour et~al.}(2000)\citenamefont{Damour, Jaranowski,
  and Schaefer}}]{Damour:2000we}
\bibinfo{author}{\bibfnamefont{T.}~\bibnamefont{Damour}},
  \bibinfo{author}{\bibfnamefont{P.}~\bibnamefont{Jaranowski}},
  \bibnamefont{and} \bibinfo{author}{\bibfnamefont{G.}~\bibnamefont{Schaefer}},
  \bibinfo{journal}{Phys. Rev.} \textbf{\bibinfo{volume}{D62}},
  \bibinfo{pages}{084011} (\bibinfo{year}{2000}), \eprint{gr-qc/0005034}.

\bibitem[{\citenamefont{Damour}(2001)}]{Damour:2001tu}
\bibinfo{author}{\bibfnamefont{T.}~\bibnamefont{Damour}},
  \bibinfo{journal}{Phys. Rev.} \textbf{\bibinfo{volume}{D64}},
  \bibinfo{pages}{124013} (\bibinfo{year}{2001}), \eprint{gr-qc/0103018}.

\bibitem[{\citenamefont{Blanchet}(2006)}]{Blanchet:2002av}
\bibinfo{author}{\bibfnamefont{L.}~\bibnamefont{Blanchet}},
  \bibinfo{journal}{Living Rev. Rel.} \textbf{\bibinfo{volume}{9}},
  \bibinfo{pages}{4} (\bibinfo{year}{2006}), \eprint{gr-qc/0202016}.

\bibitem[{\citenamefont{Brezin et~al.}(1970)\citenamefont{Brezin, Itzykson, and
  Zinn-Justin}}]{Brezin:1970zr}
\bibinfo{author}{\bibfnamefont{E.}~\bibnamefont{Brezin}},
  \bibinfo{author}{\bibfnamefont{C.}~\bibnamefont{Itzykson}}, \bibnamefont{and}
  \bibinfo{author}{\bibfnamefont{J.}~\bibnamefont{Zinn-Justin}},
  \bibinfo{journal}{Phys. Rev.} \textbf{\bibinfo{volume}{D1}},
  \bibinfo{pages}{2349} (\bibinfo{year}{1970}).

\bibitem[{\citenamefont{Damour et~al.}(1998)\citenamefont{Damour, Iyer, and
  Sathyaprakash}}]{Damour:1997ub}
\bibinfo{author}{\bibfnamefont{T.}~\bibnamefont{Damour}},
  \bibinfo{author}{\bibfnamefont{B.~R.} \bibnamefont{Iyer}}, \bibnamefont{and}
  \bibinfo{author}{\bibfnamefont{B.~S.} \bibnamefont{Sathyaprakash}},
  \bibinfo{journal}{Phys. Rev.} \textbf{\bibinfo{volume}{D57}},
  \bibinfo{pages}{885} (\bibinfo{year}{1998}), \eprint{gr-qc/9708034}.

\bibitem[{\citenamefont{Davis et~al.}(1972)\citenamefont{Davis, Ruffini, and
  Tiomno}}]{Davis:1972ud}
\bibinfo{author}{\bibfnamefont{M.}~\bibnamefont{Davis}},
  \bibinfo{author}{\bibfnamefont{R.}~\bibnamefont{Ruffini}}, \bibnamefont{and}
  \bibinfo{author}{\bibfnamefont{J.}~\bibnamefont{Tiomno}},
  \bibinfo{journal}{Phys. Rev.} \textbf{\bibinfo{volume}{D5}},
  \bibinfo{pages}{2932} (\bibinfo{year}{1972}).

\bibitem[{\citenamefont{Buonanno et~al.}(2006)\citenamefont{Buonanno, Chen, and
  Damour}}]{Buonanno:2005xu}
\bibinfo{author}{\bibfnamefont{A.}~\bibnamefont{Buonanno}},
  \bibinfo{author}{\bibfnamefont{Y.}~\bibnamefont{Chen}}, \bibnamefont{and}
  \bibinfo{author}{\bibfnamefont{T.}~\bibnamefont{Damour}},
  \bibinfo{journal}{Phys. Rev.} \textbf{\bibinfo{volume}{D74}},
  \bibinfo{pages}{104005} (\bibinfo{year}{2006}), \eprint{gr-qc/0508067}.

\bibitem[{\citenamefont{Pretorius}(2007)}]{Pretorius:2007nq}
\bibinfo{author}{\bibfnamefont{F.}~\bibnamefont{Pretorius}}
  (\bibinfo{year}{2007}), \eprint{arXiv:0710.1338 [gr-qc]}.

\bibitem[{\citenamefont{Pretorius}(2005)}]{Pretorius:2005gq}
\bibinfo{author}{\bibfnamefont{F.}~\bibnamefont{Pretorius}},
  \bibinfo{journal}{Phys. Rev. Lett.} \textbf{\bibinfo{volume}{95}},
  \bibinfo{pages}{121101} (\bibinfo{year}{2005}), \eprint{gr-qc/0507014}.

\bibitem[{\citenamefont{Campanelli et~al.}(2006)\citenamefont{Campanelli,
  Lousto, Marronetti, and Zlochower}}]{Campanelli:2005dd}
\bibinfo{author}{\bibfnamefont{M.}~\bibnamefont{Campanelli}},
  \bibinfo{author}{\bibfnamefont{C.~O.} \bibnamefont{Lousto}},
  \bibinfo{author}{\bibfnamefont{P.}~\bibnamefont{Marronetti}},
  \bibnamefont{and}
  \bibinfo{author}{\bibfnamefont{Y.}~\bibnamefont{Zlochower}},
  \bibinfo{journal}{Phys. Rev. Lett.} \textbf{\bibinfo{volume}{96}},
  \bibinfo{pages}{111101} (\bibinfo{year}{2006}), \eprint{gr-qc/0511048}.

\bibitem[{\citenamefont{Baker et~al.}(2006)\citenamefont{Baker, Centrella,
  Choi, Koppitz, and van Meter}}]{Baker05a}
\bibinfo{author}{\bibfnamefont{J.~G.} \bibnamefont{Baker}},
  \bibinfo{author}{\bibfnamefont{J.}~\bibnamefont{Centrella}},
  \bibinfo{author}{\bibfnamefont{D.-I.} \bibnamefont{Choi}},
  \bibinfo{author}{\bibfnamefont{M.}~\bibnamefont{Koppitz}}, \bibnamefont{and}
  \bibinfo{author}{\bibfnamefont{J.}~\bibnamefont{van Meter}},
  \bibinfo{journal}{Phys. Rev. Lett.} \textbf{\bibinfo{volume}{96}},
  \bibinfo{pages}{111102} (\bibinfo{year}{2006}), \eprint{gr-qc/0511103}.

\bibitem[{\citenamefont{Gonzalez et~al.}(2007)\citenamefont{Gonzalez, Sperhake,
  Bruegmann, Hannam, and Husa}}]{Gonzalez:2006md}
\bibinfo{author}{\bibfnamefont{J.~A.} \bibnamefont{Gonzalez}},
  \bibinfo{author}{\bibfnamefont{U.}~\bibnamefont{Sperhake}},
  \bibinfo{author}{\bibfnamefont{B.}~\bibnamefont{Bruegmann}},
  \bibinfo{author}{\bibfnamefont{M.}~\bibnamefont{Hannam}}, \bibnamefont{and}
  \bibinfo{author}{\bibfnamefont{S.}~\bibnamefont{Husa}},
  \bibinfo{journal}{Phys. Rev. Lett.} \textbf{\bibinfo{volume}{98}},
  \bibinfo{pages}{091101} (\bibinfo{year}{2007}), \eprint{gr-qc/0610154}.

\bibitem[{\citenamefont{Koppitz et~al.}(2007)}]{Koppitz:2007ev}
\bibinfo{author}{\bibfnamefont{M.}~\bibnamefont{Koppitz}} \bibnamefont{et~al.},
  \bibinfo{journal}{Phys. Rev. Lett.} \textbf{\bibinfo{volume}{99}},
  \bibinfo{pages}{041102} (\bibinfo{year}{2007}), \eprint{gr-qc/0701163}.

\bibitem[{\citenamefont{Damour et~al.}(2002)\citenamefont{Damour, Gourgoulhon,
  and Grandclement}}]{Damour:2002qh}
\bibinfo{author}{\bibfnamefont{T.}~\bibnamefont{Damour}},
  \bibinfo{author}{\bibfnamefont{E.}~\bibnamefont{Gourgoulhon}},
  \bibnamefont{and}
  \bibinfo{author}{\bibfnamefont{P.}~\bibnamefont{Grandclement}},
  \bibinfo{journal}{Phys. Rev.} \textbf{\bibinfo{volume}{D66}},
  \bibinfo{pages}{024007} (\bibinfo{year}{2002}), \eprint{gr-qc/0204011}.

\bibitem[{\citenamefont{Damour et~al.}(2003)\citenamefont{Damour, Iyer,
  Jaranowski, and Sathyaprakash}}]{Damour:2002vi}
\bibinfo{author}{\bibfnamefont{T.}~\bibnamefont{Damour}},
  \bibinfo{author}{\bibfnamefont{B.~R.} \bibnamefont{Iyer}},
  \bibinfo{author}{\bibfnamefont{P.}~\bibnamefont{Jaranowski}},
  \bibnamefont{and} \bibinfo{author}{\bibfnamefont{B.~S.}
  \bibnamefont{Sathyaprakash}}, \bibinfo{journal}{Phys. Rev.}
  \textbf{\bibinfo{volume}{D67}}, \bibinfo{pages}{064028}
  (\bibinfo{year}{2003}), \eprint{gr-qc/0211041}.

\bibitem[{\citenamefont{Damour and Nagar}(2007{\natexlab{a}})}]{Damour:2007xr}
\bibinfo{author}{\bibfnamefont{T.}~\bibnamefont{Damour}} \bibnamefont{and}
  \bibinfo{author}{\bibfnamefont{A.}~\bibnamefont{Nagar}},
  \bibinfo{journal}{Phys. Rev.} \textbf{\bibinfo{volume}{D76}},
  \bibinfo{pages}{064028} (\bibinfo{year}{2007}{\natexlab{a}}),
  \eprint{arXiv:0705.2519 [gr-qc]}.

\bibitem[{\citenamefont{Damour et~al.}(2001{\natexlab{a}})\citenamefont{Damour,
  Jaranowski, and Schafer}}]{Damour:2001bu}
\bibinfo{author}{\bibfnamefont{T.}~\bibnamefont{Damour}},
  \bibinfo{author}{\bibfnamefont{P.}~\bibnamefont{Jaranowski}},
  \bibnamefont{and} \bibinfo{author}{\bibfnamefont{G.}~\bibnamefont{Schafer}},
  \bibinfo{journal}{Phys. Lett.} \textbf{\bibinfo{volume}{B513}},
  \bibinfo{pages}{147} (\bibinfo{year}{2001}{\natexlab{a}}),
  \eprint{gr-qc/0105038}.

\bibitem[{\citenamefont{Buonanno et~al.}(2007{\natexlab{a}})}]{Buonanno:2007pf}
\bibinfo{author}{\bibfnamefont{A.}~\bibnamefont{Buonanno}}
  \bibnamefont{et~al.}, \bibinfo{journal}{Phys. Rev.}
  \textbf{\bibinfo{volume}{D76}}, \bibinfo{pages}{104049}
  (\bibinfo{year}{2007}{\natexlab{a}}), \eprint{arXiv:0706.3732 [gr-qc]}.

\bibitem[{\citenamefont{Damour and Nagar}(2008)}]{Damour:2007yf}
\bibinfo{author}{\bibfnamefont{T.}~\bibnamefont{Damour}} \bibnamefont{and}
  \bibinfo{author}{\bibfnamefont{A.}~\bibnamefont{Nagar}},
  \bibinfo{journal}{Phys. Rev.} \textbf{\bibinfo{volume}{D77}},
  \bibinfo{pages}{024043} (\bibinfo{year}{2008}), \eprint{arXiv:0711.2628
  [gr-qc]}.

\bibitem[{\citenamefont{Damour et~al.}(2008{\natexlab{a}})\citenamefont{Damour,
  Nagar, Dorband, Pollney, and Rezzolla}}]{Damour:2007vq}
\bibinfo{author}{\bibfnamefont{T.}~\bibnamefont{Damour}},
  \bibinfo{author}{\bibfnamefont{A.}~\bibnamefont{Nagar}},
  \bibinfo{author}{\bibfnamefont{E.~N.} \bibnamefont{Dorband}},
  \bibinfo{author}{\bibfnamefont{D.}~\bibnamefont{Pollney}}, \bibnamefont{and}
  \bibinfo{author}{\bibfnamefont{L.}~\bibnamefont{Rezzolla}},
  \bibinfo{journal}{Phys. Rev.} \textbf{\bibinfo{volume}{D77}},
  \bibinfo{pages}{084017} (\bibinfo{year}{2008}{\natexlab{a}}),
  \eprint{0712.3003}.

\bibitem[{\citenamefont{Buonanno
  et~al.}(2007{\natexlab{b}})\citenamefont{Buonanno, Cook, and
  Pretorius}}]{Buonanno:2006ui}
\bibinfo{author}{\bibfnamefont{A.}~\bibnamefont{Buonanno}},
  \bibinfo{author}{\bibfnamefont{G.~B.} \bibnamefont{Cook}}, \bibnamefont{and}
  \bibinfo{author}{\bibfnamefont{F.}~\bibnamefont{Pretorius}},
  \bibinfo{journal}{Phys. Rev.} \textbf{\bibinfo{volume}{D75}},
  \bibinfo{pages}{124018} (\bibinfo{year}{2007}{\natexlab{b}}),
  \eprint{gr-qc/0610122}.

\bibitem[{\citenamefont{Damour and Nagar}(2007{\natexlab{b}})}]{Damour:2007cb}
\bibinfo{author}{\bibfnamefont{T.}~\bibnamefont{Damour}} \bibnamefont{and}
  \bibinfo{author}{\bibfnamefont{A.}~\bibnamefont{Nagar}},
  \bibinfo{journal}{Phys. Rev.} \textbf{\bibinfo{volume}{D76}},
  \bibinfo{pages}{044003} (\bibinfo{year}{2007}{\natexlab{b}}),
  \eprint{arXiv:0704.3550 [gr-qc]}.

\bibitem[{\citenamefont{Pan et~al.}(2008)}]{Pan:2007nw}
\bibinfo{author}{\bibfnamefont{Y.}~\bibnamefont{Pan}} \bibnamefont{et~al.},
  \bibinfo{journal}{Phys. Rev.} \textbf{\bibinfo{volume}{D77}},
  \bibinfo{pages}{024014} (\bibinfo{year}{2008}), \eprint{arXiv:0704.1964
  [gr-qc]}.

\bibitem[{\citenamefont{Damour and Gopakumar}(2006)}]{Damour:2006tr}
\bibinfo{author}{\bibfnamefont{T.}~\bibnamefont{Damour}} \bibnamefont{and}
  \bibinfo{author}{\bibfnamefont{A.}~\bibnamefont{Gopakumar}},
  \bibinfo{journal}{Phys. Rev.} \textbf{\bibinfo{volume}{D73}},
  \bibinfo{pages}{124006} (\bibinfo{year}{2006}), \eprint{gr-qc/0602117}.

\bibitem[{\citenamefont{Damour et~al.}(2008{\natexlab{b}})\citenamefont{Damour,
  Jaranowski, and Schafer}}]{Damour:2008qf}
\bibinfo{author}{\bibfnamefont{T.}~\bibnamefont{Damour}},
  \bibinfo{author}{\bibfnamefont{P.}~\bibnamefont{Jaranowski}},
  \bibnamefont{and} \bibinfo{author}{\bibfnamefont{G.}~\bibnamefont{Schafer}}
  (\bibinfo{year}{2008}{\natexlab{b}}), \eprint{arXiv:0803.0915 [gr-qc]}.

\bibitem[{\citenamefont{Boyle et~al.}(2007)}]{Boyle:2007ft}
\bibinfo{author}{\bibfnamefont{M.}~\bibnamefont{Boyle}} \bibnamefont{et~al.},
  \bibinfo{journal}{Phys. Rev.} \textbf{\bibinfo{volume}{D76}},
  \bibinfo{pages}{124038} (\bibinfo{year}{2007}), \eprint{arXiv:0710.0158
  [gr-qc]}.

\bibitem[{\citenamefont{Damour et~al.}(2001{\natexlab{b}})\citenamefont{Damour,
  Iyer, and Sathyaprakash}}]{Damour:2000zb}
\bibinfo{author}{\bibfnamefont{T.}~\bibnamefont{Damour}},
  \bibinfo{author}{\bibfnamefont{B.~R.} \bibnamefont{Iyer}}, \bibnamefont{and}
  \bibinfo{author}{\bibfnamefont{B.~S.} \bibnamefont{Sathyaprakash}},
  \bibinfo{journal}{Phys. Rev.} \textbf{\bibinfo{volume}{D63}},
  \bibinfo{pages}{044023} (\bibinfo{year}{2001}{\natexlab{b}}),
  \eprint{gr-qc/0010009}.

\bibitem[{\citenamefont{Br{\"u}gmann et~al.}(2008)\citenamefont{Br{\"u}gmann,
  Gonz{\'a}lez, Hannam, Husa, Sperhake, and Tichy}}]{Bruegmann:2006at}
\bibinfo{author}{\bibfnamefont{B.}~\bibnamefont{Br{\"u}gmann}},
  \bibinfo{author}{\bibfnamefont{J.~A.} \bibnamefont{Gonz{\'a}lez}},
  \bibinfo{author}{\bibfnamefont{M.}~\bibnamefont{Hannam}},
  \bibinfo{author}{\bibfnamefont{S.}~\bibnamefont{Husa}},
  \bibinfo{author}{\bibfnamefont{U.}~\bibnamefont{Sperhake}}, \bibnamefont{and}
  \bibinfo{author}{\bibfnamefont{W.}~\bibnamefont{Tichy}},
  \bibinfo{journal}{Phys. Rev. D} \textbf{\bibinfo{volume}{77}},
  \bibinfo{pages}{024027} (\bibinfo{year}{2008}),
  \bibinfo{note}{gr-qc/0610128}.

\bibitem[{\citenamefont{Husa et~al.}(2007{\natexlab{a}})\citenamefont{Husa,
  Gonzalez, Hannam, Brugmann, and Sperhake}}]{Husa:2007hp}
\bibinfo{author}{\bibfnamefont{S.}~\bibnamefont{Husa}},
  \bibinfo{author}{\bibfnamefont{J.~A.} \bibnamefont{Gonzalez}},
  \bibinfo{author}{\bibfnamefont{M.}~\bibnamefont{Hannam}},
  \bibinfo{author}{\bibfnamefont{B.}~\bibnamefont{Brugmann}}, \bibnamefont{and}
  \bibinfo{author}{\bibfnamefont{U.}~\bibnamefont{Sperhake}}
  (\bibinfo{year}{2007}{\natexlab{a}}), \eprint{arXiv:0706.0740 [gr-qc]}.

\bibitem[{\citenamefont{Hannam et~al.}(2008{\natexlab{a}})\citenamefont{Hannam,
  Husa, Gonz{\'a}lez, Sperhake, and Br{\"u}gmann}}]{Hannam:2007ik}
\bibinfo{author}{\bibfnamefont{M.}~\bibnamefont{Hannam}},
  \bibinfo{author}{\bibfnamefont{S.}~\bibnamefont{Husa}},
  \bibinfo{author}{\bibfnamefont{J.~A.} \bibnamefont{Gonz{\'a}lez}},
  \bibinfo{author}{\bibfnamefont{U.}~\bibnamefont{Sperhake}}, \bibnamefont{and}
  \bibinfo{author}{\bibfnamefont{B.}~\bibnamefont{Br{\"u}gmann}},
  \bibinfo{journal}{Phys. Rev. D} \textbf{\bibinfo{volume}{77}},
  \bibinfo{pages}{044020} (\bibinfo{year}{2008}{\natexlab{a}}),
  \eprint{arXiv:0706.1305 [gr-qc]}.

\bibitem[{\citenamefont{Brill and Lindquist}(1963)}]{Brill:1963yv}
\bibinfo{author}{\bibfnamefont{D.~S.} \bibnamefont{Brill}} \bibnamefont{and}
  \bibinfo{author}{\bibfnamefont{R.~W.} \bibnamefont{Lindquist}},
  \bibinfo{journal}{Phys. Rev.} \textbf{\bibinfo{volume}{131}},
  \bibinfo{pages}{471} (\bibinfo{year}{1963}).

\bibitem[{\citenamefont{Beig and O'Murchadha}(1994)}]{Beig94}
\bibinfo{author}{\bibfnamefont{R.}~\bibnamefont{Beig}} \bibnamefont{and}
  \bibinfo{author}{\bibfnamefont{N.}~\bibnamefont{O'Murchadha}},
  \bibinfo{journal}{Class. Quantum Grav.} \textbf{\bibinfo{volume}{11}},
  \bibinfo{pages}{419} (\bibinfo{year}{1994}).

\bibitem[{\citenamefont{Beig and Husa}(1994)}]{Beig:1994rp}
\bibinfo{author}{\bibfnamefont{R.}~\bibnamefont{Beig}} \bibnamefont{and}
  \bibinfo{author}{\bibfnamefont{S.}~\bibnamefont{Husa}},
  \bibinfo{journal}{Phys. Rev. D} \textbf{\bibinfo{volume}{50}},
  \bibinfo{pages}{R7116} (\bibinfo{year}{1994}), \eprint{gr-qc/9410003}.

\bibitem[{\citenamefont{Brandt and Br{\"u}gmann}(1997)}]{Brandt97b}
\bibinfo{author}{\bibfnamefont{S.}~\bibnamefont{Brandt}} \bibnamefont{and}
  \bibinfo{author}{\bibfnamefont{B.}~\bibnamefont{Br{\"u}gmann}},
  \bibinfo{journal}{Phys. Rev. Lett.} \textbf{\bibinfo{volume}{78}},
  \bibinfo{pages}{3606} (\bibinfo{year}{1997}), \eprint{gr-qc/9703066}.

\bibitem[{\citenamefont{Dain and Friedrich}(2001)}]{Dain01a}
\bibinfo{author}{\bibfnamefont{S.}~\bibnamefont{Dain}} \bibnamefont{and}
  \bibinfo{author}{\bibfnamefont{H.}~\bibnamefont{Friedrich}},
  \bibinfo{journal}{Comm. Math. Phys.} \textbf{\bibinfo{volume}{222}},
  \bibinfo{pages}{569} (\bibinfo{year}{2001}), \bibinfo{note}{gr-qc/0102047}.

\bibitem[{\citenamefont{Hannam et~al.}(2007{\natexlab{a}})\citenamefont{Hannam,
  Husa, Pollney, Brugmann, and O'Murchadha}}]{Hannam:2006vv}
\bibinfo{author}{\bibfnamefont{M.}~\bibnamefont{Hannam}},
  \bibinfo{author}{\bibfnamefont{S.}~\bibnamefont{Husa}},
  \bibinfo{author}{\bibfnamefont{D.}~\bibnamefont{Pollney}},
  \bibinfo{author}{\bibfnamefont{B.}~\bibnamefont{Brugmann}}, \bibnamefont{and}
  \bibinfo{author}{\bibfnamefont{N.}~\bibnamefont{O'Murchadha}},
  \bibinfo{journal}{Phys. Rev. Lett.} \textbf{\bibinfo{volume}{99}},
  \bibinfo{pages}{241102} (\bibinfo{year}{2007}{\natexlab{a}}),
  \eprint{gr-qc/0606099}.

\bibitem[{\citenamefont{Hannam et~al.}(2007{\natexlab{b}})\citenamefont{Hannam,
  Husa, {\'O~Murchadha}, Br{\"u}gmann, Gonz{\'a}lez, and
  Sperhake}}]{Hannam:2006xw}
\bibinfo{author}{\bibfnamefont{M.}~\bibnamefont{Hannam}},
  \bibinfo{author}{\bibfnamefont{S.}~\bibnamefont{Husa}},
  \bibinfo{author}{\bibfnamefont{N.}~\bibnamefont{{\'O~Murchadha}}},
  \bibinfo{author}{\bibfnamefont{B.}~\bibnamefont{Br{\"u}gmann}},
  \bibinfo{author}{\bibfnamefont{J.~A.} \bibnamefont{Gonz{\'a}lez}},
  \bibnamefont{and} \bibinfo{author}{\bibfnamefont{U.}~\bibnamefont{Sperhake}},
  \bibinfo{journal}{Journal of Physics: Conference series} p.
  \bibinfo{pages}{012047} (\bibinfo{year}{2007}{\natexlab{b}}),
  \eprint{arXiv:gr-qc/0612097}.

\bibitem[{\citenamefont{Hannam et~al.}(2008{\natexlab{b}})\citenamefont{Hannam,
  Husa, Ohme, Brugmann, and O'Murchadha}}]{Hannam:2008sg}
\bibinfo{author}{\bibfnamefont{M.}~\bibnamefont{Hannam}},
  \bibinfo{author}{\bibfnamefont{S.}~\bibnamefont{Husa}},
  \bibinfo{author}{\bibfnamefont{F.}~\bibnamefont{Ohme}},
  \bibinfo{author}{\bibfnamefont{B.}~\bibnamefont{Brugmann}}, \bibnamefont{and}
  \bibinfo{author}{\bibfnamefont{N.}~\bibnamefont{O'Murchadha}}
  (\bibinfo{year}{2008}{\natexlab{b}}), \eprint{arXiv:0804.0628}.

\bibitem[{\citenamefont{Bowen and York}(1980)}]{Bowen80}
\bibinfo{author}{\bibfnamefont{J.~M.} \bibnamefont{Bowen}} \bibnamefont{and}
  \bibinfo{author}{\bibfnamefont{J.~W.} \bibnamefont{York}},
  \bibinfo{journal}{Phys. Rev. D} \textbf{\bibinfo{volume}{21}},
  \bibinfo{pages}{2047} (\bibinfo{year}{1980}).

\bibitem[{\citenamefont{Dain}(2001)}]{Dain00}
\bibinfo{author}{\bibfnamefont{S.}~\bibnamefont{Dain}}, \bibinfo{journal}{Phys.
  Rev. Lett.} \textbf{\bibinfo{volume}{87}}, \bibinfo{pages}{121102}
  (\bibinfo{year}{2001}), \bibinfo{note}{gr-qc/0012023}.

\bibitem[{\citenamefont{Hannam et~al.}(2007{\natexlab{c}})\citenamefont{Hannam,
  Husa, Br{\"u}gmann, Gonzalez, and Sperhake}}]{Hannam:2006zt}
\bibinfo{author}{\bibfnamefont{M.}~\bibnamefont{Hannam}},
  \bibinfo{author}{\bibfnamefont{S.}~\bibnamefont{Husa}},
  \bibinfo{author}{\bibfnamefont{B.}~\bibnamefont{Br{\"u}gmann}},
  \bibinfo{author}{\bibfnamefont{J.~A.} \bibnamefont{Gonzalez}},
  \bibnamefont{and} \bibinfo{author}{\bibfnamefont{U.}~\bibnamefont{Sperhake}},
  \bibinfo{journal}{Class. Quantum Grav.} \textbf{\bibinfo{volume}{24}},
  \bibinfo{pages}{S15} (\bibinfo{year}{2007}{\natexlab{c}}),
  \eprint{arXiv:gr-qc/0612001}.

\bibitem[{\citenamefont{Schnetter et~al.}(2006)\citenamefont{Schnetter,
  Krishnan, and Beyer}}]{Schnetter:2006yt}
\bibinfo{author}{\bibfnamefont{E.}~\bibnamefont{Schnetter}},
  \bibinfo{author}{\bibfnamefont{B.}~\bibnamefont{Krishnan}}, \bibnamefont{and}
  \bibinfo{author}{\bibfnamefont{F.}~\bibnamefont{Beyer}},
  \bibinfo{journal}{Phys. Rev. D} \textbf{\bibinfo{volume}{74}},
  \bibinfo{pages}{024028} (\bibinfo{year}{2006}), \eprint{gr-qc/0604015}.

\bibitem[{\citenamefont{Dennison et~al.}(2006)\citenamefont{Dennison,
  Baumgarte, and Pfeiffer}}]{Dennison:2006nq}
\bibinfo{author}{\bibfnamefont{K.~A.} \bibnamefont{Dennison}},
  \bibinfo{author}{\bibfnamefont{T.~W.} \bibnamefont{Baumgarte}},
  \bibnamefont{and} \bibinfo{author}{\bibfnamefont{H.~P.}
  \bibnamefont{Pfeiffer}}, \bibinfo{journal}{Phys. Rev.}
  \textbf{\bibinfo{volume}{D74}}, \bibinfo{pages}{064016}
  (\bibinfo{year}{2006}), \eprint{gr-qc/0606037}.

\bibitem[{\citenamefont{Tichy and Br{\"u}gmann}(2004)}]{Tichy:2003qi}
\bibinfo{author}{\bibfnamefont{W.}~\bibnamefont{Tichy}} \bibnamefont{and}
  \bibinfo{author}{\bibfnamefont{B.}~\bibnamefont{Br{\"u}gmann}},
  \bibinfo{journal}{Phys. Rev. D} \textbf{\bibinfo{volume}{69}},
  \bibinfo{pages}{024006} (\bibinfo{year}{2004}), \eprint{gr-qc/0307027}.

\bibitem[{\citenamefont{Christodoulou}(1970)}]{Christodoulou:1970wf}
\bibinfo{author}{\bibfnamefont{D.}~\bibnamefont{Christodoulou}},
  \bibinfo{journal}{Phys. Rev. Lett.} \textbf{\bibinfo{volume}{25}},
  \bibinfo{pages}{1596} (\bibinfo{year}{1970}).

\bibitem[{\citenamefont{Christodoulou and
  Ruffini}(1971)}]{Christodoulou:1972kt}
\bibinfo{author}{\bibfnamefont{D.}~\bibnamefont{Christodoulou}}
  \bibnamefont{and} \bibinfo{author}{\bibfnamefont{R.}~\bibnamefont{Ruffini}},
  \bibinfo{journal}{Phys. Rev.} \textbf{\bibinfo{volume}{D4}},
  \bibinfo{pages}{3552} (\bibinfo{year}{1971}).

\bibitem[{\citenamefont{Ansorg et~al.}(2004)\citenamefont{Ansorg, Br{\"u}gmann,
  and Tichy}}]{Ansorg:2004ds}
\bibinfo{author}{\bibfnamefont{M.}~\bibnamefont{Ansorg}},
  \bibinfo{author}{\bibfnamefont{B.}~\bibnamefont{Br{\"u}gmann}},
  \bibnamefont{and} \bibinfo{author}{\bibfnamefont{W.}~\bibnamefont{Tichy}},
  \bibinfo{journal}{Phys. Rev. D} \textbf{\bibinfo{volume}{70}},
  \bibinfo{pages}{064011} (\bibinfo{year}{2004}), \eprint{gr-qc/0404056}.

\bibitem[{\citenamefont{Husa et~al.}(2007{\natexlab{b}})\citenamefont{Husa,
  Hannam, Gonz{\'a}lez, Sperha~ke, and Br{\"u}gmann}}]{Husa:2007ec}
\bibinfo{author}{\bibfnamefont{S.}~\bibnamefont{Husa}},
  \bibinfo{author}{\bibfnamefont{M.}~\bibnamefont{Hannam}},
  \bibinfo{author}{\bibfnamefont{J.~A.} \bibnamefont{Gonz{\'a}lez}},
  \bibinfo{author}{\bibfnamefont{U.}~\bibnamefont{Sperha~ke}},
  \bibnamefont{and}
  \bibinfo{author}{\bibfnamefont{B.}~\bibnamefont{Br{\"u}gmann}},
  \bibinfo{journal}{Phys. Rev. D}  (\bibinfo{year}{2007}{\natexlab{b}}),
  \eprint{arXiv:0706.0904 [gr-qc]}.

\bibitem[{\citenamefont{Blanchet et~al.}(2002)\citenamefont{Blanchet, Faye,
  Iyer, and Joguet}}]{Blanchet:2001ax}
\bibinfo{author}{\bibfnamefont{L.}~\bibnamefont{Blanchet}},
  \bibinfo{author}{\bibfnamefont{G.}~\bibnamefont{Faye}},
  \bibinfo{author}{\bibfnamefont{B.~R.} \bibnamefont{Iyer}}, \bibnamefont{and}
  \bibinfo{author}{\bibfnamefont{B.}~\bibnamefont{Joguet}},
  \bibinfo{journal}{Phys. Rev.} \textbf{\bibinfo{volume}{D65}},
  \bibinfo{pages}{061501} (\bibinfo{year}{2002}), \eprint{gr-qc/0105099}.

\bibitem[{\citenamefont{Blanchet et~al.}(2004)\citenamefont{Blanchet, Damour,
  Esposito-Farese, and Iyer}}]{Blanchet:2004ek}
\bibinfo{author}{\bibfnamefont{L.}~\bibnamefont{Blanchet}},
  \bibinfo{author}{\bibfnamefont{T.}~\bibnamefont{Damour}},
  \bibinfo{author}{\bibfnamefont{G.}~\bibnamefont{Esposito-Farese}},
  \bibnamefont{and} \bibinfo{author}{\bibfnamefont{B.~R.} \bibnamefont{Iyer}},
  \bibinfo{journal}{Phys. Rev. Lett.} \textbf{\bibinfo{volume}{93}},
  \bibinfo{pages}{091101} (\bibinfo{year}{2004}), \eprint{gr-qc/0406012}.

\bibitem[{\citenamefont{Shibata and Nakamura}(1995)}]{Shibata95}
\bibinfo{author}{\bibfnamefont{M.}~\bibnamefont{Shibata}} \bibnamefont{and}
  \bibinfo{author}{\bibfnamefont{T.}~\bibnamefont{Nakamura}},
  \bibinfo{journal}{Phys. Rev. D} \textbf{\bibinfo{volume}{52}},
  \bibinfo{pages}{5428} (\bibinfo{year}{1995}).

\bibitem[{\citenamefont{Baumgarte and Shapiro}(1998)}]{Baumgarte99}
\bibinfo{author}{\bibfnamefont{T.~W.} \bibnamefont{Baumgarte}}
  \bibnamefont{and} \bibinfo{author}{\bibfnamefont{S.~L.}
  \bibnamefont{Shapiro}}, \bibinfo{journal}{Phys. Rev. D}
  \textbf{\bibinfo{volume}{59}}, \bibinfo{pages}{024007}
  (\bibinfo{year}{1998}), \eprint{gr-qc/9810065}.

\bibitem[{\citenamefont{{Bona} et~al.}(1995)\citenamefont{{Bona}, {Mass{\'o}},
  {Seidel}, and {Stela}}}]{Bona95b}
\bibinfo{author}{\bibfnamefont{C.}~\bibnamefont{{Bona}}},
  \bibinfo{author}{\bibfnamefont{J.}~\bibnamefont{{Mass{\'o}}}},
  \bibinfo{author}{\bibfnamefont{E.}~\bibnamefont{{Seidel}}}, \bibnamefont{and}
  \bibinfo{author}{\bibfnamefont{J.}~\bibnamefont{{Stela}}},
  \bibinfo{journal}{Phys. Rev. Lett.} \textbf{\bibinfo{volume}{75}},
  \bibinfo{pages}{600} (\bibinfo{year}{1995}), \eprint{gr-qc/9412071}.

\bibitem[{\citenamefont{Alcubierre et~al.}(2001)\citenamefont{Alcubierre,
  Br{\"u}gmann, Pollney, Seidel, and Takahashi}}]{Alcubierre01a}
\bibinfo{author}{\bibfnamefont{M.}~\bibnamefont{Alcubierre}},
  \bibinfo{author}{\bibfnamefont{B.}~\bibnamefont{Br{\"u}gmann}},
  \bibinfo{author}{\bibfnamefont{D.}~\bibnamefont{Pollney}},
  \bibinfo{author}{\bibfnamefont{E.}~\bibnamefont{Seidel}}, \bibnamefont{and}
  \bibinfo{author}{\bibfnamefont{R.}~\bibnamefont{Takahashi}},
  \bibinfo{journal}{Phys. Rev. D} \textbf{\bibinfo{volume}{64}},
  \bibinfo{pages}{061501(R)} (\bibinfo{year}{2001}), \eprint{gr-qc/0104020}.

\bibitem[{\citenamefont{Alcubierre et~al.}(2003)\citenamefont{Alcubierre,
  Br{\"u}gmann, Diener, Koppitz, Pollney, Seidel, and
  Takahashi}}]{Alcubierre02a}
\bibinfo{author}{\bibfnamefont{M.}~\bibnamefont{Alcubierre}},
  \bibinfo{author}{\bibfnamefont{B.}~\bibnamefont{Br{\"u}gmann}},
  \bibinfo{author}{\bibfnamefont{P.}~\bibnamefont{Diener}},
  \bibinfo{author}{\bibfnamefont{M.}~\bibnamefont{Koppitz}},
  \bibinfo{author}{\bibfnamefont{D.}~\bibnamefont{Pollney}},
  \bibinfo{author}{\bibfnamefont{E.}~\bibnamefont{Seidel}}, \bibnamefont{and}
  \bibinfo{author}{\bibfnamefont{R.}~\bibnamefont{Takahashi}},
  \bibinfo{journal}{Phys. Rev. D} \textbf{\bibinfo{volume}{67}},
  \bibinfo{pages}{084023} (\bibinfo{year}{2003}), \eprint{gr-qc/0206072}.

\bibitem[{\citenamefont{Newman and Penrose}(1962)}]{Newman62a}
\bibinfo{author}{\bibfnamefont{E.~T.} \bibnamefont{Newman}} \bibnamefont{and}
  \bibinfo{author}{\bibfnamefont{R.}~\bibnamefont{Penrose}},
  \bibinfo{journal}{J. Math. Phys.} \textbf{\bibinfo{volume}{3}},
  \bibinfo{pages}{566} (\bibinfo{year}{1962}), \bibinfo{note}{erratum in J.
  Math. Phys. 4, 998 (1963)}.

\bibitem[{\citenamefont{Stewart}(1990)}]{Stewart:1990uf}
\bibinfo{author}{\bibfnamefont{J.~M.} \bibnamefont{Stewart}},
  \emph{\bibinfo{title}{Advanced general relativity}} (\bibinfo{year}{1990}),
  \bibinfo{note}{cambridge University Press, Cambridge}.

\bibitem[{\citenamefont{Kidder}(2008{\natexlab{a}})}]{KidderJena}
\bibinfo{author}{\bibfnamefont{L.~E.} \bibnamefont{Kidder}},
  \bibinfo{journal}{{Report presented at PN2008 Jena conference,
  http://wwwsfb.tpi.uni-jena.de/Events/PN2008/PN2008\_Program.shtm }}
  (\bibinfo{year}{2008}{\natexlab{a}}).

\bibitem[{\citenamefont{Berti et~al.}(2006)\citenamefont{Berti, Cardoso, and
  Will}}]{Berti:2005ys}
\bibinfo{author}{\bibfnamefont{E.}~\bibnamefont{Berti}},
  \bibinfo{author}{\bibfnamefont{V.}~\bibnamefont{Cardoso}}, \bibnamefont{and}
  \bibinfo{author}{\bibfnamefont{C.~M.} \bibnamefont{Will}},
  \bibinfo{journal}{Phys. Rev.} \textbf{\bibinfo{volume}{D73}},
  \bibinfo{pages}{064030} (\bibinfo{year}{2006}), \eprint{gr-qc/0512160}.

\bibitem[{\citenamefont{Berti et~al.}(2007)}]{Berti:2007fi}
\bibinfo{author}{\bibfnamefont{E.}~\bibnamefont{Berti}} \bibnamefont{et~al.},
  \bibinfo{journal}{Phys. Rev.} \textbf{\bibinfo{volume}{D76}},
  \bibinfo{pages}{064034} (\bibinfo{year}{2007}), \eprint{gr-qc/0703053}.

\bibitem[{\citenamefont{Damour}(2008)}]{Damour:2008yg}
\bibinfo{author}{\bibfnamefont{T.}~\bibnamefont{Damour}}
  (\bibinfo{year}{2008}), \eprint{arXiv:0802.4047 [gr-qc]}.

\bibitem[{\citenamefont{Press and Teukolsky}(1973)}]{PT1973}
\bibinfo{author}{\bibfnamefont{W.}~\bibnamefont{Press}} \bibnamefont{and}
  \bibinfo{author}{\bibfnamefont{S.~A.} \bibnamefont{Teukolsky}},
  \bibinfo{journal}{Astrophys. J.} \textbf{\bibinfo{volume}{185}},
  \bibinfo{pages}{649} (\bibinfo{year}{1973}).

\bibitem[{\citenamefont{Hannam et~al.}(2007{\natexlab{d}})\citenamefont{Hannam,
  Husa, Br{\"u}gmann, and Gopakumar}}]{Hannam:2007wf}
\bibinfo{author}{\bibfnamefont{M.}~\bibnamefont{Hannam}},
  \bibinfo{author}{\bibfnamefont{S.}~\bibnamefont{Husa}},
  \bibinfo{author}{\bibfnamefont{B.}~\bibnamefont{Br{\"u}gmann}},
  \bibnamefont{and} \bibinfo{author}{\bibfnamefont{A.}~\bibnamefont{Gopakumar}}
  (\bibinfo{year}{2007}{\natexlab{d}}), \eprint{arXiv:0712.3787 [gr-qc]}.

\bibitem[{\citenamefont{Nagar et~al.}(2007)\citenamefont{Nagar, Damour, and
  Tartaglia}}]{Nagar:2006xv}
\bibinfo{author}{\bibfnamefont{A.}~\bibnamefont{Nagar}},
  \bibinfo{author}{\bibfnamefont{T.}~\bibnamefont{Damour}}, \bibnamefont{and}
  \bibinfo{author}{\bibfnamefont{A.}~\bibnamefont{Tartaglia}},
  \bibinfo{journal}{Class. Quant. Grav.} \textbf{\bibinfo{volume}{24}},
  \bibinfo{pages}{S109} (\bibinfo{year}{2007}), \eprint{gr-qc/0612096}.

\bibitem[{\citenamefont{Nagar and Rezzolla}(2005)}]{Nagar:2005ea}
\bibinfo{author}{\bibfnamefont{A.}~\bibnamefont{Nagar}} \bibnamefont{and}
  \bibinfo{author}{\bibfnamefont{L.}~\bibnamefont{Rezzolla}},
  \bibinfo{journal}{Class. Quant. Grav.} \textbf{\bibinfo{volume}{22}},
  \bibinfo{pages}{R167} (\bibinfo{year}{2005}), \eprint{gr-qc/0502064}.

\bibitem[{\citenamefont{Kidder}(2008{\natexlab{b}})}]{Kidder:2007rt}
\bibinfo{author}{\bibfnamefont{L.~E.} \bibnamefont{Kidder}},
  \bibinfo{journal}{Phys. Rev.} \textbf{\bibinfo{volume}{D77}},
  \bibinfo{pages}{044016} (\bibinfo{year}{2008}{\natexlab{b}}),
  \eprint{arXiv:0710.0614 [gr-qc]}.

\bibitem[{\citenamefont{Pollney et~al.}(2007)}]{Pollney:2007ss}
\bibinfo{author}{\bibfnamefont{D.}~\bibnamefont{Pollney}} \bibnamefont{et~al.},
  \bibinfo{journal}{Phys. Rev.} \textbf{\bibinfo{volume}{D76}},
  \bibinfo{pages}{124002} (\bibinfo{year}{2007}), \eprint{arXiv:0707.2559
  [gr-qc]}.

\bibitem[{\citenamefont{Schnittman et~al.}(2008)}]{Schnittman:2007ij}
\bibinfo{author}{\bibfnamefont{J.~D.} \bibnamefont{Schnittman}}
  \bibnamefont{et~al.}, \bibinfo{journal}{Phys. Rev.}
  \textbf{\bibinfo{volume}{D77}}, \bibinfo{pages}{044031}
  (\bibinfo{year}{2008}), \eprint{arXiv:0707.0301 [gr-qc]}.

\bibitem[{\citenamefont{Baker et~al.}(2007)\citenamefont{Baker, van Meter,
  McWilliams, Centrella, and Kelly}}]{Baker:2006ha}
\bibinfo{author}{\bibfnamefont{J.~G.} \bibnamefont{Baker}},
  \bibinfo{author}{\bibfnamefont{J.~R.} \bibnamefont{van Meter}},
  \bibinfo{author}{\bibfnamefont{S.~T.} \bibnamefont{McWilliams}},
  \bibinfo{author}{\bibfnamefont{J.}~\bibnamefont{Centrella}},
  \bibnamefont{and} \bibinfo{author}{\bibfnamefont{B.~J.} \bibnamefont{Kelly}},
  \bibinfo{journal}{Phys. Rev. Lett.} \textbf{\bibinfo{volume}{99}},
  \bibinfo{pages}{181101} (\bibinfo{year}{2007}), \eprint{gr-qc/0612024}.

\bibitem[{\citenamefont{Gopakumar et~al.}(2007)\citenamefont{Gopakumar, Hannam,
  Husa, and Br{\"u}gmann}}]{Gopakumar:2007vh}
\bibinfo{author}{\bibfnamefont{A.}~\bibnamefont{Gopakumar}},
  \bibinfo{author}{\bibfnamefont{M.}~\bibnamefont{Hannam}},
  \bibinfo{author}{\bibfnamefont{S.}~\bibnamefont{Husa}}, \bibnamefont{and}
  \bibinfo{author}{\bibfnamefont{B.}~\bibnamefont{Br{\"u}gmann}}
  (\bibinfo{year}{2007}), \eprint{arXiv:0712.3737 [gr-qc]}.

\bibitem[{\citenamefont{Arun et~al.}(2004)\citenamefont{Arun, Blanchet, Iyer,
  and Qusailah}}]{Arun:2004ff}
\bibinfo{author}{\bibfnamefont{K.~G.} \bibnamefont{Arun}},
  \bibinfo{author}{\bibfnamefont{L.}~\bibnamefont{Blanchet}},
  \bibinfo{author}{\bibfnamefont{B.~R.} \bibnamefont{Iyer}}, \bibnamefont{and}
  \bibinfo{author}{\bibfnamefont{M.~S.~S.} \bibnamefont{Qusailah}},
  \bibinfo{journal}{Class. Quant. Grav.} \textbf{\bibinfo{volume}{21}},
  \bibinfo{pages}{3771} (\bibinfo{year}{2004}), \eprint{gr-qc/0404085}.

\bibitem[{\citenamefont{Cutler et~al.}(1993)\citenamefont{Cutler, Poisson,
  Sussman, and Finn}}]{Cutler:1993vq}
\bibinfo{author}{\bibfnamefont{C.}~\bibnamefont{Cutler}},
  \bibinfo{author}{\bibfnamefont{E.}~\bibnamefont{Poisson}},
  \bibinfo{author}{\bibfnamefont{G.~J.} \bibnamefont{Sussman}},
  \bibnamefont{and} \bibinfo{author}{\bibfnamefont{L.~S.} \bibnamefont{Finn}},
  \bibinfo{journal}{Phys. Rev.} \textbf{\bibinfo{volume}{D47}},
  \bibinfo{pages}{1511} (\bibinfo{year}{1993}).

\bibitem[{\citenamefont{Poisson}(1995)}]{Poisson:1995vs}
\bibinfo{author}{\bibfnamefont{E.}~\bibnamefont{Poisson}},
  \bibinfo{journal}{Phys. Rev.} \textbf{\bibinfo{volume}{D52}},
  \bibinfo{pages}{5719} (\bibinfo{year}{1995}), \eprint{gr-qc/9505030}.

\bibitem[{\citenamefont{Brady et~al.}(1998)\citenamefont{Brady, Creighton, and
  Thorne}}]{Brady:1998du}
\bibinfo{author}{\bibfnamefont{P.~R.} \bibnamefont{Brady}},
  \bibinfo{author}{\bibfnamefont{J.~D.~E.} \bibnamefont{Creighton}},
  \bibnamefont{and} \bibinfo{author}{\bibfnamefont{K.~S.}
  \bibnamefont{Thorne}}, \bibinfo{journal}{Phys. Rev.}
  \textbf{\bibinfo{volume}{D58}}, \bibinfo{pages}{061501}
  (\bibinfo{year}{1998}), \eprint{gr-qc/9804057}.

\end{thebibliography}


\end{document}